\documentclass[english,onecolumn,draftclsnofoot,12pt]{IEEEtran}
\usepackage[T1]{fontenc}
\usepackage[latin9]{inputenc}
\usepackage{babel}
\usepackage{array}
\usepackage{verbatim}
\usepackage{float}
\usepackage{url}
\usepackage{multirow}
\usepackage{amsmath}
\usepackage{amsthm}
\usepackage{amssymb}
\usepackage{graphicx}
\usepackage[unicode=true,pdfusetitle,
 bookmarks=true,bookmarksnumbered=false,bookmarksopen=false,
 breaklinks=false,pdfborder={0 0 1},backref=false,colorlinks=false]
 {hyperref}

\makeatletter

\providecommand{\tabularnewline}{\\}

\theoremstyle{plain}
\newtheorem{thm}{\protect\theoremname}
\theoremstyle{remark}
\newtheorem{rem}{\protect\remarkname}
\theoremstyle{plain}
\newtheorem{fact}{\protect\factname}
\theoremstyle{plain}
\newtheorem{claim}{\protect\claimname}

\usepackage{bbm}

\makeatother

\providecommand{\claimname}{Claim}
\providecommand{\factname}{Fact}
\providecommand{\remarkname}{Remark}
\providecommand{\theoremname}{Theorem}

\begin{document}
\global\long\def\expect#1{\mathbb{E}\left[#1\right]}%

\global\long\def\abs#1{\left\lvert #1\right\lvert }%

\global\long\def\twonorm#1{\left\Vert #1\right\Vert }%

\global\long\def\brac#1{\left(#1\right)}%

\global\long\def\lgbrac#1{\log\left(#1\right)}%

\global\long\def\lnbrac#1{\ln\left(#1\right)}%

\global\long\def\lghbrac#1{\log\left(2\pi e\brac{#1}\right)}%

\global\long\def\cbrac#1{\left\{  #1\right\}  }%

\global\long\def\rline#1{\left.#1\right| }%

\global\long\def\sbrac#1{\left[#1\right] }%

\global\long\def\Det#1{\left|#1\right|}%

\global\long\def\prob#1{\mathbb{P}\brac{#1}}%

\global\long\def\eqdof{\doteq}%

\global\long\def\leqdof{\overset{.}{\leq}}%

\global\long\def\geqdof{\overset{.}{\geq}}%

\global\long\def\union{\bigcup}%

\global\long\def\inter{\bigcap}%

\global\long\def\real{\mathbb{R}}%

\global\long\def\tran{\mathsf{Tran}}%

\global\long\def\idty{\mathbbm{1}}%

\global\long\def\snr{\mathsf{SNR}}%

\global\long\def\inr{\mathsf{INR}}%

\global\long\def\X{\boldsymbol{X}}%

\global\long\def\U{\boldsymbol{U}}%

\global\long\def\Y{\boldsymbol{Y}}%

\global\long\def\W{\boldsymbol{W}}%

\global\long\def\Z{\boldsymbol{Z}}%

\global\long\def\S{\boldsymbol{S}}%

\global\long\def\x{\boldsymbol{x}}%

\global\long\def\y{\boldsymbol{y}}%

\global\long\def\u{\boldsymbol{u}}%

\global\long\def\w{\boldsymbol{w}}%

\global\long\def\z{\boldsymbol{z}}%

\global\long\def\G{\boldsymbol{G}}%

\global\long\def\g{\boldsymbol{g}}%

\global\long\def\ts{\boldsymbol{\Lambda}}%

\global\long\def\Q{\boldsymbol{Q}}%

\global\long\def\Z{\boldsymbol{Z}}%

\global\long\def\pderiv#1#2{\frac{\partial#1}{\partial#2}}%

\newcommand{\etal}{{\it et al.}~}

\allowdisplaybreaks

\newif\ifarxiv

\arxivtrue
\title{On the Generalized Degrees of Freedom of the Noncoherent Interference
Channel\thanks{This work was supported in part by NSF grant 1514531, UC-NL grant
LFR-18-548554 and a gift from Guru Krupa Foundation.}}
\author{Joyson Sebastian, Suhas Diggavi}
\maketitle
\begin{abstract}
We study the generalized degrees of freedom (gDoF) of the block-fading
\emph{noncoherent} $2$-user interference channel (IC) with a coherence
time of $T$ symbol durations and symmetric fading statistics. We
demonstrate that a standard training-based scheme for the noncoherent
IC is suboptimal in several regimes. We study and analyze several
alternate schemes: the first is a new noncoherent scheme using rate-splitting.
We also consider a scheme that treats interference-as-noise (TIN)
and a time division multiplexing (TDM) scheme. We show that a standard
training-based scheme for the noncoherent IC is outperformed by one
of these schemes in several regimes: our results demonstrate that
in the very weak interference regime, the TIN scheme is the best;
in the strong interference regime, the TDM scheme and the noncoherent
rate-splitting scheme give better performance; in other cases either
of the TIN, TDM or noncoherent rate-splitting scheme could be preferred.
We also study the noncoherent IC with feedback and propose another
noncoherent rate-splitting scheme. Again for the feedback case, our
results demonstrate that a standard training-based scheme can be outperformed
by other schemes.
\end{abstract}

\section{Introduction\label{sec:Introduction}}

Noncoherent wireless channels where neither the transmitter nor the
receiver knows the channel \cite{marzetta1999capacity,Abou_Faycal_noncoherent,Zheng_Tse_Grassmann_MIMO,Koch2013,Joyson_noncoh_diamond_journal}
have been studied for point-to-point communication systems. To the
best of our knowledge, the \emph{noncoherent} interference channel
(IC) has not been studied from an information theoretic viewpoint.
In this paper, we consider the noncoherent $2$-user IC with symmetric
statistics and study the generalized degrees of freedom (gDoF) region
as a first step towards understanding its capacity region.
\begin{figure}
\begin{centering}
\includegraphics[scale=0.6]{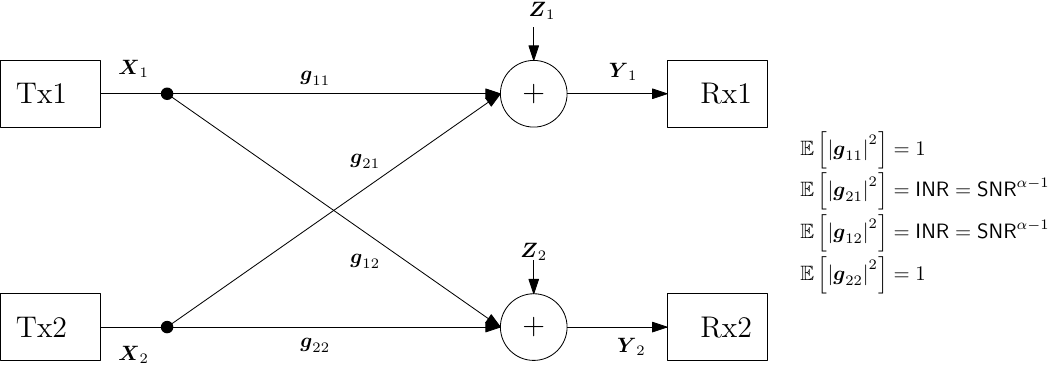}
\par\end{centering}
\caption{\label{fig:nonfeedback_ic}The channel model without feedback.}
\end{figure}

Our model is illustrated in Figure \ref{fig:nonfeedback_ic}. We have
two transmitters, each with its own intended receiver. The transmitted
signals are multiplied by random fading channels. Each receiver receives
a sum of signals from both  transmitters and with additive white Gaussian
noise. In our model, the power $P$ at transmitters is set to be equal
to the average signal-to-noise ratio\footnote{We use the abbreviation SNR for the \emph{average} signal-to-noise
ratio in the context of fading channels and not for the (instantaneous)
signal-to-noise ratio. Similarly, we use the abbreviation INR for
the \emph{average }interference-to-noise ratio.} ($\snr$), the direct channel links are set to be of unit power,
and the additive noise at the receivers are also set to be of unit
power. We consider the symmetric IC interfering links with average
power equal to the average interference-to-noise ratio ($\inr$) with
$\inr=\snr^{\alpha-1}$. The parameter $\alpha$ (interference level)
is used to capture the relative strength of the interference at the
receiver:
\[
\alpha=\frac{\lgbrac{P\times\expect{\boldsymbol{g}_{21}}^{2}}}{\lgbrac{P\times\expect{\boldsymbol{g}_{11}}^{2}}}=\frac{\lgbrac{P\times\expect{\boldsymbol{g}_{12}}^{2}}}{\lgbrac{P\times\expect{\boldsymbol{g}_{22}}^{2}}}=\frac{\lgbrac{P\times\inr}}{\lgbrac{P\times1}}.
\]
The interference level is typically less than one. When $\alpha<1/2$,
we have the very weak interference regime. When $1/2<\alpha<1$, we
have the weak interference regime. For mathematical completeness,
we also consider $\alpha>1$ which is the strong interference regime.
We consider a block fading model where the channels remain constant
for a coherence time of $T$ symbol durations and hence we model our
system with vectors of size $T$. We have
\begin{eqnarray}
\Y_{1} & = & \boldsymbol{g}_{11}\X_{1}+\boldsymbol{g}_{21}\X_{2}+\boldsymbol{Z}_{1},\\
\Y_{2} & = & \boldsymbol{g}_{12}\X_{1}+\boldsymbol{g}_{22}\X_{2}+\boldsymbol{Z}_{2},
\end{eqnarray}
where the $\X_{i}$, $\Y_{i}$, $\boldsymbol{Z}_{i}$ with $i\in\{1,2\}$
are transmitted symbols, received symbols and noise at receivers respectively.
The variables $\X_{i}$, $\Y_{i}$, $\boldsymbol{Z}_{i}$ with $i\in\{1,2\}$
are $1\times T$ vectors. The noise $\boldsymbol{Z}_{1},\boldsymbol{Z}_{2}$
are independent of each other and their realizations are i.i.d. across
time. The entries of the vector $\boldsymbol{Z}_{i}$ are i.i.d. according
to $\mathcal{CN}\brac{0,1}$. The fading channels are indicated by
scalar random variables $\boldsymbol{g}_{ij}$ with $i,j\in\{1,2\}$.
The realizations of $\boldsymbol{g}_{ij}$ for any fixed $\brac{i,j}$,
$i,j\in\{1,2\}$ are i.i.d. across time, and the realizations for
different $\brac{i,j}$ are independent. We consider the case with
symmetric fading statistics $\boldsymbol{g}_{11}\sim\boldsymbol{g}_{22}\sim\mathcal{CN}\brac{0,1}$,
$\boldsymbol{g}_{12}\sim\boldsymbol{g}_{21}\sim\mathcal{CN}\brac{0,\inr},\inr=\snr^{\alpha-1}$.
Neither the receivers nor the transmitters have knowledge of any of
the realizations of $\boldsymbol{g}_{ij}$, but the channel statistics
are known to all the receivers and transmitters. The average power
constraint on the transmitted signals gives
\begin{equation}
\frac{1}{T}\expect{\abs{\X_{i}}^{2}}=P=\snr\label{eq:power_constraint}
\end{equation}
for $i\in\{1,2\}$.
\begin{figure}
\begin{centering}
\includegraphics[scale=0.6]{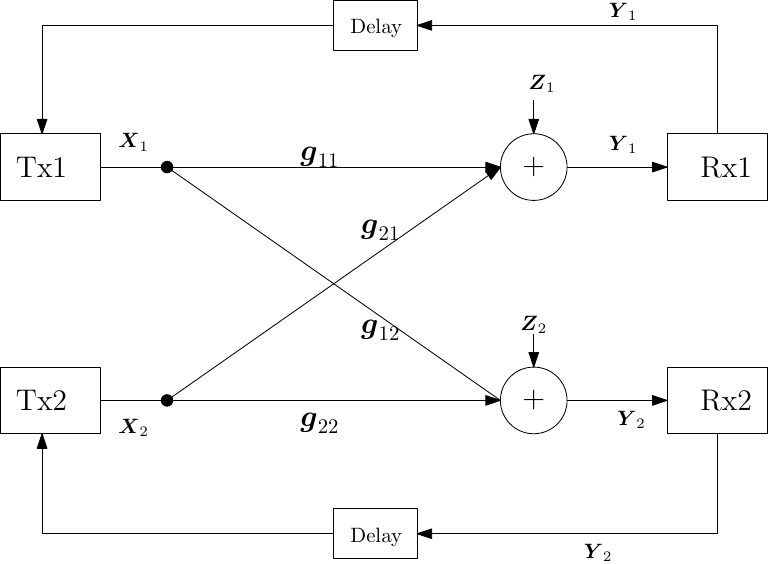}
\par\end{centering}
\caption{\label{fig:feedback_ic}The channel model with feedback.}
\end{figure}

We also consider a feedback model (Figure \ref{fig:feedback_ic}),
where each receiver reliably feeds back the received symbols\footnote{The IC with rate limited feedback is considered in \cite{VahidSuh_12_ratelimited}
where outputs are quantized and fed back. Our schemes can also be
extended for such cases.} to the corresponding transmitter. We consider the feedback of symbols
in blocks of $T$: the symbols are fed back after all the symbols
in a coherence period of $T$ are received. However, the results that
we derive are valid even if the feedback is performed after receiving
each symbol.

The main metric used in this paper to evaluate the performance of
achievability schemes is the gDoF. It is desirable to have an exact
capacity characterization, but this is unknown even for the coherent
IC except for some regimes \cite{etkin_tse_no_fb_IC,Shang_Kramer_Chen_IC}.
For the noncoherent case, the exact capacity characterization is open
even for point-to-point channels. Hence, we give a characterization
for the noncoherent IC in terms of the asymptotic approximation of
the capacity region defined by the gDoF region. For this, we consider
a series of channels with a fixed interference level $\alpha$ and
letting $\snr,\ \inr\rightarrow\infty$. Let $\mathcal{C}\brac{\snr,\inr}$
denote the capacity region of the channel and let $\mathcal{\tilde{D}}_{C}$
be a scaled version of $\mathcal{C}\brac{\snr,\inr}$ given by \sloppy$\mathcal{\tilde{D}}_{C}\brac{\snr,\inr}=\cbrac{\brac{R_{1}/\lgbrac{\snr},R_{2}/\lgbrac{\snr}}:\brac{R_{1},R_{2}}\in\mathcal{C}\brac{\snr,\inr}}$.
Following \cite{etkin_tse_no_fb_IC}, we define the generalized degrees
of freedom region as the asymptote of the scaled capacity region:
\begin{equation}
\mathcal{D}_{C}\brac{\alpha}=\underset{\alpha\text{ fixed}}{\lim_{\snr,\ \inr\rightarrow\infty}}\mathcal{\tilde{D}}_{C}\brac{\snr,\inr}.
\end{equation}
In other words, $\mathcal{D}_{C}\brac{\alpha}$ contains elements
$\brac{d_{1},d_{2}}$ iff $\brac{d_{1},d_{2}}$ lies within $\mathcal{\tilde{D}}_{C}\brac{\snr,\inr}$
in the asymptotic case of $\snr,\ \inr\rightarrow\infty$ with fixed
$\alpha$. This can be formally stated as:

\begin{equation}
\mathcal{D}_{C}\brac{\alpha}=\cbrac{\brac{d_{1},d_{2}}:\underset{\alpha\text{ fixed}}{\lim_{\snr,\ \inr\rightarrow\infty}}\brac{\underset{\brac{y_{1},y_{2}}\in\mathcal{\tilde{D}}_{C}\brac{\snr,\inr}}{\min}\left|\brac{d_{1},d_{2}}-\brac{y_{1},y_{2}}\right|}=0}.
\end{equation}
If we have any rate region $\mathcal{R}\brac{\snr,\inr}$, we can
similarly define a prelog region $\mathcal{D}_{R}\brac{\alpha}$ in
the following manner:

\[
\mathcal{\tilde{D}}_{R}\brac{\snr,\inr}=\cbrac{\brac{R_{1}/\lgbrac{\snr},R_{2}/\lgbrac{\snr}}:\brac{R_{1},R_{2}}\in\mathcal{R}\brac{\snr,\inr}}
\]
\begin{equation}
\mathcal{D}_{R}\brac{\alpha}=\cbrac{\brac{d_{1},d_{2}}:\underset{\alpha\text{ fixed}}{\lim_{\snr,\ \inr\rightarrow\infty}}\brac{\underset{\brac{y_{1},y_{2}}\in\mathcal{\tilde{D}}_{R}\brac{\snr,\inr}}{\min}\left|\brac{d_{1},d_{2}}-\brac{y_{1},y_{2}}\right|}=0}.
\end{equation}
An achievable prelog region is the prelog region derived from an achievable
rate region. Just as the capacity region is the maximal achievable
rate region, so the gDoF region is the maximal achievable prelog region\footnote{For point-to-point channels, the prelog of an achievable rate $R$
from a scheme can be defined as $\lim_{\snr\rightarrow\infty}R/\lgbrac{\snr}$
and the DoF is the maximal achievable prelog.}. Our scaling process in the above definitions not only scales the
input power, but also scales the interfering link simultaneously to
obtain an asymptotic region that is dependent on the interference
level. If we scale only the transmit power $P$, we would just obtain
an asymptotic region that corresponds to the case with the interfering
signal at the same strength as the desired signal at the receivers,
and we would not be able to capture the effect of different interference
levels. This method of characterization was first used in \cite{etkin_tse_no_fb_IC}
to characterize the asymptotic behavior of the capacity region of
a 2-user symmetric IC for high SNR. In \cite{etkin_tse_no_fb_IC},
the received signal strengths through the four links of the IC were
set to scale as $\snr,\snr^{\alpha},\snr^{\alpha},\snr$. The method
of scaling the received signal strengths on different links with different
SNR exponents to obtain the gDoF region is also used in other works
like \cite{gDoF_K_user_IC,gDoF_MIMO_IC,Joyson_2x2_mimo_journal,Joyson_noncoh_diamond_journal}.
In this paper, we also assume that $T\geq2$, since if $T=1$, the
gDoF region of the IC is null following the result for the noncoherent
multiple-input multiple-output (MIMO) channel \cite{Taricco_Elia_97,lapidoth2003capacity,Joyson_2x2_mimo_journal}.

A standard training-based scheme estimates the channel at the receiver
using known training symbols sent from the transmitter and uses the
estimate to operate a coherent decoder. Such a scheme is known to
be DoF optimal for the noncoherent single-user MIMO channel \cite{Zheng_Tse_Grassmann_MIMO}.
A natural question to ask is whether operating the noncoherent IC
with such a standard training-based scheme achieves the gDoF region.
The main observation in this paper is that we can improve the prelog
region of the standard training-based coherent schemes in several
regimes for the noncoherent IC.

\begin{figure}
\begin{minipage}[c][1\totalheight][t]{0.45\textwidth}%
\begin{center}
\includegraphics[scale=0.6]{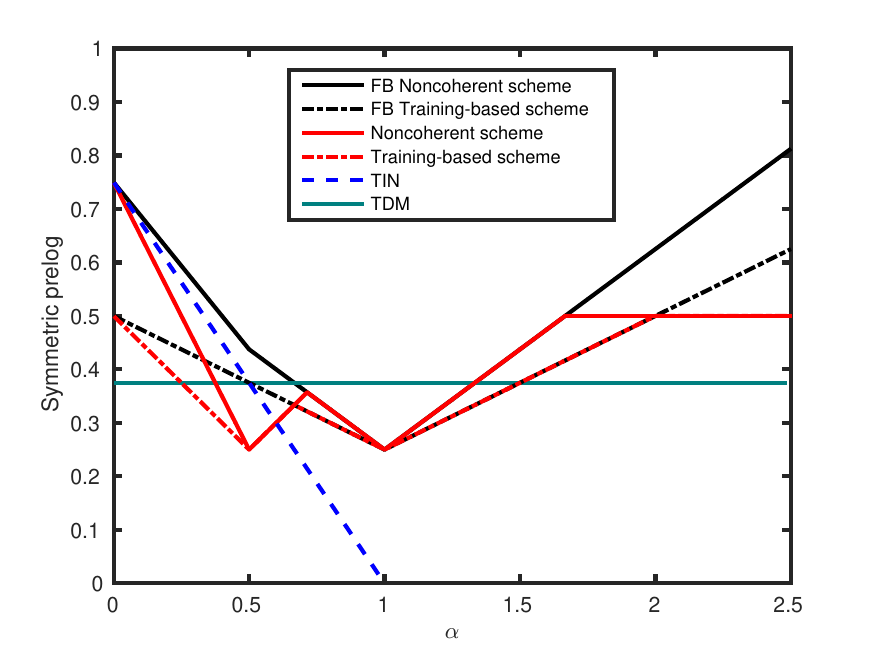}
\par\end{center}
\caption{\label{fig:Symmetric-gDoF_T=00003D4}Symmetric achievable prelog of
the noncoherent IC for coherence time $T=4$.}
\end{minipage}\hfill{}%
\begin{minipage}[c][1\totalheight][t]{0.45\textwidth}%
\begin{center}
\includegraphics[scale=0.6]{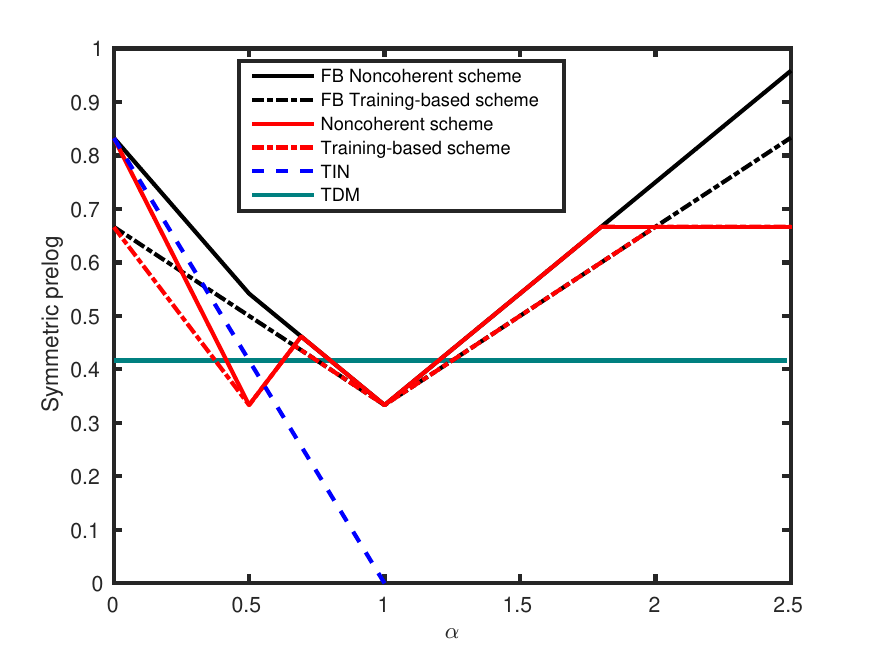}
\par\end{center}
\caption{\label{fig:Symmetric-gDoF_T=00003D6}Symmetric achievable prelog of
the noncoherent IC for coherence time $T=6$.}
\end{minipage}
\end{figure}

We provide the coding schemes and analysis for the following schemes.
\begin{enumerate}
\item We develop a noncoherent version of the simplified Han-Kobayashi scheme
from \cite{etkin_tse_no_fb_IC} for the 2-user IC, where the transmitters
use superposition coding, rate-splitting their messages into common
and private parts based on the INR. Each receiver noncoherently decodes
its own private message and the common messages from both users.
\item Similar to the previous scheme, another noncoherent scheme is developed
for the 2-user IC with feedback extending the coherent scheme from
\cite{suh_tse_fb_gaussian}. This scheme involves $B$ blocks. In
the first block, each transmitter splits its own message into common
and private parts and then sends a codeword superimposing the common
and private messages. In subsequent blocks, the common message from
the other user is decoded at the transmitter using the feedback. Each
transmitter generates new common and private messages, conditioned
on the previous common messages from both users. After a total of
$B$ blocks, each receiver performs backward decoding. Each decoding
step in this scheme is performed noncoherently.
\item A training-based scheme is analyzed for the noncoherent IC without
feedback. The first two symbols in every coherence period of $T$
symbols is used for estimating the channels\footnote{As we are considering high SNR behavior, one training symbol is sufficient
for each link.}. The rest of the symbols are used for transmitting data. The part
of data transmission is performed according to a rate-splitting scheme
\cite{etkin_tse_no_fb_IC} for the coherent IC: the transmitters use
superposition coding, rate-splitting their messages into common and
private parts based on the INR. Each receiver uses the channel estimates
and decodes its own private message and the common messages from both
users.
\item A training-based scheme is analyzed for the noncoherent IC with feedback.
The first two symbols in every coherence period of $T$ symbols is
used for estimating the channels. The rest of the symbols are used
for transmitting data. The part of data transmission is performed
according to a rate-splitting scheme \cite{suh_tse_fb_gaussian} for
the coherent IC with feedback. This is similar to the scheme 2) above
that we described for the noncoherent case, except that the decoding
is performed coherently using the estimated channel values.
\item We consider a scheme which treats interference-as-noise (TIN) where
each receiver treats the symbols from the other user as interference.
The first symbol in every coherence period is used for estimating
the channels. Each user estimates its own channel while treating the
other user as interference.
\item We also consider a time division multiplexing (TDM) between single-user
transmissions with equal time-sharing between the users. Alternate
blocks of length $T$ are used by alternate users. For each user,
the first symbol in the block of length $T$ is used for estimating
its channel.
\end{enumerate}
The TIN and TDM schemes are implemented using one training symbol
in each coherence period, as there is only one channel coefficient
to be estimated for each user. The TIN and TDM schemes can also be
implemented in a noncoherent manner without training symbols, but
it can be verified that the prelog performance remains the same. We
evaluate the achievable prelog region with the above schemes and compare
the performance. Our main results on the prelog of the noncoherent
IC are illustrated in Figure \ref{fig:Symmetric-gDoF_T=00003D4} and
Figure \ref{fig:Symmetric-gDoF_T=00003D6}.

When the INR is much lower than the SNR in the absence of feedback,
the TIN scheme is better than other schemes that decode part of the
interfering message. In contrast, for the case when the channel is
perfectly known, the TIN scheme has the same performance as a rate-splitting
scheme without feedback when the INR is much lower than the SNR. However,
for the noncoherent case, rate-splitting schemes without feedback
have lower prelog. We believe that this is due to the added uncertainty
in the interfering link along with the uncertainty of the interfering
message to be decoded. Due to this added uncertainty, it also is better
to avoid interference using the TDM scheme when the interference level
$\alpha$ is close to 1.%

In general, the noncoherent schemes perform better than the standard
training-based schemes. The schemes with feedback have larger prelog
than the corresponding schemes without feedback. With feedback, the
performance of noncoherent rate-splitting schemes is in general better
than the TIN scheme. However, the TDM scheme is still the best around
$\alpha=1$.

We also provide some numerical results to show that our results can
provide improvements in the rates compared to the standard training-based
schemes at finite SNRs, the rate-SNR points are given in Table \ref{tab:Comparison-of-rates}
on page \pageref{tab:Comparison-of-rates}.

\subsection{Related Work}

To the best of our knowledge, the capacity of the noncoherent interference
channel has not received much attention in the literature. Hence,
we give an overview of the existing works on noncoherent wireless
networks and the related work on the interference channels. The noncoherent
wireless model for the MIMO channel was studied by Marzetta and Hochwald
\cite{marzetta1999capacity}. In their model, neither the receiver
nor the transmitter knows the fading coefficients and the fading gains
remain constant within a block of length $T$ symbol periods. Across
the blocks, the fading gains are independent and identically distributed
(i.i.d.) according to a Rayleigh distribution. The capacity behavior
at high SNR for the noncoherent MIMO channel was studied by Zheng
and Tse in \cite{Zheng_Tse_Grassmann_MIMO}. The main conclusion of
that work was that a standard training-based scheme was DoF optimal
for the noncoherent MIMO channels, a message distinct from our conclusions
in this paper for the noncoherent IC. Some works have specifically
studied the case with $T=1$ \cite{Taricco_Elia_97,Abou_Faycal_noncoherent,lapidoth2003capacity}.
In \cite{Abou_Faycal_noncoherent}, it was demonstrated that for $T=1$,
the capacity is achieved by a distribution with a finite number of
mass points, but the number of mass points grows with the SNR. The
capacity for the case with $T=1$ was shown to behave double-logarithmically
 in \cite{lapidoth2003capacity}.

There have been other works that studied noncoherent relay channels.
The noncoherent single relay network was studied in \cite{Koch2013},
where the authors considered identical link strengths and unit coherence
time. They showed that under certain conditions on the fading statistics,
the relay does not increase the capacity at high SNR. In \cite{Gohary_non_coherent_2014},
similar observations were made for the noncoherent MIMO full-duplex
single relay channel with block-fading. The authors showed that Grassmanian
signaling can achieve the DoF without using the relay. Also for certain
regimes, decode-and-forward with Grassmanian signaling was shown to
approximately achieve the capacity at high SNR.

The above works considered a DoF framework for the noncoherent model
in the sense that for high SNR, the link strengths are not significantly
different, \emph{i.e.,} the links scale with the same SNR-exponent.
The gDoF framework for the noncoherent MIMO channel was considered
in \cite{Joyson_2x2MIMO_isit,Joyson_2x2_mimo_journal} and it was
shown that several insights from the DoF framework may not carry on
to the gDoF framework. It was shown that a standard  training-based
scheme is not gDoF optimal and that all antennas may have to be used
for achieving the gDoF, even when the coherence time is low, in contrast
to the results for the MIMO channel with i.i.d. links. In \cite{Joyson_noncoh_diamond_journal},
the gDoF of the 2-relay diamond network was studied. The standard
training-based schemes were proved to be sub-optimal and a new scheme
was proposed, which partially trains the network and performs a scaling
and  quantize-map-forward operation \cite{ozgur_diggavi_2010_isit,ozgur_diggavi_2013,avest_det}
at the relays.

In this work, we study the noncoherent 2-user IC with symmetric statistics.
This, we believe, is the first information theoretic analysis of noncoherent
channels in multiple unicast networks with interference. The capacity
of the (coherent) 2-user Gaussian IC is well studied \cite{han_kobayashi,chong2008han,etkin_tse_no_fb_IC,suh_tse_fb_gaussian}
when the channels are perfectly known at the receivers and transmitters.
The capacity region of the 2-user IC without feedback was characterized
in \cite{etkin_tse_no_fb_IC}, to within 1 bit per user. In \cite{suh_tse_fb_gaussian},
a similar result was derived for the 2-user Gaussian IC with feedback,
obtaining the capacity region within 2 bits per user. In \cite{joyson_fading_TCOM},
the approximate capacity region (within a constant additive gap) for
2-user fast fading interference channels (FF-IC), with no instantaneous
CSIT but with perfect channel knowledge at the receiver, was derived.
There, the authors used a rate-splitting scheme based on the average
interference-to-noise ratio, extending the existing rate-splitting
schemes for the IC \cite{etkin_tse_no_fb_IC,suh_tse_fb_gaussian}.
The approximate capacity region was derived for the FF-IC without
feedback and also for the case with feedback; the feedback improves
the capacity region for the FF-IC, similar to the case for the static
IC \cite{suh_tse_fb_gaussian}. In this work, we extend the results
from \cite{joyson_fading_TCOM} for the FF-IC (where the receivers
know the channel, but not the transmitters) to the case when both
transmitters and receivers do not know the channel, \emph{i.e., }the
noncoherent IC.

The paper is organized as follows. In Section \ref{sec:Notation},
we explain the notations used. In Section \ref{sec:NC_IC_noFB}, we
discuss our results on the noncoherent IC without feedback and in
Section \ref{sec:noncoh_FFIC_FB}, we discuss the noncoherent IC with
feedback. In Section \ref{sec:Conclusions-and-remarks}, we give the
conclusions and remarks. Some of the proofs for the analysis are deferred
to the appendices.

\section{Notational Conventions\label{sec:Notation}}

We use the notation $\mathcal{CN}\brac{\mu,\sigma^{2}}$ for circularly
symmetric complex Gaussian distribution with mean $\mu$ and variance
$\sigma^{2}$. The logarithm to base 2 is denoted by $\lgbrac{}$.
We use the symbol $\sim$ with overloaded meanings: one to indicate
that a random variable has a given distribution and second to indicate
that two random variables have the same distribution. We use the notation
$\doteq$ for order equality, \emph{i.e.}, we say $f_{1}\brac{\mathsf{SNR}}\doteq f_{2}\brac{\mathsf{SNR}}$
if
\begin{equation}
\underset{\snr\rightarrow\infty}{\text{lim}}\frac{f_{1}\brac{\mathsf{SNR}}}{\lgbrac{\mathsf{SNR}}}=\underset{\snr\rightarrow\infty}{\text{lim}}\frac{f_{2}\brac{\mathsf{SNR}}}{\lgbrac{\mathsf{SNR}}}.
\end{equation}
The use of symbols $\leqdof,\geqdof$ are defined analogously. When
we have an $\snr$-dependent term $t_{1}$ in evaluating the rate
of a scheme, we have the prelog of the term $t_{1}$ as the limit
$\underset{\snr\rightarrow\infty}{\text{lim}}t_{1}/\lgbrac{\snr}$.
Similarly, an upper bound in prelog can be defined. We use a bold
script for random variables and the normal script for deterministic
variables. We use small letters for scalars, capital letters for vectors
and capital letter with underline for matrices. The following capital
letters being a common notation are used for scalars: $P$ for power,
$B$ for number of codeblocks, $T$ for the coherence time, $R$ for
rate and $C$ for capacity. The special script of the form $\mathcal{A},\mathcal{E}$
is used to indicate sets. The notation $\mathcal{CN}\brac{\mu,\sigma^{2}}$
is reserved for complex Gaussian distribution with mean $\mu$ and
variance $\sigma^{2}$. The notation $\x_{i,j}$ indicates the $j^{\text{th}}$
element of $\boldsymbol{X}_{i}$. Similar definitions follow for $\boldsymbol{y}_{i,j}$
and $\boldsymbol{z}_{i,j}$. The random variables $\boldsymbol{g}_{ij}$
with $i,j\in\{1,2\}$ are scalar random variables to capture the block
fading.

\section{Noncoherent IC without feedback\label{sec:NC_IC_noFB}}

In this section, we provide our results for the noncoherent IC without
feedback. We compare the achievable prelog using a standard training-based
scheme to our noncoherent rate-splitting scheme and we also compare
it with the TIN and TDM schemes.
\begin{thm}
\label{thm:noncoh_IC_noFB}Using a noncoherent rate-splitting scheme,
the prelog region given in Table \ref{tab:gdof_no_FB} is achievable.

\begin{table}[H]
\centering{}\caption{Achievable prelog region for different regimes of $\alpha$.\label{tab:gdof_no_FB}}
\begin{tabular}{|c|c|c|}
\hline
 $\alpha<1/2$ &  $1/2\leq\alpha\leq1$  & $\alpha>1$\tabularnewline
\hline
\hline
$\begin{array}{c}
d_{1}\leq\brac{1-1/T}-\alpha/T\\
d_{2}\leq\brac{1-1/T}-\alpha/T\\
d_{1}+d_{2}\leq2\brac{1-1/T}-2\alpha
\end{array}$ & $\begin{array}{c}
d_{1}+d_{2}\leq\brac{2-3/T}-\alpha\brac{1-1/T}\\
d_{1}+d_{2}\leq2\brac{1-2/T}\alpha\\
2d_{1}+d_{2}\leq\brac{2-3/T}-\alpha/T\\
d_{1}+2d_{2}\leq\brac{2-3/T}-\alpha/T
\end{array}$ & $\begin{array}{c}
d_{1}\leq\brac{1-2/T}\\
d_{2}\leq\brac{1-2/T}\\
d_{1}+d_{2}\leq\brac{1-1/T}\alpha-1/T
\end{array}$\tabularnewline
\hline
\end{tabular}
\end{table}
\end{thm}
\begin{IEEEproof}
The proof follows by analyzing a Han-Kobayashi scheme \cite{han_kobayashi,chong2008han}
with rate-splitting based on the average interference-to-noise ratio
\cite{joyson_fading_TCOM}. The message for User 1 is split into two
parts, a common message $w_{\text{c}1}$ at rate $R_{\text{c}1}$
and a private message $w_{\text{p}1}$ at rate $R_{\text{p}1}$. The
common message $w_{\text{c}1}$ is mapped into Gaussian vector symbols
represented by $\U_{1}$ and private message $w_{\text{p}1}$ is mapped
into Gaussian vector symbols represented by $\X_{\text{p}1}$ where
$\U_{1},\X_{\text{p}1}$ are independent. The vectors are of size
$T$. The transmitted symbols at Transmitter 1 are of the form $\X_{1}=\U_{1}+\X_{\text{p}1}$.
The power allocation to the symbols are determined based on the average
interference-to-noise ratio. The power of each element of $\X_{\text{p}1}$
is $1/\inr$ and the power of each element of $\U_{1}$ is $P-1/\inr$.

Similarly at Transmitter 2, we have a common message $w_{\text{c}2}$
at rate $R_{\text{c}2}$ and a private message $w_{\text{p}2}$ at
rate $R_{\text{p}2}$. The common message $w_{\text{c}2}$ is mapped
into Gaussian vector symbols represented by $\U_{2}$ and private
message is mapped into Gaussian vector symbols represented by $\X_{\text{p}2},$
where $\U_{2},\X_{\text{p}2}$ are independent. The transmitted symbols
at Transmitter 2 are of the form $\X_{2}=\U_{2}+\X_{\text{p}2}$.
The power of each element of $\X_{\text{p}2}$ is $1/\inr$ and the
power of each element of $\U_{2}$ is $P-1/\inr$.

Each receiver, in a noncoherent manner jointly decodes its own private
message and the common messages from both users, \emph{i.e.}, receiver
1 decodes $w_{\text{c}1},w_{\text{c}2},w_{\text{p}1}$ and receiver
1 decodes $w_{\text{c}1},w_{\text{c}2},w_{\text{p}2}$. The details
of the coding scheme and its analysis are in Section \ref{subsec:no_fb}.
\end{IEEEproof}
We now compare our achievable prelog with that of a standard training-based
scheme.
\begin{thm}
\label{thm:training_noFB}A standard training-based scheme for the
noncoherent IC can achieve the prelog region described in Table \ref{tab:gdof_no_FB_with_training}.

\begin{table}[H]
\centering{}\caption{Achievable prelog region for different regimes of $\alpha$.\label{tab:gdof_no_FB_with_training}}
\begin{tabular}{|c|c|c|}
\hline
$\alpha<1/2$ & $1/2\leq\alpha\leq1$ & $\alpha>1$\tabularnewline
\hline
\hline
$\begin{array}{c}
d_{1}\leq\brac{1-2/T}\\
d_{2}\leq\brac{1-2/T}\\
d_{1}+d_{2}\leq2\brac{1-2/T}\brac{1-\alpha}
\end{array}$ & $\begin{array}{c}
d_{1}+d_{2}\leq\brac{1-2/T}\brac{2-\alpha}\\
d_{1}+d_{2}\leq2\brac{1-2/T}\alpha\\
2d_{1}+d_{2}\leq2\brac{1-2/T}\\
d_{1}+2d_{2}\leq2\brac{1-2/T}
\end{array}$ & $\begin{array}{c}
d_{1}\leq\brac{1-2/T}\\
d_{2}\leq\brac{1-2/T}\\
d_{1}+d_{2}\leq\brac{1-2/T}\alpha
\end{array}$\tabularnewline
\hline
\end{tabular}
\end{table}
\end{thm}
\begin{IEEEproof}
With two users, in every coherence period of $T$ symbols, we need
at least two symbols for training. For training, the first transmitter
can send a known symbol while the second transmitter remains turned
off. With this, both receivers can estimate the channels from the
first transmitter. Next the second transmitter can send a known symbol
while the first transmitter remains turned off. With this, both receivers
can estimate the channels from the second transmitter. The rest of
the symbols in every coherence period of $T$ symbols can be used
to transmit data using a Han-Kobayashi scheme scheme similar to that
described in Theorem \ref{thm:noncoh_IC_noFB}. The detailed analysis
for obtaining the prelog is given in \ifarxiv  Appendix \ref{app: Training Scheme NoFB}\else  \cite[Appendix C]{Joyson_noncoh_IC_arxiv}\fi.
\end{IEEEproof}
\begin{rem}
The capacity region of the coherent FF-IC is known within a constant
gap from \cite{joyson_fading_TCOM} and hence its gDoF region is known.
The prelog region from the above theorem is the same as the gDoF region
for the coherent FF-IC with a multiplication factor of $\brac{1-2/T}$.
Hence the prelog region obtained in Theorem \ref{thm:training_noFB}
is the best among any scheme that uses two symbols for training in
every coherence period of $T$ symbols.
\end{rem}
We also consider the strategy of treating-interference-as-noise (TIN)
with Gaussian codebooks. Transmitter $i$ sends a message $w_{i}$
at rate $R_{i}$ using vector Gaussian symbols $\X_{i}$ of length
$T$, $i\in\cbrac{1,2}$. Each receiver $i$ decodes $w_{i}$, treating
the symbols from the other transmitter as noise. Using standard analysis,
it can be shown that the prelog region\begin{subequations}
\begin{eqnarray}
d_{1} & \leq & \brac{1-1/T}\brac{1-\alpha}\\
d_{2} & \leq & \brac{1-1/T}\brac{1-\alpha}
\end{eqnarray}
\end{subequations}is achievable by the TIN scheme.

Another strategy is time division multiplexing (TDM). Again transmitter
$i$ can send a message $w_{i}$ at rate $R_{i}$ using vector Gaussian
symbols $\X_{i}$ of length $T$, $i\in\cbrac{1,2}$. For the TDM
case, each transmitter transmits in every alternate time periods of
length $T$. While one transmitter is ON, the other is OFF. Each reciever
obtains symbols only from the intended transmitter and can perform
typicality decoding. Using standard analysis we can obtain that the
prelog region\begin{subequations}
\begin{eqnarray}
d_{1} & \leq & \brac{1/2}\brac{1-1/T},\\
d_{2} & \leq & \brac{1/2}\brac{1-1/T},
\end{eqnarray}
\end{subequations}is achievable.

\subsection{Discussion\label{subsec:Discussion_noFB}}

\begin{figure}
\centering{}\includegraphics[scale=0.6]{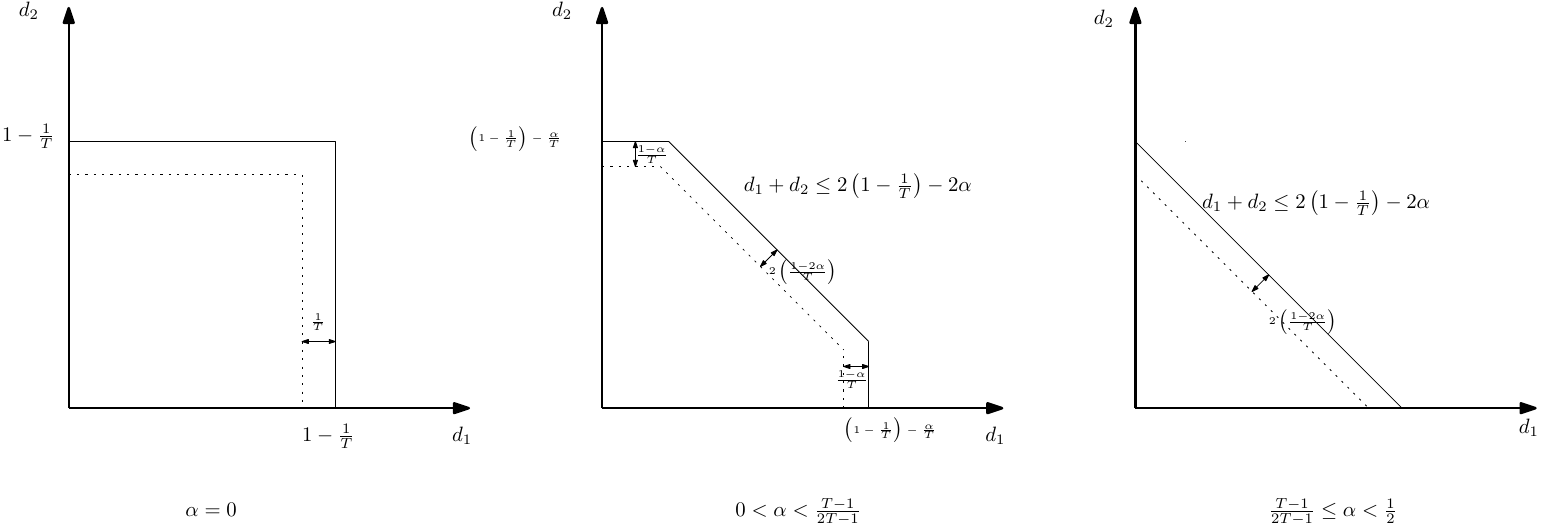}\caption{Prelog region for $\alpha<1/2$,$T\protect\geq2$. The solid line
gives the prelog region achievable for a noncoherent scheme and the
dotted line gives the prelog for the scheme that uses 2 symbols for
training.\label{fig:gDoF-nofb-set1}}
\end{figure}

\begin{figure}
\centering{}\includegraphics[scale=0.55]{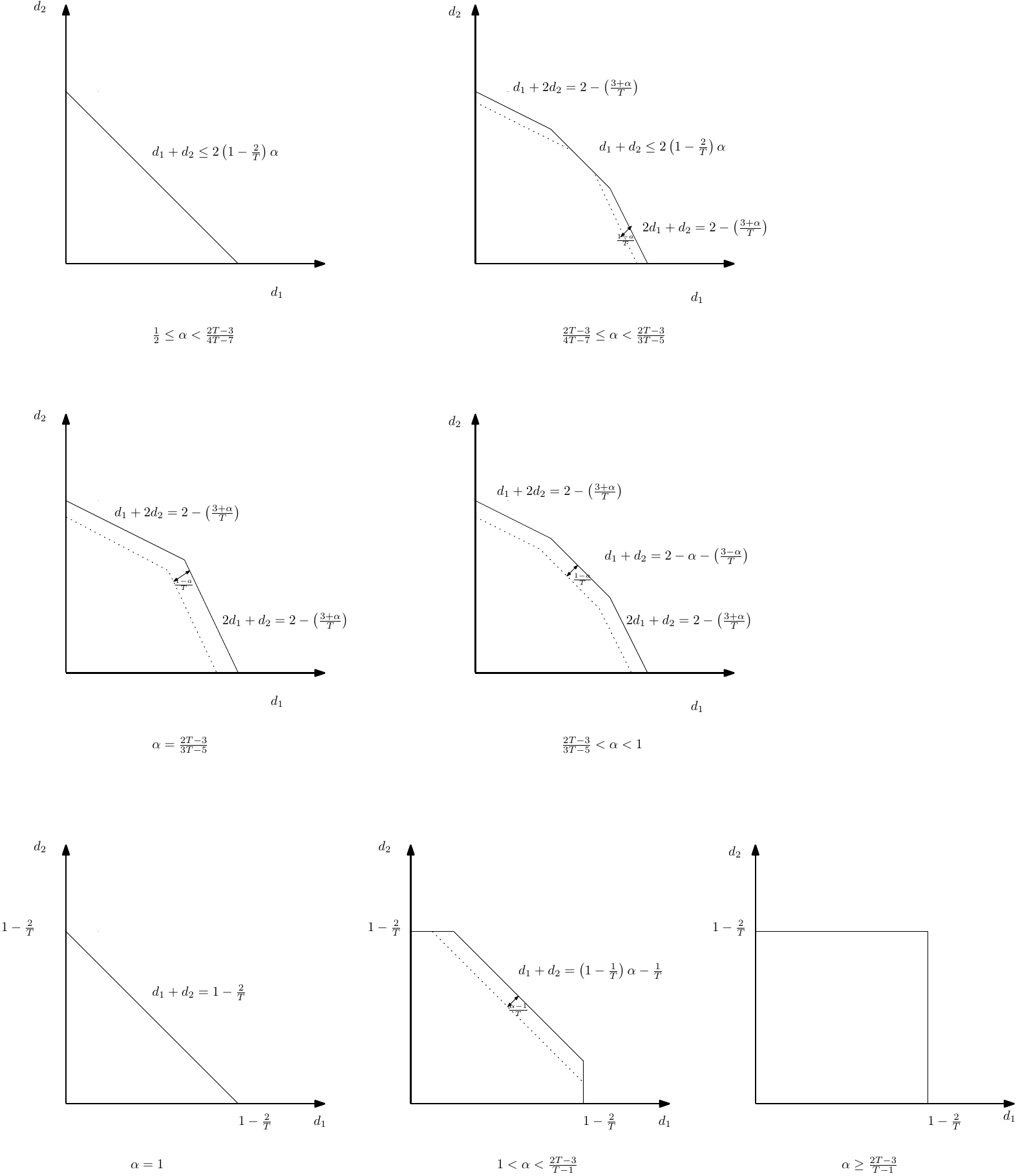}\caption{Prelog region for $1/2<\alpha$, $T\protect\geq3$. For $T=2$ no
prelog region is achievable using known schemes. The solid line gives
the prelog region achievable for a noncoherent scheme and the dotted
line gives the prelog region for a scheme that uses 2 symbols for
training.\label{fig:gDoF-nofb-set2}}
\end{figure}

In Figure \ref{fig:gDoF-nofb-set1} and Figure \ref{fig:gDoF-nofb-set2},
the prelog region achievable using our noncoherent scheme is compared
with the prelog region achievable using the aforementioned training-based
scheme. It can be observed that our noncoherent scheme outperforms
the standard training-based scheme.

\begin{figure}
\begin{minipage}[c][1\totalheight][t]{0.45\textwidth}%
\begin{center}
\includegraphics[scale=0.6]{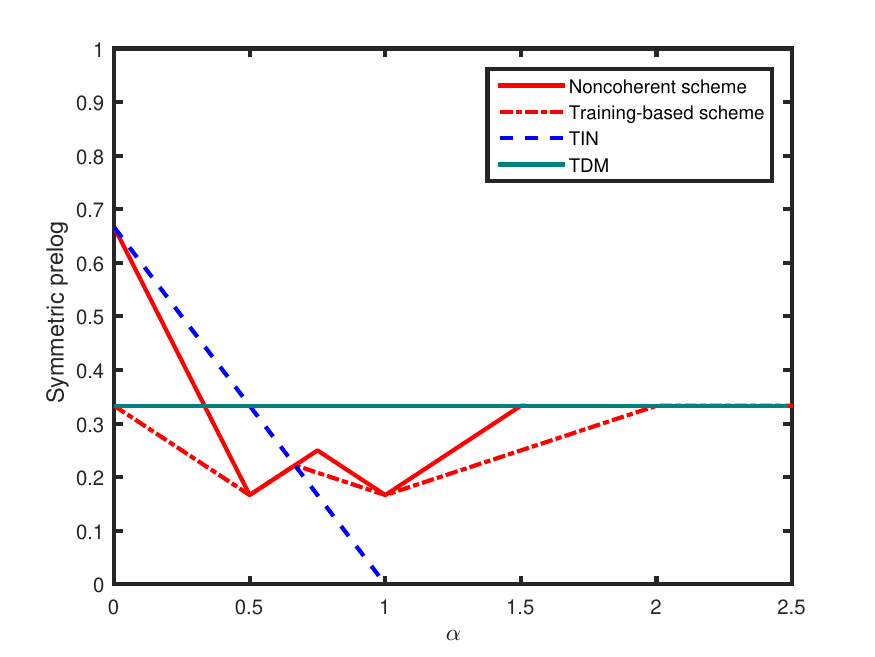}\caption{\label{fig:Symmetric_gDoF_T=00003D3}Symmetric achievable prelog for
coherence time $T=3$: feedback and nonfeedback cases.}
\par\end{center}%
\end{minipage}\hfill{}%
\begin{minipage}[c][1\totalheight][t]{0.45\textwidth}%
\begin{center}
\includegraphics[scale=0.6]{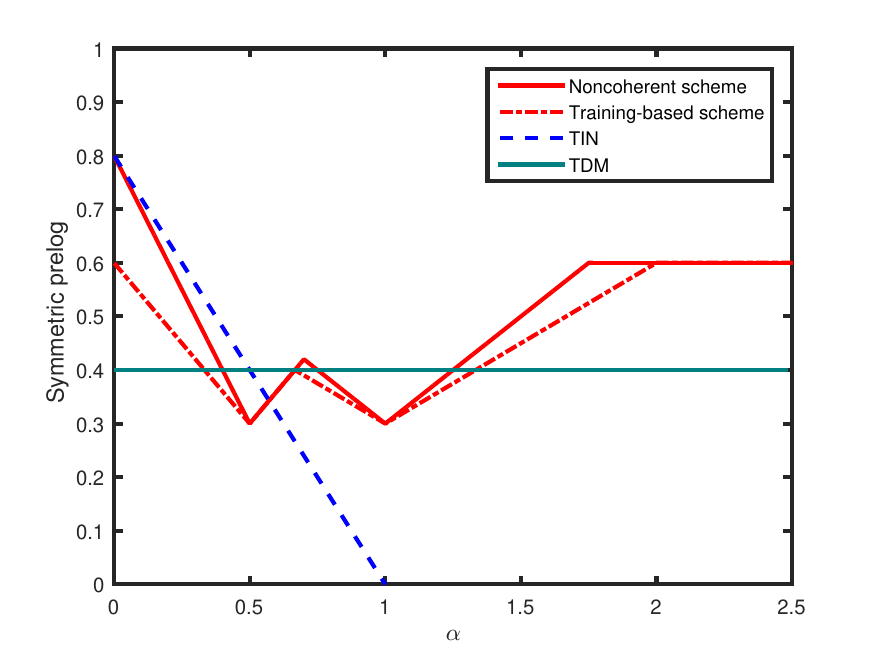}\caption{\label{fig:Symmetric-gDoF_T=00003D5}Symmetric achievable prelog for
coherence time $T=5$. Training-based scheme uses $2$ symbols for
training.}
\par\end{center}%
\end{minipage}
\end{figure}

In Figure \ref{fig:Symmetric_gDoF_T=00003D3} and Figure \ref{fig:Symmetric-gDoF_T=00003D5},
we give the achievable symmetric prelog with coherence time $T=3$
and $T=5$ respectively for the strategies that we discussed. In the
overview of our results in Section I, we had noticed that TIN outperforms
rate-splitting schemes. In fact, it can be calculated from our prelog
regions that the TIN scheme outperforms TDM scheme also for very weak
interference level ($\alpha<1/2$).

For a broad region of $\alpha$, the TDM scheme outperforms the noncoherent
rate-splitting scheme. This can be clearly seen by looking at the
points with $\alpha=.5$ and $\alpha=1$. For these values of $\alpha$,
the noncoherent rate-splitting scheme gives a prelog of $\brac{1/2}\brac{1-2/T}$
and the TDM scheme gives a prelog of $\brac{1/2}\brac{1-1/T}$. Hence,
for $\alpha=.5$ and $\alpha=1$, the noncoherent scheme effectively
behaves as a TDM scheme that uses two training symbols per coherence
period, where actually the TDM scheme can be implemented with only
one training symbol per coherence period.

Although our main results are on the prelog of the system, we can
provide guidelines for specific scenarios depending on the interference
level. For example with transmit SNR $6$ dB, coherence time $T=5$,
and all the links with average strength $1$, using the TDM scheme
can improve the rate by $6\%$ compared to the standard training-based
schemes used with rate-splitting. More rate points are illustrated
in Table \ref{tab:Comparison-of-rates}. The rates for training-based
scheme is obtained by numerically evaluating the expression in (\ref{eq:ach_nofb_training_simplified})
for the point in the rate region where both users have the same rate.
The expressions used for obtaining the rates for the TDM scheme is
given in \ifarxiv  Appendix \ref{app:Numerical-Calculation}\else  \cite[Appendix E]{Joyson_noncoh_IC_arxiv}\fi.
We also provide a Mathematica code at \url{https://arxiv.org/src/1812.03579/anc/Noncoh_IC_rates.nb}
for calculating the rate points.
\begin{table}[H]
\centering{}\caption{\label{tab:Comparison-of-rates}Comparison of rates (bits per user)
achievable with different schemes for $T=5$, $\alpha=1$}
\begin{tabular}{|c|c|c|}
\hline
\multirow{2}{*}{$\snr$ dB} & \multicolumn{2}{c|}{Rates for different schemes}\tabularnewline
\cline{2-3} \cline{3-3}
 & 2 symbol training & TDM\tabularnewline
\hline
6 & 0.47 & 0.50\tabularnewline
\hline
7 & 0.54 & 0.57\tabularnewline
\hline
8 & 0.61 & 0.66\tabularnewline
\hline
9 & 0.69 & 0.75\tabularnewline
\hline
10 & 0.77 & 0.84\tabularnewline
\hline
\end{tabular}
\end{table}

\subsection{Coding Scheme and Analysis of the Rate Region\label{subsec:no_fb}}

We describe the coding scheme starting with a general input distribution
and then we evaluate the prelog region for Gaussian inputs.

\noindent \textbf{Encoding:} \sloppy We consider a fixed distribution
$p\brac{\U_{1}}p\brac{\U_{2}}p\brac{\X_{1}\middle|\U_{1}}p\brac{\X_{2}\middle|\U_{2}}$
where $\U_{1},\U_{2},\X_{1},\X_{2}$ are vectors of length $T$. For
transmitter 1, generate $2^{NTR_{\text{c}1}}$ codewords $\U_{1}^{N}\brac i$
with $i\in\cbrac{1,\ldots,2^{NTR_{\text{c}1}}}$ according to $\prod_{l=1}^{N}p\brac{\U_{1(l)}}$.
For each $\U_{1}^{N}\brac i$, generate $2^{NTR_{\text{p}1}}$ codewords
$\X_{1}^{N}\brac{i,j}$, with $j\in\cbrac{1,\ldots,2^{NTR_{\text{p}1}}}$,
according to $\prod_{l=1}^{N}p\brac{\X_{1(l)}\middle|\U_{1(l)}}$.
Similarly for transmitter 2, generate $2^{NTR_{\text{c}2}}$ codewords
$\U_{2}^{N}\brac i$, with $i\in\cbrac{1,\ldots,2^{NTR_{\text{c}2}}}$,
according to $\prod_{l=1}^{N}p\brac{\U_{2(l)}}$. For each $\U_{2}^{N}\brac j$,
generate $2^{NTR_{\text{p}2}}$ codewords $\X_{2}^{N}\brac{i,j}$,
with $j\in\cbrac{1,\ldots,2^{NTR_{\text{p}2}}}$, according to $\prod_{l=1}^{N}p\brac{\X_{2(l)}\middle|\U_{2(l)}}$.

Transmitter 1 has uniformly random messages $w_{\text{c}1}\in\cbrac{1,\ldots,2^{NTR_{\text{c}1}}},w_{\text{p}1}\in\cbrac{1,\ldots,2^{NTR_{\text{p}1}}}$
to transmit and transmitter 2 has uniformly random messages $w_{\text{c}2}\in\cbrac{1,\ldots,2^{NTR_{\text{c}2}}},w_{\text{p}2}\in\cbrac{1,\ldots,2^{NTR_{\text{p}2}}}$
to transmit. Transmitter 1 sends the symbols $\X_{1}^{N}\brac{w_{\text{c}1},w_{\text{p}1}}$
and transmitter 2 sends the symbols $\X_{1}^{N}\brac{w_{\text{c}2},w_{\text{p}2}}.$

\noindent \textbf{Decoding: }For decoding, receiver 1 finds a triplet
$\brac{\hat{i},\hat{j},\hat{k}}$ requiring $\hat{i},\hat{j}$ to
be unique with
\[
\brac{\X_{1}^{N}\brac{\hat{i},\hat{j}},\U_{1}^{N}\brac{\hat{i}},\U_{2}^{N}\brac{\hat{k}},\Y_{1}^{N}}\in\mathcal{A}_{\epsilon}^{\brac N}.
\]
 Similarly receiver 2 finds a triplet $\brac{\hat{i},\hat{j},\hat{k}}$
requiring $\hat{i},\hat{j}$ to be unique with
\[
\brac{\X_{2}^{N}\brac{\hat{i},\hat{j}},\U_{2}^{N}\brac{\hat{i}},\U_{1}^{N}\brac{\hat{k}},\Y_{2}^{N}}\in\mathcal{A}_{\epsilon}^{\brac N},
\]
where $\mathcal{A}_{\epsilon}^{\brac N}$ indicates the set of jointly
typical sequences.

\noindent \textbf{Error Analysis:} We give the sketch of analysis
for the error probability at receiver 1 assuming $\brac{i,j,k}=\brac{1,1,1}$.
Let $\mathcal{E}_{ijk}$ be the event $\cbrac{\brac{\X_{1}^{N}\brac{i,j},\U_{1}^{N}\brac i,\U_{2}^{N}\brac k,\Y_{1}^{N}}\in\mathcal{A}_{\epsilon}^{\brac N}}$
for a given $i,j,k$. By asymptotic equipartition property (AEP),
the probability of $\text{Pr}\brac{\union_{k}\mathcal{E}_{11k}}$
approaches unity. The error probability at receiver 1 is then captured
by the following:

\begin{eqnarray*}
\text{Pr}\brac{\union_{(i,j)\neq\brac{1,1}}\mathcal{E}_{ijk}} & \leq & \brac{\sum_{i\neq1,j\neq1,k\neq1}\text{Pr}\brac{\mathcal{E}_{ijk}}+\sum_{i\neq1,j=1,k\neq1}\text{Pr}\brac{\mathcal{E}_{ijk}}}\\
 &  & {+}\:\brac{\sum_{i\neq1,j\neq1,k=1}\text{Pr}\brac{\mathcal{E}_{ijk}}+\sum_{i\neq1,j=1,k=1}\text{Pr}\brac{\mathcal{E}_{ijk}}}\\
 &  & {+}\:\sum_{i=1,j\neq1,k=1}\text{Pr}\brac{\mathcal{E}_{ijk}}+\sum_{i=1,j\neq1,k\neq1}\text{Pr}\brac{\mathcal{E}_{ijk}}\\
 & \leq & 2^{N\brac{TR_{\text{c}1}+TR_{\text{c}2}+TR_{\text{p}1}-I\brac{\X_{1},\U_{2};\Y_{1}}+\epsilon}}+2^{N\brac{TR_{\text{p}1}+TR_{\text{c}1}-I\brac{\X_{1};\Y_{1}|\U_{2}}+\epsilon}}\\
 &  & {+}\:2^{N\brac{TR_{\text{p}1}-I\brac{\rline{\X_{1};\Y_{1}}\U_{1},\U_{2}}+\epsilon}}+2^{N\brac{TR_{\text{c}2}+TR_{\text{p}1}-I\brac{\rline{\X_{1},\U_{2};\Y_{1}}\U_{1}}+\epsilon}}.
\end{eqnarray*}
The details of the simplification in the last step can be followed,
for example from \cite[Ch 6]{gamal2011network}. Requiring the average
error probability to vanish at receiver 1 and receiver 2, we get the
following equations as a sufficient condition:\begin{subequations}
\begin{eqnarray}
TR_{\text{c}1}+TR_{\text{c}2}+TR_{\text{p}1} & \leq & I\brac{\X_{1},\U_{2};\Y_{1}},\\
TR_{\text{p}1}+TR_{\text{c}1} & \leq & I\brac{\X_{1};\Y_{1}\middle|\U_{2}},\\
TR_{\text{p}1} & \leq & I\brac{\X_{1};\Y_{1}\middle|\U_{1},\U_{2}},\\
TR_{\text{c}2}+TR_{\text{p}1} & \leq & I\brac{\X_{1},\U_{2};\Y_{1}\middle|\U_{1}},\\
TR_{\text{c}1}+TR_{\text{c}2}+TR_{\text{p}2} & \leq & I\brac{\X_{2},\U_{1};\Y_{2}},\\
TR_{\text{p}2}+TR_{\text{c}2} & \leq & I\brac{\X_{2};\Y_{2}\middle|\U_{1}},\\
TR_{\text{p}2} & \leq & I\brac{\X_{2};\Y_{2}\middle|\U_{1},\U_{2}},\\
TR_{\text{c}1}+TR_{\text{p}2} & \leq & I\brac{\X_{2},\U_{1};\Y_{2}\middle|\U_{2}}.
\end{eqnarray}
\end{subequations}After Fourier-Motzkin elimination, the following
equations are obtained for achievability, with $R_{1}=R_{\text{c}1}+R_{\text{p}1},R_{2}=R_{\text{c}2}+R_{\text{p}2}$:\begin{subequations}\label{eq:ach_nofb}

\begin{eqnarray}
TR_{1} & \leq & I\brac{\X_{1};\Y_{1}\middle|\U_{2}},\label{eq:ach_nofb1}\\
TR_{2} & \leq & I\brac{\X_{2};\Y_{2}\middle|\U_{1}},\label{eq:ach_nofb2}\\
T\brac{R_{1}+R_{2}} & \leq & I\brac{\X_{2},\U_{1};\Y_{2}}+I\brac{\X_{1};\Y_{1}\middle|\U_{1},\U_{2}},\label{eq:ach_nofb3}\\
T\brac{R_{1}+R_{2}} & \leq & I\brac{\X_{1},\U_{2};\Y_{1}}+I\brac{\X_{2};\Y_{2}\middle|\U_{1},\U_{2}},\label{eq:ach_nofb4}\\
T\brac{R_{1}+R_{2}} & \leq & I\brac{\X_{1},\U_{2};\Y_{1}|\U_{1}}+I\brac{\X_{2},\U_{1};\Y_{2}\middle|\U_{2}},\label{eq:ach_nofb5}\\
T\brac{2R_{1}+R_{2}} & \leq & I\brac{\X_{1},\U_{2};\Y_{1}}+I\brac{\X_{1};\Y_{1}\middle|\U_{1},\U_{2}}+I\brac{\X_{2},\U_{1};\Y_{2}\middle|\U_{2}},\label{eq:ach_nofb6}\\
T\brac{R_{1}+2R_{2}} & \leq & I\brac{\X_{2},\U_{1};\Y_{2}}+I\brac{\X_{2};\Y_{2}\middle|\U_{1},\U_{2}}+I\brac{\X_{1},\U_{2};\Y_{1}\middle|\U_{1}}.\label{eq:ach_nofb7}
\end{eqnarray}
\end{subequations}For power splitting, we adapt the idea of the simplified
Han-Kobayashi scheme where the power allocation is such that the private
signal is seen below the noise level at the other receiver. Similar
to \cite{etkin_tse_no_fb_IC,joyson_fading_TCOM}, we choose $\U_{k}$
as a vector of length $T$ with i.i.d. $\mathcal{CN}\brac{0,\lambda_{\text{c}}}$
elements and $\X_{\text{p}k}$ as a vector of length $T$ with i.i.d.
$\mathcal{CN}\brac{0,\lambda_{\text{p}}}$ elements for $k\in\cbrac{1,2}$.
The random variables are chosen independent of each other so that
the set $\cbrac{\U_{1},\U_{2},\X_{\text{p}1},\X_{\text{p}2}}$ is
mutually independent. We use $\X_{1}=\U_{1}+\X_{\text{p}1},\ \X_{2}=\U_{2}+\X_{\text{p}2}$,
$\lambda_{\text{c}}+\lambda_{\text{p}}=P$ and $\lambda_{\text{p}}=\min\brac{1/\inr,P}$.
For prelog characterization, we can assume $P\cdot\inr=\snr\cdot\snr^{\alpha-1}=\snr^{\alpha}>1$.
Hence, we have $\lambda_{\text{p}}=1/\inr.$  The prelog results in
Table \ref{tab:gdof_no_FB} can be obtained by evaluating the rate
region (\ref{eq:ach_nofb}) for our choice of input distribution.

\noindent \textbf{Preliminaries for prelog evaluation:} We give some
preliminary results that can be used in obtaining prelog region from
our achievability region.
\begin{fact}
For an exponentially distributed random variable $\boldsymbol{\xi}$
with mean $\mu_{\xi}$ and with given constants $a\geq0,b>0$, we
have
\begin{equation}
\lgbrac{a+b\mu_{\xi}}-\gamma\lgbrac e\leq\expect{\lgbrac{a+b\boldsymbol{\xi}}}\leq\lgbrac{a+b\mu_{\xi}},
\end{equation}
 where $\gamma$ is Euler's constant.\label{fact:Jensens_gap}
\end{fact}
\begin{IEEEproof}
This is given in \cite[Section III-B]{joyson_fading_TCOM}.
\end{IEEEproof}
We now simplify the region (\ref{eq:ach_nofb}) by considering the
terms in it one by one.
\begin{claim}
\label{claim:h_Y1_U2_X1}The term $h\brac{\Y_{1}|\U_{2},\X_{1}}$
is upper bounded at high SNR as
\[
h\brac{\Y_{1}\middle|\U_{2},\X_{1}}\leqdof\lgbrac{\snr+\snr^{\alpha}}+\lgbrac{\min\brac{\snr,\snr^{\alpha}}}
\]
\end{claim}
\begin{IEEEproof}
The outline of the proof is as follows: with $\y_{1,i}$ as the components
of $\Y_{1}$, we expand $h\brac{\Y_{1}\middle|\U_{2},\X_{1}}=\sum_{i}h\brac{\y_{1,i}\middle|\U_{2},\X_{1},\cbrac{\y_{1,j}}_{j=1}^{i}}$.
The first term $h\brac{\y_{1,1}\middle|\U_{2},\X_{1}}$ gives rise
to the term $\lgbrac{\snr+\snr^{\alpha}}$ with uncertainty from both
 incoming channels.

Let us consider the term $h\brac{\y_{1,2}\middle|\U_{2},\X_{1},\y_{1,1}}$.
In $\y_{1,2}$, the contribution to uncertainty is from the channels
as well as from the symbols. When conditioned on $\U_{2},\X_{1}$,
the contribution of uncertainty from these symbols can be removed.
The uncertainty from $\X_{p2}$ in $\y_{1,2}$ can be neglected in
prelog calculation due to the power allocation strategy that we use.
The term $\y_{1,1}$ is a linear combination of the symbols as well
as the channels. Using this single linear combination given in the
conditioning, the uncertainty from one of the channels can be removed.
Thus $h\brac{\y_{1,2}\middle|\U_{2},\X_{1},\y_{1,1}}$ gives rise
to $\lgbrac{\min\brac{P\times1,P\times\inr}}=\lgbrac{\min\brac{\snr,\snr^{\alpha}}}$,
with either the uncertainty from the direct channel removed or the
uncertainty from the interfering channel removed.

In terms $h\brac{\y_{1,i}\middle|\U_{2},\X_{1},\cbrac{\y_{1,j}}_{j=1}^{i}}$
with $i\geq3$, we can follow the same procedure as stated in the
above paragraph. However with $\cbrac{\y_{1,j}}_{j=1}^{i}$ available
in the conditioning, we have more than a single linear combination
of the channels available. Using these, the contribution from both
 channels can be removed, and hence $h\brac{\y_{1,i}\middle|\U_{2},\X_{1},\cbrac{\y_{1,j}}_{j=1}^{i}}$
do not contribute to the prelog. The detailed proof is in \ifarxiv  Appendix \ref{app:proof_h_Y1_U2_X1}\else  \cite[Appendix B]{Joyson_noncoh_IC_arxiv}\fi.
\end{IEEEproof}
\begin{claim}
\label{claim:h(Y1|U1,U2)}The term $h\brac{\Y_{1}\middle|\U_{1},\U_{2}}$
is lower bounded at high SNR as
\begin{align*}
h\brac{\Y_{1}\middle|\U_{1},\U_{2}}\overset{}{\geqdof}\: & \lgbrac{\snr+\snr^{\alpha}}+\lgbrac{\snr^{1-\alpha}+\min\brac{\snr,\snr^{\alpha}}}\\
 & \:{+}\brac{T-2}\lgbrac{1+\snr^{1-\alpha}}.
\end{align*}
\end{claim}
\begin{IEEEproof}
We expand $h\brac{\Y_{1}\middle|\U_{1},\U_{2}}=\sum_{i}h\brac{\y_{1,i}\middle|\U_{1},\U_{2},\cbrac{\y_{1,j}}_{j=1}^{i}}$.
One way to lower bound $h\brac{\y_{1,i}\middle|\U_{1},\U_{2},\cbrac{\y_{1,j}}_{j=1}^{i}}$
is to condition on the channel strengths and reduce the term to that
for a coherent channel. Another way to lower bound $h\brac{\y_{1,i}\middle|\U_{1},\U_{2},\cbrac{\y_{1,j}}_{j=1}^{i}}$
is to give all the transmit signals in the conditioning and reduce
the entropy to that of a (conditionally) joint Gaussian distribution.
These two techniques help us prove the claim. See Appendix \ref{app:h(Y1|U1,U2)proof}
{} for more details.
\end{IEEEproof}
\noindent \textbf{Bounding mutual information terms:} In the following
four claims, we obtain the lower bounds for four mutual information
terms in the achievability region (\ref{eq:ach_nofb}). We need to
bound only four terms and the other terms can be bounded by using
symmetry of the setup.
\begin{claim}
\label{claim:I(X1;Y1|U2)_lower_bound}The term $I\brac{\X_{1};\Y_{1}\middle|\U_{2}}$
is lower bounded at high SNR as
\[
I\brac{\X_{1};\Y_{1}\middle|\U_{2}}\geqdof\brac{T-1}\lgbrac{\snr}-\lgbrac{\min\brac{\snr,\snr^{\alpha}}}.
\]
\end{claim}
\begin{IEEEproof}
We have
\begin{eqnarray}
h\brac{\Y_{1}|\U_{2}} & = & h\brac{\boldsymbol{g}_{11}\X_{1}+\boldsymbol{g}_{21}\X_{2}+\Z_{1}\middle|\U_{2}}\\
 & = & \sum_{i=1}^{T}h\brac{\boldsymbol{g}_{11}\x_{1,i}+\boldsymbol{g}_{21}\x_{2,i}+\z_{1,i}\middle|\cbrac{\boldsymbol{g}_{11}\x_{1,j}+\boldsymbol{g}_{21}\x_{2,j}+\z_{1,j}}_{j=1}^{i-1},\U_{2}}\\
 & \overset{\brac i}{\geq} & h\brac{\boldsymbol{g}_{11}\x_{1,1}+\boldsymbol{g}_{21}\x_{2,1}+\z_{1,1}\middle|\x_{1,1},\x_{2,1},\U_{2}}\nonumber \\
 &  & {+}\:\sum_{i=2}^{T}h\brac{\boldsymbol{g}_{11}\x_{1,i}+\boldsymbol{g}_{21}\x_{2,i}+\z_{1,i}\middle|\u_{2,i},\boldsymbol{g}_{21},\boldsymbol{g}_{11}}\\
 & \overset{\brac{ii}}{\geqdof} & \lgbrac{\snr+\snr^{\alpha}}+\brac{T-1}\lgbrac{\snr},\label{eq:y1_given_u2}
\end{eqnarray}
where $\brac i$ is \sloppy due to the fact that conditioning reduces
entropy and Markovity $\brac{\boldsymbol{g}_{11}\x_{1,i}+\boldsymbol{g}_{21}\x_{2,i}+\z_{1,i}}-\brac{\u_{2,i},\boldsymbol{g}_{21},\boldsymbol{g}_{11}}-\brac{\cbrac{\boldsymbol{g}_{11}\x_{1,j}+\boldsymbol{g}_{21}\x_{2,j}+\z_{1,j}}_{j=1}^{i-1},\U_{2}}$.
The step $\brac{ii}$ is using the property of Gaussians for the terms
$h\brac{\boldsymbol{g}_{11}\x_{1,1}+\boldsymbol{g}_{21}\x_{2,1}+\z_{1,1}\middle|\x_{1,1},\x_{2,1},\U_{2}}$,
$h\brac{\boldsymbol{g}_{11}\x_{1,i}+\boldsymbol{g}_{21}\x_{2,i}+\z_{1,i}\middle|\u_{2,i},\boldsymbol{g}_{21},\boldsymbol{g}_{11}}$
$=h\brac{\boldsymbol{g}_{11}\x_{1,i}+\boldsymbol{g}_{21}\x_{\text{p}2,i}+\z_{1,i}\middle|\boldsymbol{g}_{21},\boldsymbol{g}_{11}}$
and using Fact \ref{fact:Jensens_gap}. Using (\ref{eq:y1_given_u2})
and Claim \ref{claim:h_Y1_U2_X1} completes the proof.
\end{IEEEproof}
\begin{claim}
\label{claim:I(X2,U1;Y2)_lower_bound}The term $I\brac{\X_{2},\U_{1};\Y_{2}}$
is lower bounded at high SNR as
\[
I\brac{\X_{2},\U_{1};\Y_{2}}\geqdof\brac{T-1}\lgbrac{\snr+\snr^{\alpha}}-\lgbrac{\min\brac{\snr,\snr^{\alpha}}}.
\]
\end{claim}
\begin{IEEEproof}
We have
\begin{align}
h\brac{\Y_{2}} & \geqdof T\lgbrac{\snr+\snr^{\alpha}}.
\end{align}
Using Claim \ref{claim:h_Y1_U2_X1} for $h\brac{\boldsymbol{Y}_{1}\middle|\X_{1},\U_{2}}$
and using symmetry we get,
\begin{align}
h\brac{\Y_{2}\middle|\X_{2},\U_{1}} & \leqdof\lgbrac{\snr+\snr^{\alpha}}+\lgbrac{\min\brac{\snr,\snr^{\alpha}}}.
\end{align}
 Combining the last two equations completes the proof.
\end{IEEEproof}

\begin{claim}
\label{claim:I(X1;Y1|U1,U2)}The term $I\brac{\X_{1};\Y_{1}\middle|\U_{1},\U_{2}}$
is lower bounded at high SNR as
\begin{eqnarray*}
I\brac{\X_{1};\Y_{1}\middle|\U_{1},\U_{2}} & \geqdof & \lgbrac{\snr^{1-\alpha}+\min\brac{\snr,\snr^{\alpha}}}+\brac{T-2}\lgbrac{1+\snr^{1-\alpha}}\\
 &  & {-}\:\lgbrac{\min\brac{\snr,\snr^{\alpha}}}.
\end{eqnarray*}
\end{claim}
\begin{IEEEproof}
This follows by using
\[
h\brac{\Y_{1}\middle|\U_{1},\U_{2}}\overset{}{\geqdof}\lgbrac{\snr+\snr^{\alpha}}+\lgbrac{\snr^{1-\alpha}+\min\brac{\snr,\snr^{\alpha}}}+\brac{T-2}\lgbrac{1+\snr^{1-\alpha}}
\]
from Claim \ref{claim:h(Y1|U1,U2)} and
\begin{align*}
h\brac{\Y_{1}\middle|\X_{1},\U_{1},\U_{2}} & \leq h\brac{\Y_{1}\middle|\X_{1},\U_{2}}\leqdof\lgbrac{\snr+\snr^{\alpha}}+\lgbrac{\min\brac{\snr,\snr^{\alpha}}}
\end{align*}
from Claim \ref{claim:h_Y1_U2_X1}.
\end{IEEEproof}
\begin{claim}
\label{claim:I(X1,U2;Y1|U1)_Lower_bound}The term $I\brac{\X_{1},\U_{2};\Y_{1}\middle|\U_{1}}$
is lower bounded at high SNR as
\[
I\brac{\X_{1},\U_{2};\Y_{1}\middle|\U_{1}}\geqdof\brac{T-1}\lgbrac{\snr^{1-\alpha}+\snr^{\alpha}}-\lgbrac{\min\brac{\snr,\snr^{\alpha}}}.
\]
\end{claim}
\begin{IEEEproof}
We have
\begin{eqnarray}
h\brac{\Y_{1}|\U_{1}} & = & h\brac{\boldsymbol{g}_{11}\X_{1}+\boldsymbol{g}_{21}\X_{2}+\Z_{1}\middle|\U_{1}}\nonumber \\
 & = & \sum_{i}h\brac{\boldsymbol{g}_{11}\x_{1,i}+\boldsymbol{g}_{21}\x_{2,i}+\z_{1,i}\middle|\cbrac{\boldsymbol{g}_{11}\x_{1,j}+\boldsymbol{g}_{21}\x_{2,j}+\z_{1,j}}_{j=1}^{i-1},\U_{1}}\\
 & \overset{\brac i}{\geqdof} & h\brac{\boldsymbol{g}_{11}\x_{1,1}+\boldsymbol{g}_{21}\x_{2,1}+\z_{1,1}\middle|\U_{1},\x_{1,1},\x_{2,1}}\nonumber \\
 &  & {+}\:\sum_{i=2}^{T}h\brac{\boldsymbol{g}_{11}\x_{1,i}+\boldsymbol{g}_{21}\x_{2,i}+\z_{1,i}\middle|\u_{1,i},\boldsymbol{g}_{21},\boldsymbol{g}_{11}}\\
 & \overset{\brac{ii}}{\geqdof} & \lgbrac{1+\snr+\snr^{\alpha}}+\brac{T-1}\lgbrac{1+\snr^{1-\alpha}+\snr^{\alpha}},\label{eq:h_Y1_given_U1}
\end{eqnarray}
where \sloppy $\brac i$ is due to the fact that conditioning reduces
entropy and Markovity $\brac{\boldsymbol{g}_{11}\x_{1,i}+\boldsymbol{g}_{21}\x_{2,i}+\z_{1,i}}-\brac{\u_{1,i},\boldsymbol{g}_{21},\boldsymbol{g}_{11}}-\brac{\cbrac{\boldsymbol{g}_{11}\x_{1,j}+\boldsymbol{g}_{21}\x_{2,j}+\z_{1,j}}_{j=1}^{i-1},\U_{1}}$.
In step $\brac{ii}$ we removed the contribution of $\boldsymbol{g}_{11}\u_{1,i}$
from the second term and used the structure $\x_{1,i}=\u_{1,i}+\x_{1\text{p},i}$,
where $\u_{1,i},\x_{1\text{p},i}$ are independent Gaussian random
variables and $\x_{1\text{p},i}$ has variance $1/\inr=\snr^{1-\alpha}$.
We also used Fact \ref{fact:Jensens_gap} together with the fact that
the channels are Gaussian distributed.

We also have
\begin{align}
h\brac{\Y_{1}\middle|\U_{2},\U_{1},\X_{1}} & \leq h\brac{\Y_{1}\middle|\U_{2},\X_{1}}\leqdof\lgbrac{\snr+\snr^{\alpha}}+\lgbrac{\min\brac{\snr,\snr^{\alpha}}},\label{eq:I(X1,U_2;Y1|U1)}
\end{align}
where the last step is using Claim \ref{claim:h_Y1_U2_X1} for $h\brac{\boldsymbol{Y}_{1}\middle|\X_{1},\U_{2}}$.
Using (\ref{eq:I(X1,U_2;Y1|U1)}) and (\ref{eq:h_Y1_given_U1}) completes
the proof.
\end{IEEEproof}
We collect the results from Claim \ref{claim:I(X1;Y1|U2)_lower_bound},
Claim \ref{claim:I(X2,U1;Y2)_lower_bound}, Claim \ref{claim:I(X1;Y1|U1,U2)}
and Claim \ref{claim:I(X1,U2;Y1|U1)_Lower_bound} in the second column
of Table \ref{tab:gDoF-inner-bounds_noFB_prelog}. In the third column
of Table \ref{tab:gDoF-inner-bounds_noFB_prelog}, we obtain the prelog
for the lower bounds.
\begin{table}[H]
\centering{}\caption{\label{tab:gDoF-inner-bounds_noFB_prelog}Lower bounds at high SNR
for the terms in the achievability region and their prelog}
\begin{tabular}{|c|c|c|c|c|}
\hline
\multirow{2}{*}{Term} & \multirow{2}{*}{Lower bound at high SNR} & \multicolumn{3}{c|}{Prelog of lower bound}\tabularnewline
\cline{3-5} \cline{4-5} \cline{5-5}
 &  & $\alpha<1/2$ & $1/2<\alpha<1$ & $\alpha>1$\tabularnewline
\hline
\hline
$I\brac{\X_{1};\Y_{1}\middle|\U_{2}}$ & $\begin{array}{c}
\brac{T-1}\lgbrac{\snr}\\
-\lgbrac{\min\brac{\snr,\snr^{\alpha}}}
\end{array}$ & $\brac{T-1}-\alpha$ & $\brac{T-1}-\alpha$ & $\brac{T-2}$\tabularnewline
\hline
$I\brac{\X_{2},\U_{1};\Y_{2}}$ & $\begin{array}{c}
\brac{T-1}\lgbrac{\snr+\snr^{\alpha}}\\
-\lgbrac{\min\brac{\snr,\snr^{\alpha}}}
\end{array}$ & $\brac{T-1}-\alpha$ & $\brac{T-1}-\alpha$ & $\brac{T-1}\alpha-1$\tabularnewline
\hline
$I\brac{\X_{1};\Y_{1}\middle|\U_{1},\U_{2}}$ & $\begin{array}{c}
\lgbrac{\snr^{1-\alpha}+\min\brac{\snr,\snr^{\alpha}}}\\
+\brac{T-2}\lgbrac{1+\snr^{1-\alpha}}\\
-\lgbrac{\min\brac{\snr,\snr^{\alpha}}}
\end{array}$ & $\brac{T-1}\brac{1-\alpha}-\alpha$ & $\brac{T-2}\brac{1-\alpha}$ & $0$\tabularnewline
\hline
$I\brac{\X_{1},\U_{2};\Y_{1}\middle|\U_{1}}$ & $\begin{array}{c}
\brac{T-1}\lgbrac{\snr^{1-\alpha}+\snr^{\alpha}}\\
-\lgbrac{\min\brac{\snr,\snr^{\alpha}}}
\end{array}$ & $\brac{T-1}\brac{1-\alpha}-\alpha$ & $\brac{T-2}\alpha$ & $\brac{T-1}\alpha-1$\tabularnewline
\hline
\end{tabular}
\end{table}
Using the prelog of the lower bounds from Table \ref{tab:gDoF-inner-bounds_noFB_prelog}
in (\ref{eq:ach_nofb}), using symmetry of the terms and using only
the active inequalities, it can be verified that the prelog region
in Table \ref{tab:gdof_no_FB} is achievable.

\section{Noncoherent IC with feedback\label{sec:noncoh_FFIC_FB}}

In this section, we provide our results for the noncoherent rate-splitting
scheme for the noncoherent IC with feedback and compare the achievable
prelog with a standard training-based scheme. We also compare the
performance with the TIN and TDM schemes.
\begin{thm}
For a noncoherent IC with feedback, the prelog region given in Table
\ref{tab:gdof_FB} is achievable:\label{thm:noncoh_IC_FB}
\begin{table}[H]
\centering{}\caption{Achievable prelog region for different regimes of $\alpha$.\label{tab:gdof_FB}}
\begin{tabular}{|c|c|c|}
\hline
 $\alpha<1/2$ &  $1/2\leq\alpha\leq1$  & $\alpha>1$\tabularnewline
\hline
\hline
$\begin{array}{c}
d_{1}\leq\brac{1-1/T}-2\alpha/T\\
d_{2}\leq\brac{1-1/T}-2\alpha/T\\
d_{1}+d_{2}\leq2\brac{1-1/T}-\alpha\brac{1+1/T}
\end{array}$ & $\begin{array}{c}
d_{1}\leq\brac{1-2/T}\\
d_{2}\leq\brac{1-2/T}\\
d_{1}+d_{2}\leq\brac{2-3/T}-\alpha\brac{1-1/T}
\end{array}$ & $\begin{array}{c}
d_{1}+d_{2}\leq\brac{1-1/T}\alpha-1/T\end{array}$\tabularnewline
\hline
\end{tabular}
\end{table}
\end{thm}
\begin{IEEEproof}
This is obtained using the block Markov scheme of \cite[Lemma 1]{suh_tse_fb_gaussian}
for the noncoherent case. We use a rate-splitting scheme based on
the average interference-to-noise ratio and noncoherent decoding at
the receivers. We use the block Markov scheme from \cite[Lemma 1]{suh_tse_fb_gaussian}
with a total size of blocks $B$.

In  block $b$, the message for User 1 is split into two parts, a
common message $w_{\text{c}1}^{\brac b}$ at rate $R_{\text{c}1}$
and a private message $w_{\text{p}1}^{\brac b}$ at rate $R_{\text{p}1}$.
The transmitted vector symbols at Transmitter 1 are of the form $\X_{1}=\U_{1}+\X_{\text{p}1}$
where $\U_{1},\X_{\text{p}1}$ are independent Gaussian vectors of
length $T$. The power of each element of $\X_{\text{p}1}$ is $1/\inr$
and the power of each element of $\U_{1}$ is $P-1/\inr$. The Transmitter
1 is able to decode $w_{\text{c}2}^{\brac{b-1}}$ using feedback.
The messages $w_{\text{c}2}^{\brac{b-1}}$ , $w_{\text{c}1}^{\brac{b-1}}$
and $w_{\text{c}1}^{\brac b}$ are mapped into $\U_{1}$ in  $b^{\text{th}}$
block. The private message $w_{\text{p}1}^{\brac b}$ is mapped into
$\X_{\text{p}1}$.

For User 2, in block $b$, we have a common message $w_{\text{c}2}^{\brac b}$
at rate $R_{\text{c}2}$ and a private message $w_{\text{p}2}^{\brac b}$
at rate $R_{\text{p}2}$. The transmitted vector symbols at Transmitter
2 are of the form $\X_{2}=\U_{2}+\X_{\text{p}2}$ where $\U_{2},\X_{\text{p}2}$
are independent Gaussian vectors of length $T$. The power of each
element of $\X_{\text{p}2}$ is $1/\inr$ and the power of each element
of $\U_{2}$ is $P-1/\inr$. The Transmitter 2 is able to decode $w_{\text{c}1}^{\brac{b-1}}$
using feedback. The messages $w_{\text{c}1}^{\brac{b-1}}$ , $w_{\text{c}2}^{\brac{b-1}}$
and $w_{\text{c}2}^{\brac b}$ are mapped into $\U_{2}$ in  $b^{\text{th}}$
block. The private message $w_{\text{p}2}^{\brac b}$ is mapped into
$\X_{\text{p}2}$.

The messages $w_{\text{c}1}^{\brac b},w_{\text{p}1}^{\brac b},w_{\text{c}2}^{\brac b},w_{\text{p}2}^{\brac b}$
with $b=0,B$ are set to be fixed and known to all transmitters and
 receivers. After $B$ blocks, the receivers perform noncoherent backward
decoding. Receiver 1 uses the symbols received in block $b$ and decodes
$w_{\text{c}1}^{\brac{b-1}},w_{\text{c}2}^{\brac{b-1}},w_{\text{p}2}^{\brac b}$
assuming $w_{\text{c}1}^{\brac b},w_{\text{c}2}^{\brac b}$ are decoded
from the symbols received in block $b+1$. Receiver 2 uses the symbols
received in block $b$ and decodes $w_{\text{c}1}^{\brac{b-1}},w_{\text{c}2}^{\brac{b-1}},w_{\text{p}1}^{\brac b}$
assuming $w_{\text{c}1}^{\brac b},w_{\text{c}2}^{\brac b}$ are decoded
from the symbols received in block $b+1$. The details of the coding
scheme and its analysis are in Section \ref{subsec:noncoh_IC_FB}.
\end{IEEEproof}
We now obtain the prelog of a standard training-based scheme for the
noncoherent IC with feedback.
\begin{thm}
\label{thm:training_FB}A standard training-based scheme for the noncoherent
IC with feedback can achieve the prelog region described in Table
\ref{tab:gdof_FB_with_training}.
\begin{table}[H]
\centering{}\caption{Achievable prelog region for different regimes of $\alpha$.\label{tab:gdof_FB_with_training}}
\begin{tabular}{|c|c|}
\hline
$\alpha\leq1$ & $\alpha>1$\tabularnewline
\hline
\hline
$\begin{array}{c}
d_{1}\leq\brac{1-2/T}\\
d_{2}\leq\brac{1-2/T}\\
d_{1}+d_{2}\leq\brac{1-2/T}\brac{2-\alpha}
\end{array}$ & $\begin{array}{c}
d_{1}+d_{2}\leq\brac{1-2/T}\alpha\end{array}$\tabularnewline
\hline
\end{tabular}
\end{table}
\end{thm}
\begin{IEEEproof}
For training, in every coherence period of $T$ symbols, the first
transmitter can send a known symbol while the second transmitter remains
turned off; with this both receivers can estimate the channels from
the first transmitter. Next the second transmitter can send a known
symbol while the first transmitter remains turned off; with this both
receivers can estimate the channels from the second transmitter. The
rest of the symbols can be used to transmit data using a block Markov
scheme similar to that described in Theorem \ref{thm:noncoh_IC_FB}.
The detailed analysis for obtaining the prelog is given in \ifarxiv  Appendix \ref{app:Trainin_Scheme_FB}\else  \cite[Appendix D]{Joyson_noncoh_IC_arxiv}\fi.
\end{IEEEproof}
\begin{rem}
The capacity region of the coherent FF-IC with feedback is known within
a constant gap from \cite{joyson_fading_TCOM} and hence its gDoF
region is known. The prelog region from the above theorem is the same
as the gDoF region for the coherent case with a multiplication factor
of $\brac{1-2/T}$. Hence the prelog obtained in Theorem \ref{thm:training_FB}
is the best among any scheme that uses two symbols for training in
every coherence period of $T$ symbols.
\end{rem}

\subsection{Discussion}

\begin{figure}
\centering{}\includegraphics[scale=0.6]{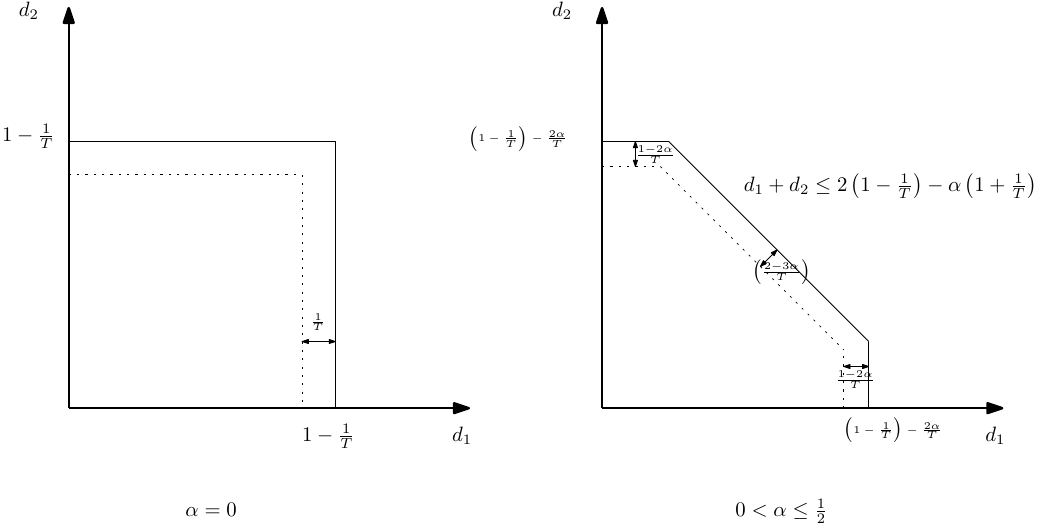}\caption{Prelog region for $\alpha<1/2$,$T\protect\geq2$. The solid line
gives the prelog region achievable for a noncoherent scheme and the
dotted line gives the prelog region for the scheme that uses 2 symbols
for training.\label{fig:gDoF-fb-set1}}
\end{figure}

\begin{figure}
\centering{}\includegraphics[scale=0.6]{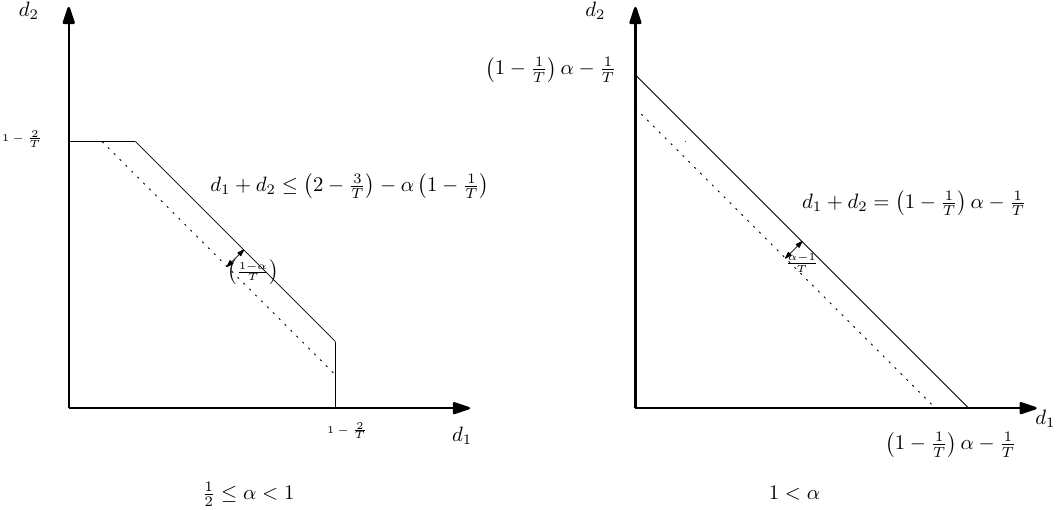}\caption{Prelog region for $1/2<\alpha$, $T\protect\geq2$. The solid line
is prelog achievable for a noncoherent scheme and the dotted line
is the prelog for a scheme that uses 2 symbols for training.\label{fig:gDoF-fb-set2}}
\end{figure}

In Figure \ref{fig:gDoF-fb-set1} and Figure \ref{fig:gDoF-fb-set2},
the prelog region achievable using our noncoherent scheme is compared
with the prelog region achievable using the aforementioned training-based
scheme. It can be observed that our noncoherent scheme outperforms
the standard training-based scheme.

\begin{figure}
\begin{minipage}[c][1\totalheight][t]{0.45\textwidth}%
\begin{center}
\includegraphics[scale=0.6]{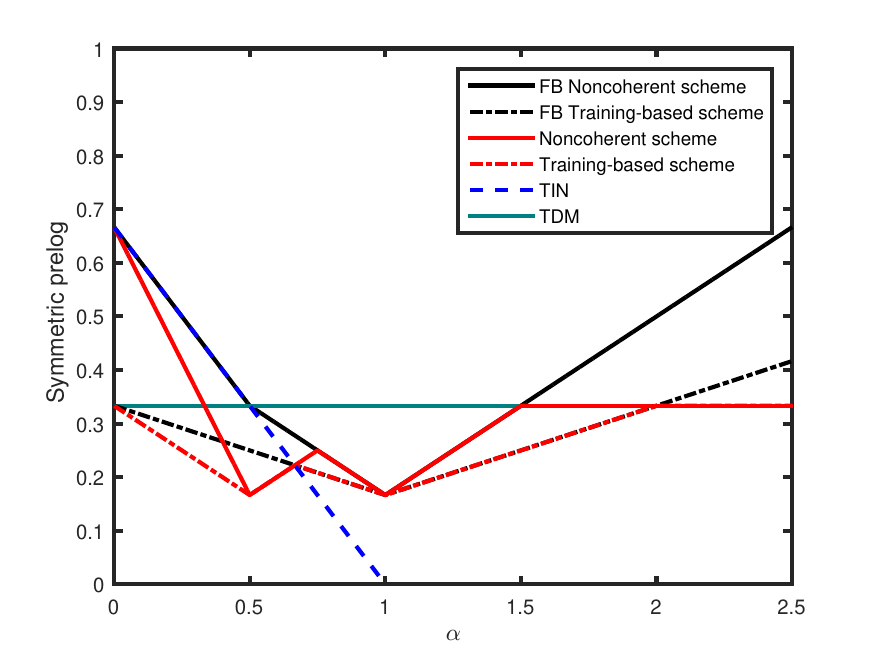}\caption{\label{fig:Symmetric_gDoF_fb_T=00003D3}Symmetric achievable prelog
for coherence time $T=3$: feedback and nonfeedback cases.}
\par\end{center}%
\end{minipage}\hfill{}%
\begin{minipage}[c][1\totalheight][t]{0.45\textwidth}%
\begin{center}
\includegraphics[scale=0.6]{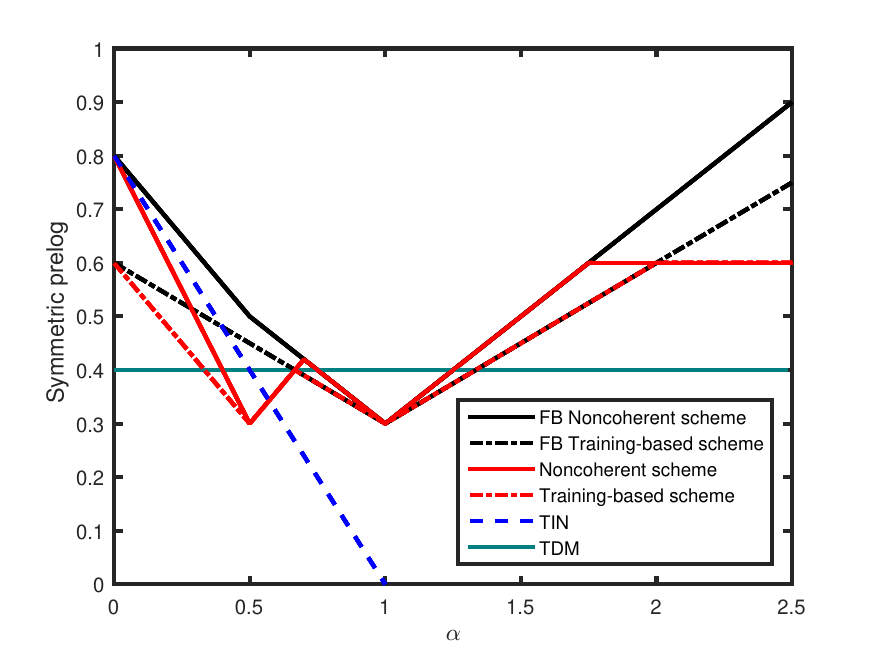}\caption{\label{fig:Symmetric_gDoF_fb}Symmetric achievable prelog for coherence
time $T=5$: feedback and nonfeedback cases.}
\par\end{center}%
\end{minipage}
\end{figure}

In Figure \ref{fig:Symmetric_gDoF_fb_T=00003D3}, we give the achievable
symmetric prelog with coherence time $T=3$ for our noncoherent rate-splitting
scheme and the aforementioned training-based scheme for the feedback
case. We give similar plots in Figure \ref{fig:Symmetric_gDoF_fb}
for coherence time $T=5$. We also include the prelog of the nonfeedback
schemes from Section \ref{sec:NC_IC_noFB} in the figures. We had
noticed in Section I that with feedback, the performance of noncoherent
rate-splitting schemes is in general better than the TIN scheme. There
are a few exceptions: when $T=2$ and $\alpha<1$, it can be calculated
from Table \ref{tab:gdof_FB} and (\ref{eq:training_dof_fb}) that
the TIN scheme outperforms our noncoherent strategy with feedback.
With $T=3$ and $\alpha\leq.5$, our noncoherent rate-splitting strategy
in the presence of feedback has same prelog as the TIN scheme.

The noncoherent rate-splitting scheme attempts to decode part of the
interfering message at the transmitter, and use it in subsequent transmissions.
The rate that can be decoded at the transmitter using the feedback
increases with $T$. For very weak interference level, the noncoherent
rate-splitting scheme has a disadvantage as we explained in the discussion
in Section \ref{sec:Introduction} together with Figures \ref{fig:Symmetric-gDoF_T=00003D4}
and \ref{fig:Symmetric-gDoF_T=00003D6}. The advantage gained by decoding
at the transmitter outweighs this disadvantage when $T\geq3$.

The TDM scheme outperforms other schemes for a region of $\alpha$
close to $1$. This behavior can be explained similar to what we did
in Section \ref{subsec:Discussion_noFB}. When $\alpha=1$, the noncoherent
rate-splitting scheme gives a prelog of $\brac{1/2}\brac{1-2/T}$
and the TDM scheme gives a prelog of $\brac{1/2}\brac{1-1/T}$. Hence
for $\alpha=1$, the noncoherent scheme effectively behaves as a TDM
scheme that uses two symbols to train, but the TDM scheme can actually
be implemented with only one training symbol.

\subsection{Coding Scheme and Analysis of the Rate Region \label{subsec:noncoh_IC_FB}}

We describe the coding scheme starting with a general input distribution
and then we evaluate the prelog region for Gaussian inputs.

\noindent \textbf{Encoding: }Fix a joint distribution $p\brac{\U_{1}}p\brac{\U_{2}}p\brac{\X_{1}\middle|\U_{1}}p\brac{\X_{2}\middle|\U_{2}}$
where $\U_{1},\U_{2},\X_{1},\X_{2}$ are vectors of length $T$. Generate
$2^{NT\brac{2R_{\text{c}1}+R_{\text{c}2}}}$ codewords $\U_{1}^{N}\brac{i,j,k}$
with $i,k\in\cbrac{1,\ldots,2^{NTR_{\text{c}1}}}$, $j\in\cbrac{1,\ldots,2^{NTR_{\text{c}2}}}$
according to $\prod_{l=1}^{N}p\brac{\U_{1(l)}}$. For each codeword
$\U_{1}^{N}\brac{i,j,k}$, generate $2^{NTR_{\text{p}1}}$ codewords
$\X_{1}^{N}\brac{i,j,k,l}$ with $l\in\cbrac{1,\ldots,2^{NTR_{\text{p}1}}}$
according to $\prod_{l=1}^{N}p\brac{\X_{1(l)}\middle|\U_{1(l)}}$.

Similarly generate $2^{NT\brac{2R_{\text{c}2}+R_{\text{c}1}}}$ codewords
$\U_{2}^{N}\brac{j,i,r}$ with $j,r\in\cbrac{1,\ldots,2^{NTR_{\text{c}2}}}$,
$i\in\cbrac{1,\ldots,2^{NTR_{\text{c}1}}}$. For each codeword $\U_{2}^{N}\brac{j,i,r}$,
generate $2^{NTR_{\text{p}2}}$ codewords $\X_{2}^{N}\brac{j,i,r,s}$
with $s\in\cbrac{1,\ldots,2^{NTR_{\text{p}2}}}$ according to $\prod_{s=1}^{N}p\brac{\X_{2(s)}\middle|\U_{2(s)}}$.

At \sloppy block $b$, transmitter 1 has uniformly random messages
$w_{\text{c}1}^{\brac b}\in\cbrac{1,\ldots,2^{NTR_{\text{c}1}}},w_{\text{p}1}^{\brac b}\in\cbrac{1,\ldots,2^{NTR_{\text{p}1}}}$
to transmit and transmitter 2 has uniformly random messages $w_{\text{c}2}^{\brac b}\in\cbrac{1,\ldots,2^{NTR_{\text{c}2}}},w_{\text{p}2}^{\brac b}\in\cbrac{1,\ldots,2^{NTR_{\text{p}2}}}$
to transmit. Using the symbols $\Y_{1}^{N,\brac{b-1}}$ obtained through
feedback, transmitter 1 tries to noncoherently decode $\hat{w}_{2c}^{\brac{b-1}}=\hat{k}$
by finding unique $\hat{k}$ such that
\begin{align*}
 & \Big(\U_{1}^{N}\big(w_{\text{c}1}^{\brac{b-2}},w_{\text{c}2}^{\brac{b-2}},w_{\text{c}1}^{\brac{b-1}}\big),\X_{1}^{N}\big(w_{\text{c}1}^{\brac{b-2}},w_{\text{c}2}^{\brac{b-2}},w_{\text{c}1}^{\brac{b-1}},w_{\text{p}1}^{\brac{b-1}}\big),\\
 & \quad\U_{2}^{N}\big(w_{\text{c}2}^{\brac{b-2}},w_{\text{c}1}^{\brac{b-2}},\hat{k}\big),\Y_{1}^{N,\brac{b-1}}\Big)\in\mathcal{A}_{\epsilon}^{\brac N}.
\end{align*}
 where $\mathcal{A}_{\epsilon}^{\brac N}$ indicates the set of jointly
typical sequences. Transmitter \sloppy 1 already knows $w_{\text{c}1}^{\brac{b-2}},w_{\text{c}1}^{\brac{b-1}},w_{\text{p}1}^{\brac{b-1}}$.
Also $w_{\text{c}2}^{\brac{b-2}}$ is assumed to be correctly decoded
in the previous block at transmitter 1 and $w_{\text{c}1}^{\brac{b-2}}$
is assumed to be correctly decoded in the previous block at transmitter
2. The current noncoherent decoding at transmitter 1 is performed
with vanishing error probability if
\begin{equation}
TR_{\text{c}2}\leq I\brac{\U_{2};\Y_{1}\middle|\X_{1}}.\label{eq:fb_markov_eq1}
\end{equation}

Based on $\hat{w}_{2c}^{\brac{b-1}}$, transmitter 1 then sends $\X_{1}^{N}\big(w_{\text{c}1}^{\brac{b-1}},\hat{w}_{2c}^{\brac{b-1}},w_{\text{c}1}^{\brac b},w_{\text{p}1}^{\brac b}\big)$.
Similarly transmitter 2 decodes $\hat{w}_{1c}^{\brac{b-1}}$ and sends
$\X_{2}^{N}\big(w_{\text{c}2}^{\brac{b-1}},\hat{w}_{1c}^{\brac{b-1}},w_{\text{c}2}^{\brac b},w_{\text{p}2}^{\brac b}\big)$.
The messages $w_{\text{c}1}^{\brac b},w_{\text{p}1}^{\brac b},w_{\text{c}2}^{\brac b},w_{\text{p}2}^{\brac b}$
for $b=0,B$ can be set to be fixed and known to all transmitters
and receivers.

\noindent \textbf{Decoding:} After receiving $B$ blocks, each receiver
performs backward decoding. At receiver 1, block $b$ is decoded assuming
block $b+1$ is correctly decoded. From block $b+1$, $w_{\text{c}2}^{\brac b},w_{1c}^{\brac b}$
is assumed to be available at receiver 1 after successful decoding.
Now using the symbols from block $b$, receiver 1 finds unique triplet
$\brac{\hat{i},\hat{j},\hat{l}}$ such that
\[
\brac{\U_{1}^{N}\big(\hat{i},\hat{j},w_{\text{c}1}^{\brac b}\big),\X_{1}^{N}\big(\hat{i},\hat{j},w_{\text{c}1}^{\brac b},\hat{l}\big),\U_{2}^{N}\big(\hat{j,}\hat{i},w_{\text{c}2}^{\brac b}\big),\Y_{1}^{N,\brac b}}\in\mathcal{A}_{\epsilon}^{\brac N}.
\]
Similarly receiver 2 finds unique triplet $\brac{\hat{j},\hat{i},\hat{s}}$
such that
\[
\brac{\U_{2}^{N}\big(\hat{j},\hat{i},w_{\text{c}1}^{\brac b}\big),\X_{2}^{N}\big(\hat{j},\hat{i},w_{\text{c}2}^{\brac b},\hat{s}\big),\U_{1}^{N}\big(\hat{i},\hat{j,}w_{\text{c}1}^{\brac b}\big),\Y_{2}^{N,\brac b}}\in\mathcal{A}_{\epsilon}^{\brac N}.
\]

\noindent \textbf{Error analysis: }We give the sketch of error analysis
at receiver 1 assuming $\big(w_{\text{c}1}^{\brac{b-1}},w_{\text{c}2}^{\brac{b-1}},w_{\text{p}1}^{\brac b}\big)=\brac{1,1,1}$
was sent through block $b-1$ and block $b$. We assume that there
was no backward decoding error, \emph{i.e.}, $\big(w_{\text{c}1}^{\brac b},w_{\text{c}2}^{\brac b}\big)$
was correctly decoded. Let $\mathcal{E}_{ijl}$ be the event $\cbrac{\brac{\U_{1}^{N}\big(\hat{i},\hat{j},w_{\text{c}1}^{\brac b}\big),\X_{1}^{N}\big(\hat{i},\hat{j},w_{\text{c}1}^{\brac b},\hat{k}\big),\U_{2}^{N}\big(\hat{j,}\hat{i},w_{\text{c}2}^{\brac b}\big),\Y_{1}^{N,\brac b}}\in\mathcal{A}_{\epsilon}^{\brac N}}$
for given $i,j,l$. By AEP, the probability of $\mathcal{E}_{111}$
approaches unity. The error probability is thus captured by the following
equation using standard analysis similar to that in \cite[Appendix B]{suh_tse_fb_gaussian}.
\begin{eqnarray}
\text{Pr}\Big(\union_{(i,j,l)\neq\brac{1,,1,1}}\mathcal{E}_{ijl}\Big) & \leq & \sum_{i\neq1,j\neq1,l\neq1}\text{Pr}\brac{\mathcal{E}_{ijl}}+\sum_{i=1,j=1,l\neq1}\text{Pr}\brac{\mathcal{E}_{ijl}}+\sum_{i=1,j\neq1,l=1}\text{Pr}\brac{\mathcal{E}_{ijl}}\nonumber \\
 &  & {+}\:\sum_{i\neq1,j=1,l=1}\text{Pr}\brac{\mathcal{E}_{ijl}}+\sum_{i\neq1,j\neq1,l=1}\text{Pr}\brac{\mathcal{E}_{ijl}}+\sum_{i\neq1,j=1,l\neq1}\text{Pr}\brac{\mathcal{E}_{ijl}}\nonumber \\
 &  & {+}\:\sum_{i=1,j\neq1,l\neq1}\text{Pr}\brac{\mathcal{E}_{ijl}}\nonumber \\
 & \leq & 2^{N\brac{TR_{\text{c}1}+TR_{\text{c}2}+TR_{\text{p}1}-I\brac{\X_{1},\U_{2};\Y_{1}}+\epsilon}}+2^{N\brac{TR_{\text{p}1}-I\brac{\X_{1};\Y_{1}|\U_{1},\U_{2}}+\epsilon}}\nonumber \\
 &  & {+}\:2^{N\brac{TR_{\text{c}2}-I\brac{\X_{1},\U_{2};\Y_{1}}+\epsilon}}+2^{N\brac{TR_{\text{c}1}-I\brac{\X_{1},\U_{2};\Y_{1}}+\epsilon}}\nonumber \\
 &  & {+}\:2^{N\brac{TR_{\text{c}1}+TR_{\text{c}2}-I\brac{\X_{1},\U_{2};\Y_{1}}+\epsilon}}+2^{N\brac{TR_{\text{c}1}+TR_{\text{p}1}-I\brac{\X_{1},\U_{2};\Y_{1}}+\epsilon}}\nonumber \\
 &  & {+}\:2^{N\brac{TR_{\text{c}2}+TR_{\text{p}1}-I\brac{\X_{1},\U_{2};\Y_{1}}+\epsilon}}\label{eq:fb_markov_eq2}
\end{eqnarray}
Combining (\ref{eq:fb_markov_eq1}) and (\ref{eq:fb_markov_eq2}),
and considering similar analysis for user 2, we get the following
equations for achievability:\begin{subequations}\label{eq:ach_fb_markov}
\begin{eqnarray}
TR_{\text{c}2} & \leq & I\brac{\U_{2};\Y_{1}\middle|\X_{1}},\label{eq:ach_fb_markov1}\\
TR_{\text{p}1} & \leq & I\brac{\X_{1};\Y_{1}\middle|\U_{1},\U_{2}},\label{eq:ach_fb_markov2}\\
T\brac{R_{\text{c}1}+R_{\text{c}2}+R_{\text{p}1}} & \leq & I\brac{\X_{1},\U_{2};\Y_{1}},\label{eq:ach_fb_markov3}\\
TR_{\text{c}1} & \leq & I\brac{\U_{1};\Y_{2}\middle|\X_{2}},\label{eq:ach_fb_markov4}\\
TR_{\text{p}2} & \leq & I\brac{\X_{2};\Y_{2}\middle|\U_{2},\U_{1}},\label{eq:ach_fb_markov5}\\
T\brac{R_{\text{c}1}+R_{\text{c}2}+R_{\text{p}2}} & \leq & I\brac{\X_{2},\U_{1};\Y_{2}}.\label{eq:ach_fb_markov6}
\end{eqnarray}
\end{subequations}

After performing Fourier-Motzkin elimination similar to that in \cite[Appendix B]{suh_tse_fb_gaussian},
we obtain the following achievability region with $R_{1}=R_{\text{c}1}+R_{\text{p}1},R_{2}=R_{\text{c}2}+R_{\text{p}2}$:\begin{subequations}\label{eq:ach_fb}
\begin{eqnarray}
TR_{1} & \leq & I\brac{\X_{1},\U_{2};\Y_{1}},\label{eq:ach_fb1}\\
TR_{1} & \leq & I\brac{\U_{1};\Y_{2}\middle|\X_{2}}+I\brac{\X_{1};\Y_{1}\middle|\U_{1},\U_{2}},\label{eq:ach_fb2}\\
TR_{2} & \leq & I\brac{\X_{2},\U_{1};\Y_{2}},\label{eq:ach_fb3}\\
TR_{2} & \leq & I\brac{\U_{2};\Y_{1}\middle|\X_{1}}+I\brac{\X_{2};\Y_{2}\middle|\U_{1},\U_{2}},\label{eq:ach_fb4}\\
T\brac{R_{1}+R_{2}} & \leq & I\brac{\X_{1};\Y_{1}\middle|\U_{1},\U_{2}}+I\brac{\X_{2},\U_{1};\Y_{2}},\label{eq:ach_fb5}\\
T\brac{R_{1}+R_{2}} & \leq & I\brac{\X_{2};\Y_{2}\middle|\U_{1},\U_{2}}+I\brac{\X_{1},\U_{2};\Y_{1}}.\label{eq:ach_fb6}
\end{eqnarray}
\end{subequations} We choose $\U_{k}$ as a vector of length $T$
with i.i.d. $\mathcal{CN}\brac{0,\lambda_{\text{c}}}$ elements and
$\X_{\text{p}k}$ as a vector of length $T$ with i.i.d. $\mathcal{CN}\brac{0,\lambda_{\text{p}}}$
elements for $k\in\cbrac{1,2}$. The random variables are chosen independent
of each other so that the set $\cbrac{\U_{1},\U_{2},\X_{\text{p}1},\X_{\text{p}2}}$
is mutually independent. We use $\X_{1}=\U_{1}+\X_{\text{p}1},\quad\X_{2}=\U_{2}+\X_{\text{p}2}$
where $\lambda_{\text{c}}+\lambda_{\text{p}}=P$ and $\lambda_{\text{p}}=\min\brac{1/\inr,P}$
similar to \cite{suh_tse_fb_gaussian,joyson_fading_TCOM}. For prelog
characterization, we can assume $P\cdot\inr=\snr\cdot\snr^{\alpha-1}=\snr^{\alpha}>1$.
Hence, we have $\lambda_{\text{p}}=1/\inr$.

The prelog results in Table \ref{tab:gdof_FB} can be obtained by
evaluating the rate region (\ref{eq:ach_fb}) for our choice of input
distribution. Note that the joint distribution of $\brac{\X_{1},\Y_{1},\U_{1},\X_{2},\Y_{2},\U_{2}}$
in its \emph{single letter form} is the same as that for the nonfeedback
case in Section \ref{subsec:no_fb}, hence we can carry over the inequalities
for the single letter mutual information terms from Section \ref{subsec:no_fb}.
We will use Claim \ref{claim:I(X2,U1;Y2)_lower_bound} and Claim \ref{claim:I(X1;Y1|U1,U2)}
from Section \ref{subsec:no_fb} to bound $I\brac{\X_{2},\U_{1};\Y_{2}}$
and $I\brac{\X_{1};\Y_{1}|\U_{1},\U_{2}}$ respectively. We bound
the term $I\brac{\U_{2};\Y_{1}|\X_{1}}$ with the following claim.
\begin{claim}
\label{claim:simplifyI(U2;Y1|X1)feedback}The term $I\brac{\U_{2};\Y_{1}\middle|\X_{1}}$
is lower bounded at high SNR as
\[
I\brac{\U_{2};\Y_{1}\middle|\X_{1}}\geqdof\brac{T-1}\lgbrac{\snr^{\alpha}}-\lgbrac{\min\brac{\snr,\snr^{\alpha}}}.
\]
\end{claim}
\begin{IEEEproof}
We have
\begin{eqnarray}
h\brac{\rline{\Y_{1}}\X_{1}} & = & h\brac{\boldsymbol{g}_{11}\X_{1}+\boldsymbol{g}_{21}\X_{2}+\Z_{1}\middle|\X_{1}}\\
 & = & \sum_{i}h\brac{\boldsymbol{g}_{11}\x_{1,i}+\boldsymbol{g}_{21}\x_{2,i}+\z_{1,i}\middle|\cbrac{\boldsymbol{g}_{11}\x_{1,j}+\boldsymbol{g}_{21}\x_{2,j}+\z_{1,j}}_{j=1}^{i-1},\X_{1}}\\
 & \overset{\brac i}{\geqdof} & h\brac{\boldsymbol{g}_{11}\x_{1,1}+\boldsymbol{g}_{21}\x_{2,1}+\z_{1,1}\middle|\x_{2,1},\X_{1}}\nonumber \\
 &  & {+}\:\sum_{i=2}^{T}h\brac{\boldsymbol{g}_{11}\x_{1,i}+\boldsymbol{g}_{21}\x_{2,i}+\z_{1,i}\middle|\X_{1},\boldsymbol{g}_{21},\boldsymbol{g}_{11}}\\
 & \overset{\brac{ii}}{\geqdof} & \lgbrac{\snr+\snr^{\alpha}}+\brac{T-1}\lgbrac{\snr^{\alpha}},\label{eq:hY1|X1}
\end{eqnarray}
where $\brac i$ is due to  the fact that conditioning reduces entropy
and Markovity $\brac{\boldsymbol{g}_{11}\x_{1,i}+\boldsymbol{g}_{21}\x_{2,i}+\z_{1,i}}-\brac{\X_{1},\boldsymbol{g}_{21},\boldsymbol{g}_{11}}-\brac{\cbrac{\boldsymbol{g}_{11}\x_{1,j}+\boldsymbol{g}_{21}\x_{2,j}+\z_{1,j}}_{j=1}^{i-1},\X_{1}}$
and $\brac{ii}$ is using the property of Gaussians for the terms
$h\brac{\boldsymbol{g}_{11}\x_{1,1}+\boldsymbol{g}_{21}\x_{2,1}+\z_{1,1}\middle|\x_{2,1},\X_{1}}$,
$h\brac{\boldsymbol{g}_{11}\x_{1,i}+\boldsymbol{g}_{21}\x_{2,i}+\z_{1,i}\middle|\X_{1},\boldsymbol{g}_{21},\boldsymbol{g}_{11}}$
and using Fact \ref{fact:Jensens_gap}. Using (\ref{eq:hY1|X1}) and
$h\brac{\Y_{1}\middle|\U_{2},\X_{1}}\leqdof\lgbrac{\snr+\snr^{\alpha}}+\lgbrac{\min\brac{\snr,\snr^{\alpha}}}$
from Claim \ref{claim:h_Y1_U2_X1} completes the proof.
\end{IEEEproof}

Using Claim \ref{claim:I(X2,U1;Y2)_lower_bound}, Claim \ref{claim:I(X1;Y1|U1,U2)}
and Claim \ref{claim:simplifyI(U2;Y1|X1)feedback}, we have the lower
bounds for the terms in the achievability region in the second column
of Table \ref{tab:gDoF-inner-bounds_FB_prelog}. In the third column
of Table \ref{tab:gDoF-inner-bounds_FB_prelog}, we obtain the prelog
for the lower bounds.
\begin{table}[H]
\centering{}\caption{\label{tab:gDoF-inner-bounds_FB_prelog}Lower bounds at high SNR for
the terms in the achievability region and their prelog}
\begin{tabular}{|c|c|c|c|c|}
\hline
\multirow{2}{*}{Term} & \multirow{2}{*}{Lower bound at high SNR} & \multicolumn{3}{c|}{Prelog of lower bound}\tabularnewline
\cline{3-5} \cline{4-5} \cline{5-5}
 &  & $\alpha<1/2$ & $1/2<\alpha<1$ & $\alpha>1$\tabularnewline
\hline
\hline
$I\brac{\X_{2},\U_{1};\Y_{2}}$ & $\begin{array}{c}
\brac{T-1}\lgbrac{\snr+\snr^{\alpha}}\\
-\lgbrac{\min\brac{\snr,\snr^{\alpha}}}
\end{array}$ & $\brac{T-1}-\alpha$ & $\brac{T-1}-\alpha$ & $\brac{T-1}\alpha-1$\tabularnewline
\hline
$I\brac{\X_{1};\Y_{1}\middle|\U_{1},\U_{2}}$ & $\begin{array}{c}
\lgbrac{\snr^{1-\alpha}+\min\brac{\snr,\snr^{\alpha}}}\\
+\brac{T-2}\lgbrac{1+\snr^{1-\alpha}}\\
-\lgbrac{\min\brac{\snr,\snr^{\alpha}}}
\end{array}$ & $\brac{T-1}\brac{1-\alpha}-\alpha$ & $\brac{T-2}\brac{1-\alpha}$ & $0$\tabularnewline
\hline
$I\brac{\U_{2};\Y_{1}\middle|\X_{1}}$ & $\begin{array}{c}
\brac{T-1}\lgbrac{\snr^{\alpha}}\\
-\lgbrac{\min\brac{\snr,\snr^{\alpha}}}
\end{array}$ & $\brac{T-2}\alpha$ & $\brac{T-2}\alpha$ & $\brac{T-1}\alpha-1$\tabularnewline
\hline
\end{tabular}
\end{table}
Using the prelog of the lower bounds from Table \ref{tab:gDoF-inner-bounds_FB_prelog}
in (\ref{eq:ach_fb}), using symmetry of the terms and using only
the active inequalities, it can be verified that the prelog region
in Table \ref{tab:gdof_FB} is achievable.

\section{Conclusions and remarks \label{sec:Conclusions-and-remarks}}

We studied the 2-user noncoherent IC with different achievability
strategies. We observed that a standard training-based scheme is suboptimal
in terms of prelog. Depending on the level of interference, a noncoherent
scheme or a TIN scheme or a TDM scheme can give superior performance
than the standard training-based scheme. Thus, the result for single
user noncoherent channels that training-based schemes are DoF optimal
does not extend to the gDoF of the noncoherent IC. Our current results
are on inner bounds, outer bounds are still open.

\bibliographystyle{IEEEtran}
\bibliography{C:/Users/joyso/OneDrive/Desktop/Project/Bibtex/references}

\appendices{}

\section{Proof of Claim \ref{claim:h(Y1|U1,U2)} \label{app:h(Y1|U1,U2)proof}}

In this appendix, we prove that
\begin{align*}
h\brac{\Y_{1}\middle|\U_{1},\U_{2}} & \overset{}{\geqdof}\lgbrac{\snr+\snr^{\alpha}}+\lgbrac{\snr^{1-\alpha}+\min\brac{\snr,\snr^{\alpha}}}+\brac{T-2}\lgbrac{1+\snr^{1-\alpha}}.
\end{align*}

We have
\begin{eqnarray}
h\brac{\Y_{1}\middle|\U_{1},\U_{2}} & = & h\brac{\boldsymbol{g}_{11}\X_{1}+\boldsymbol{g}_{21}\X_{2}+\Z_{1}\middle|\U_{1},\U_{2}}\nonumber \\
 & = & \sum_{i}h\brac{\boldsymbol{g}_{11}\x_{1,i}+\boldsymbol{g}_{21}\x_{2,i}+\z_{1,i}\middle|\cbrac{\boldsymbol{g}_{11}\x_{1,j}+\boldsymbol{g}_{21}\x_{2,j}+\z_{1,j}}_{j=1}^{i-1},\U_{1},\U_{2}}\nonumber \\
 & \overset{\brac i}{\geq} & h\brac{\boldsymbol{g}_{11}\x_{1,1}+\boldsymbol{g}_{21}\x_{2,1}+\z_{1,1}\middle|\x_{1,1},\x_{2,1},\U_{1},\U_{2}}\nonumber \\
 &  & {+}\:h\brac{\boldsymbol{g}_{11}\x_{1,2}+\boldsymbol{g}_{21}\x_{2,2}+\z_{1,2}\middle|\boldsymbol{g}_{11}\x_{1,1}+\boldsymbol{g}_{21}\x_{2,1}+\z_{1,1},\U_{1},\U_{2}}\nonumber \\
 &  & {+}\:\sum_{i=3}^{T}h\brac{\boldsymbol{g}_{11}\x_{1,i}+\boldsymbol{g}_{21}\x_{2,i}+\z_{1,i}\middle|\u_{1,i},\u_{2,i},\boldsymbol{g}_{21},\boldsymbol{g}_{11}}\nonumber \\
 & \overset{\brac{ii}}{\geqdof} & \lgbrac{1+\snr+\snr^{\alpha}}\nonumber \\
 &  & {+}\:h\brac{\boldsymbol{g}_{11}\x_{1,2}+\boldsymbol{g}_{21}\x_{2,2}+\z_{1,2}\middle|\boldsymbol{g}_{11}\x_{1,1}+\boldsymbol{g}_{21}\x_{2,1}+\z_{1,1},\U_{1},\U_{2}}\nonumber \\
 &  & {+}\:\brac{T-2}\lgbrac{1+\snr^{1-\alpha}},\label{eq:h_Y1_given_U1_U2_part1}
\end{eqnarray}
where $\brac i$ is due to the fact that conditioning reduces entropy
and Markovity $\brac{\boldsymbol{g}_{11}\x_{1,i}+\boldsymbol{g}_{21}\x_{2,i}+\z_{1,i}}-\brac{\u_{1,i},\u_{2,i},\boldsymbol{g}_{21},\boldsymbol{g}_{11}}-\brac{\cbrac{\boldsymbol{g}_{11}\x_{1,j}+\boldsymbol{g}_{21}\x_{2,j}+\z_{1,j}}_{j=1}^{i-1},\U_{1},\U_{2}}$
and $\brac{ii}$ is using the property of Gaussians and using Fact
\ref{fact:Jensens_gap}. In $\brac{ii}$ for the last term, we use
\begin{eqnarray}
h\brac{\boldsymbol{g}_{11}\x_{1,i}+\boldsymbol{g}_{21}\x_{2,i}+\z_{1,i}\middle|\u_{1,i},\u_{2,i},\boldsymbol{g}_{21},\boldsymbol{g}_{11}} & \overset{\brac i}{=} & h\brac{\boldsymbol{g}_{11}\x_{\text{p}1,i}+\boldsymbol{g}_{21}\x_{\text{p}2,i}+\z_{1,i}\middle|\boldsymbol{g}_{21},\boldsymbol{g}_{11}}\nonumber \\
 & = & \expect{\lgbrac{\pi e\brac{1+\frac{\abs{\boldsymbol{g}_{11}}^{2}}{\inr}+\frac{\abs{\boldsymbol{g}_{21}}^{2}}{\inr}}}}\nonumber \\
 & \overset{\brac{ii}}{\geqdof} & \lgbrac{1+\snr^{1-\alpha}}.\label{eq:h_y1i|u1i,u2i,g21,g11}
\end{eqnarray}
$\brac i$ is by removing $\boldsymbol{g}_{11}\boldsymbol{u}_{1,i}+\boldsymbol{g}_{21}\boldsymbol{u}_{2,i}$
that is available in the conditioning and because the private message
parts $\x_{\text{p}1,i}$, $\x_{\text{p}2,i}$ are independent of
the common message parts $\u_{1,i},\u_{2,i}$. The step $\brac{ii}$
is using Fact \ref{fact:Jensens_gap}. Now,
\begin{align}
 & h\brac{\boldsymbol{g}_{11}\x_{1,2}+\boldsymbol{g}_{21}\x_{2,2}+\z_{1,2}\middle|\boldsymbol{g}_{11}\x_{1,1}+\boldsymbol{g}_{21}\x_{2,1}+\z_{1,1},\U_{1},\U_{2}}\nonumber \\
 & \geq h\brac{\boldsymbol{g}_{11}\x_{1,2}+\boldsymbol{g}_{21}\x_{2,2}+\z_{1,2}\middle|\boldsymbol{g}_{11}\x_{1,1}+\boldsymbol{g}_{21}\x_{2,1}+\z_{1,1},\X_{1},\X_{2},\U_{1},\U_{2}}\nonumber \\
 & =h\brac{\boldsymbol{g}_{11}\x_{1,2}+\boldsymbol{g}_{21}\x_{2,2}+\z_{1,2},\boldsymbol{g}_{11}\x_{1,1}+\boldsymbol{g}_{21}\x_{2,1}+\z_{1,1}\middle|\X_{1},\X_{2},\U_{1},\U_{2}}\nonumber \\
 & \qquad{-}\:h\brac{\boldsymbol{g}_{11}\x_{1,1}+\boldsymbol{g}_{21}\x_{2,1}+\z_{1,1}\middle|\X_{1},\X_{2},\U_{1},\U_{2}}\\
 & \overset{\brac i}{=}\expect{\lgbrac{\pi e\left|\begin{array}{cc}
\abs{\x_{1,2}}^{2}+\inr\abs{\x_{2,2}}^{2}+1 & \x_{1,2}\x_{1,1}^{\dagger}+\inr\x_{2,2}\x_{2,1}^{\dagger}\\
\brac{\x_{1,2}\x_{1,1}^{\dagger}+\inr\x_{2,2}\x_{2,1}^{\dagger}}^{\dagger} & \abs{\x_{1,1}}^{2}+\inr\abs{\x_{2,1}}^{2}+1
\end{array}\right|}}\nonumber \\
 & \qquad{-}\:\expect{\lgbrac{1+\abs{\x_{2,1}}^{2}\inr+\abs{\x_{1,1}}^{2}}}\\
 & \geq\expect{\lgbrac{\inr\brac{\abs{\x_{1,1}}^{2}\abs{\x_{2,2}}^{2}+\abs{\x_{1,2}}^{2}\abs{\x_{2,1}}^{2}-2\text{Re}\brac{\x_{1,2}\x_{1,1}^{\dagger}\x_{2,2}^{\dagger}\x_{2,1}}}}}\nonumber \\
 & \qquad{-}\:\lgbrac{1+P\cdot\inr+P}\\
 & =\lgbrac{\inr}+\expect{\lgbrac{\abs{\x_{1,1}\x_{2,2}-\x_{1,2}\x_{2,1}}^{2}}}-\lgbrac{1+P\cdot\inr+P}\nonumber \\
 & \overset{\brac{ii}}{\eqdof}\lgbrac{\inr}+\lgbrac{P^{2}}-\lgbrac{1+P\cdot\inr+P}\\
 & \overset{\brac{iii}}{=}\lgbrac{\snr^{\alpha-1}\cdot\snr^{2}}-\lgbrac{1+\snr^{\alpha}+\snr}\label{eq:h_y12_given_y11_u1_u2_secondlaststep}\\
 & \overset{}{\eqdof}\lgbrac{\min\brac{\snr,\snr^{\alpha}}},\label{eq:h_y12_given_y11_u1_u2_part1}
\end{align}
where $\brac i$ is using the property of Gaussian random variables,
$\brac{ii}$ is using Fact \ref{fact:Jensens_gap} on page \pageref{fact:Jensens_gap}
and Tower property of expectation for $\expect{\lgbrac{\abs{\x_{1,1}\x_{2,2}-\x_{1,2}\x_{2,1}}^{2}}}$,
$\brac{iii}$ is using our system setting with $P=\snr\cdot\inr=\snr^{\alpha-1}$.
Also
\begin{align}
 & h\brac{\boldsymbol{g}_{11}\x_{1,2}+\boldsymbol{g}_{21}\x_{2,2}+\z_{1,2}\middle|\boldsymbol{g}_{11}\x_{1,1}+\boldsymbol{g}_{21}\x_{2,1}+\z_{1,1},\U_{1},\U_{2}}\nonumber \\
 & \overset{\brac i}{\geq}h\brac{\boldsymbol{g}_{11}\x_{1,2}+\boldsymbol{g}_{21}\x_{2,2}+\z_{1,2}\middle|\boldsymbol{g}_{11}\x_{1,1}+\boldsymbol{g}_{21}\x_{2,1}+\z_{1,1},\U_{1},\U_{2},\boldsymbol{g}_{11},\boldsymbol{g}_{21}}\nonumber \\
 & \overset{\brac{ii}}{=}h\brac{\boldsymbol{g}_{11}\x_{1,2}+\boldsymbol{g}_{21}\x_{2,2}+\z_{1,2}\middle|\U_{1},\U_{2},\boldsymbol{g}_{11},\boldsymbol{g}_{21}}\nonumber \\
 & \overset{\brac{ii}}{\geqdof}\lgbrac{1+\snr^{1-\alpha}},\label{eq:h_y12_given_y11_u1_u2_part2}
\end{align}
where $\brac i$ is using the fact that conditioning reduces entropy,
$\brac{ii}$ is due to the Markov chain $\brac{\boldsymbol{g}_{11}\x_{1,2}+\boldsymbol{g}_{21}\x_{2,2}+\z_{1,2}}-\brac{\U_{1},\U_{2},\boldsymbol{g}_{21},\boldsymbol{g}_{11}}-\brac{\boldsymbol{g}_{11}\x_{1,1}+\boldsymbol{g}_{21}\x_{2,1}+\z_{1,1},\U_{1},\U_{2}}$,
$\brac{iii}$ is following similar steps as for (\ref{eq:h_y1i|u1i,u2i,g21,g11}).
Now combining (\ref{eq:h_y12_given_y11_u1_u2_part2}), (\ref{eq:h_y12_given_y11_u1_u2_part1}),
we get
\begin{align}
h\brac{\boldsymbol{g}_{11}\x_{1,2}+\boldsymbol{g}_{21}\x_{2,2}+\z_{1,2}\middle|\boldsymbol{g}_{11}\x_{1,1}+\boldsymbol{g}_{21}\x_{2,1}+\z_{1,1},\U_{1},\U_{2}}\nonumber \\
\geqdof\lgbrac{1+\snr^{1-\alpha}+\min\brac{\snr,\snr^{\alpha}}}.
\end{align}
Hence, substituting the above equation in (\ref{eq:h_Y1_given_U1_U2_part1}),
we get
\begin{eqnarray*}
h\brac{\Y_{1}\middle|\U_{1},\U_{2}} & \overset{}{\geqdof} & \lgbrac{1+\snr+\snr^{\alpha}}+\lgbrac{1+\snr^{1-\alpha}+\min\brac{\snr,\snr^{\alpha}}}\\
 &  & {+}\:\brac{T-2}\lgbrac{1+\snr^{1-\alpha}}.
\end{eqnarray*}

\section{Proof of Claim \ref{claim:h_Y1_U2_X1}\label{app:proof_h_Y1_U2_X1}}

In this appendix, we prove that $h\brac{\Y_{1}\middle|\U_{2},\X_{1}}\leqdof\lgbrac{\snr+\snr^{\alpha}}+\lgbrac{\min\brac{\snr,\snr^{\alpha}}}$.
We have
\begin{align}
 & h\brac{\Y_{1}\middle|\U_{2},\X_{1}}\nonumber \\
 & =h\brac{\boldsymbol{g}_{11}\X_{1}+\boldsymbol{g}_{21}\X_{2}+\Z_{1}\middle|\U_{2},\X_{1}}\\
 & \leq h\brac{\boldsymbol{g}_{11}\x_{1,1}+\boldsymbol{g}_{21}\x_{2,1}+\z_{1,1}\middle|\u_{2,1},\x_{1,1}}\nonumber \\
 & \quad\:+h\brac{\boldsymbol{g}_{11}\x_{1,2}+\boldsymbol{g}_{21}\x_{2,2}+\z_{1,2}\middle|\boldsymbol{g}_{11}\x_{1,1}+\boldsymbol{g}_{21}\x_{2,1}+\z_{1,1},\U_{2},\X_{1}}\nonumber \\
 & \quad\:+\sum_{i=3}^{T}h\brac{\boldsymbol{g}_{11}\x_{1,i}+\boldsymbol{g}_{21}\x_{2,i}+\z_{1,i}\middle|\boldsymbol{g}_{11}\x_{1,1}+\boldsymbol{g}_{21}\x_{2,1}+\z_{1,1},\boldsymbol{g}_{11}\x_{1,2}+\boldsymbol{g}_{21}\x_{2,2}+\z_{1,2},\U_{2},\X_{1}}\nonumber \\
 & \leqdof\lgbrac{1+\snr+\inr}+h\brac{\boldsymbol{g}_{11}\x_{1,2}+\boldsymbol{g}_{21}\x_{2,2}+\z_{1,2}\middle|\boldsymbol{g}_{11}\x_{1,1}+\boldsymbol{g}_{21}\x_{2,1}+\z_{1,1},\U_{2},\X_{1}}\nonumber \\
 & \quad\:+\sum_{i=3}^{T}h\left(\boldsymbol{g}_{11}\x_{1,i}+\boldsymbol{g}_{21}\x_{2,i}+\z_{1,i}\middle|\right.\nonumber \\
 & \hphantom{\quad\:+\sum_{i=3}^{T}h\Big(}\left.\boldsymbol{g}_{11}\x_{1,1}+\boldsymbol{g}_{21}\x_{2,1}+\z_{1,1},\boldsymbol{g}_{11}\x_{1,2}+\boldsymbol{g}_{21}\x_{2,2}+\z_{1,2},\U_{2},\X_{1}\right),\label{eq:y1_given_u2_x1_expand}
\end{align}
Considering the second term in the previous expression,
\begin{align}
 & h\brac{\boldsymbol{g}_{11}\x_{1,2}+\boldsymbol{g}_{21}\x_{2,2}+\z_{1,2}\middle|\boldsymbol{g}_{11}\x_{1,1}+\boldsymbol{g}_{21}\x_{2,1}+\z_{1,1},\U_{2},\X_{1}}\nonumber \\
 & =h\brac{\boldsymbol{g}_{11}\x_{1,1}\x_{1,2}+\boldsymbol{g}_{21}\x_{1,1}\x_{2,2}+\x_{1,1}\z_{1,2}\middle|\boldsymbol{g}_{11}\x_{1,1}+\boldsymbol{g}_{21}\x_{2,1}+\z_{1,1},\U_{2},\X_{1}}\nonumber \\
 & \quad\:-\expect{\lgbrac{\abs{\x_{1,1}}^{2}}}\nonumber \\
 & \overset{\brac i}{\leq}h\brac{\boldsymbol{g}_{11}\x_{1,1}\x_{1,2}+\boldsymbol{g}_{21}\x_{1,1}\x_{2,2}+\x_{1,1}\z_{1,2}-\x_{1,2}\brac{\boldsymbol{g}_{11}\x_{1,1}+\boldsymbol{g}_{21}\x_{2,1}+\z_{1,1}}}\nonumber \\
 & \quad\:-\expect{\lgbrac{\abs{\x_{1,1}}^{2}}}\nonumber \\
 & =h\brac{\boldsymbol{g}_{21}\brac{\x_{1,1}\x_{2,2}-\x_{2,1}\x_{1,2}}+\x_{1,1}\z_{1,2}-\x_{1,2}\z_{1,1}}-\expect{\lgbrac{\abs{\x_{1,1}}^{2}}}\label{eq:elim1}\\
 & \leq\lgbrac{\pi e\expect{\abs{\boldsymbol{g}_{21}\brac{\x_{1,1}\x_{2,2}-\x_{2,1}\x_{1,2}}+\x_{1,1}\z_{1,2}-\x_{1,2}\z_{1,1}}^{2}}}-\expect{\lgbrac{\abs{\x_{1,1}}^{2}}}\nonumber \\
 & \overset{\brac{ii}}{\eqdof}\lgbrac{\brac{\inr\brac{2P^{2}}+2P}}-\lgbrac P\nonumber \\
 & =\lgbrac{P\times\inr}\nonumber \\
 & =\lgbrac{\snr^{\alpha}},\label{eq:y1_given_u2_x1_part1}
\end{align}
where $\brac i$ is by subtracting $\x_{1,2}\brac{\boldsymbol{g}_{11}\x_{1,1}+\boldsymbol{g}_{21}\x_{2,1}+\z_{1,1}}$
which is available from the conditioning and then using the fact that
 conditioning reduces entropy, $\brac{ii}$ is by using the  property
of Gaussian random variables for i.i.d. $\boldsymbol{g}_{21},\x_{1,1},\x_{2,2},\x_{2,1},\x_{1,2},\z_{1,2},\z_{1,1}$
and Fact \ref{fact:Jensens_gap} from page \pageref{fact:Jensens_gap}
for $\expect{\lgbrac{\abs{\x_{1,1}}^{2}}}$. Note that $\abs{\x_{1,1}}^{2}$
is exponentially distributed with mean $P=\snr$. We can also use
\begin{align}
 & h\brac{\boldsymbol{g}_{11}\x_{1,2}+\boldsymbol{g}_{21}\x_{2,2}+\z_{1,2}\middle|\boldsymbol{g}_{11}\x_{1,1}+\boldsymbol{g}_{21}\x_{2,1}+\z_{1,1},\U_{2},\X_{1}}\nonumber \\
 & =h\brac{\boldsymbol{g}_{11}\u_{2,1}\x_{1,2}+\boldsymbol{g}_{21}\u_{2,1}\x_{2,2}+\u_{2,1}\z_{1,2}\middle|\boldsymbol{g}_{11}\x_{1,1}+\boldsymbol{g}_{21}\x_{2,1}+\z_{1,1},\U_{2},\X_{1}}-\expect{\lgbrac{\abs{\u_{2,1}}^{2}}}\nonumber \\
 & \overset{\brac i}{\leq}h\brac{\boldsymbol{g}_{11}\u_{2,1}\x_{1,2}+\boldsymbol{g}_{21}\u_{2,1}\x_{2,2}+\u_{2,1}\z_{1,2}-\u_{2,2}\brac{\boldsymbol{g}_{11}\x_{1,1}+\boldsymbol{g}_{21}\x_{2,1}+\z_{1,1}}}-\expect{\lgbrac{\abs{\u_{2,1}}^{2}}}\nonumber \\
 & =h\brac{\boldsymbol{g}_{11}\brac{\u_{2,1}\x_{1,2}-\x_{1,1}\u_{2,2}}+\boldsymbol{g}_{21}\brac{\u_{2,1}\x_{\text{p}2,2}-\u_{2,2}\x_{\text{p}2,1}}+\u_{2,1}\z_{1,2}-\u_{2,2}\z_{1,1}}\nonumber \\
 & \quad\:-\expect{\lgbrac{\abs{\u_{2,1}}^{2}}}\nonumber \\
 & \leqdof\lgbrac{\expect{\abs{\boldsymbol{g}_{11}\brac{\u_{2,1}\x_{1,2}-\x_{1,1}\u_{2,2}}+\boldsymbol{g}_{21}\brac{\u_{2,1}\x_{\text{p}2,2}-\u_{2,2}\x_{\text{p}2,1}}+\u_{2,1}\z_{1,2}-\u_{2,2}\z_{1,1}}^{2}}}\nonumber \\
 & \quad\:-\expect{\lgbrac{\abs{\u_{2,1}}^{2}}}\nonumber \\
 & \overset{\brac{ii}}{\eqdof}\lgbrac{\brac{\brac{2P\brac{P-1/\inr}}+\inr\brac{2\brac{P-1/\inr}\brac{1/\inr}}+2\brac{P-1/\inr}}}-\lgbrac{P-1/\inr}\\
 & \eqdof\lgbrac{\brac{2P\brac{P-1/\inr}}+4\brac{P-1/\inr}}-\lgbrac{P-1/\inr}\\
 & \eqdof\lgbrac{\snr},\label{eq:h_y1_given_u2_x1_highsnr_part1}
\end{align}
where $\brac i$ is by subtracting $\x_{1,2}\brac{\boldsymbol{g}_{11}\x_{1,1}+\boldsymbol{g}_{21}\x_{2,1}+\z_{1,1}}$
which is available from the conditioning and then using  the fact
that  conditioning reduces entropy, $\brac{ii}$ is by using properties
of i.i.d. Gaussian random variables to evaluate the second moments
and Fact \ref{fact:Jensens_gap} for $\expect{\lgbrac{\abs{\u_{2,1}}^{2}}}$.

Using (\ref{eq:y1_given_u2_x1_part1}) and (\ref{eq:h_y1_given_u2_x1_highsnr_part1})
in (\ref{eq:y1_given_u2_x1_expand}), we get
\begin{eqnarray*}
h\brac{\Y_{1}|\U_{2},\X_{1}} & \leqdof & \lgbrac{\snr+\snr^{\alpha}}+\lgbrac{\min\brac{\snr,\snr^{\alpha}}}\\
 &  & {+}\:\sum_{i=3}^{T}h\left(\boldsymbol{g}_{11}\x_{1,i}+\boldsymbol{g}_{21}\x_{2,i}+\z_{1,i}\middle|\right.\\
 &  & \hphantom{{+}\:\sum_{i=3}^{T}h\Big(}\left.\boldsymbol{g}_{11}\x_{1,1}+\boldsymbol{g}_{21}\x_{2,1}+\z_{1,1},\boldsymbol{g}_{11}\x_{1,2}+\boldsymbol{g}_{21}\x_{2,2}+\z_{1,2},\U_{2},\X_{1}\right)
\end{eqnarray*}

Now for $i\geq3$, we will show that
\begin{align}
 & h\left(\boldsymbol{g}_{11}\x_{1,i}+\boldsymbol{g}_{21}\x_{2,i}+\z_{1,i}\middle|\right.\nonumber \\
 & \quad\left.\boldsymbol{g}_{11}\x_{1,1}+\boldsymbol{g}_{21}\x_{2,1}+\z_{1,1},\boldsymbol{g}_{11}\x_{1,2}+\boldsymbol{g}_{21}\x_{2,2}+\z_{1,2},\U_{2},\X_{1}\right)\leqdof0.\label{eq:h(y1_given_u2_x1)_zeroterm}
\end{align}
and will complete our proof. For (\ref{eq:h(y1_given_u2_x1)_zeroterm}),
similar to the elimination done in (\ref{eq:elim1}), we have
\begin{align}
 & h\brac{\boldsymbol{g}_{11}\x_{1,i}+\boldsymbol{g}_{21}\x_{2,i}+\z_{1,i}\middle|\boldsymbol{g}_{11}\x_{1,1}+\boldsymbol{g}_{21}\x_{2,1}+\z_{1,1},\boldsymbol{g}_{11}\x_{1,2}+\boldsymbol{g}_{21}\x_{2,2}+\z_{1,2},\U_{2},\X_{1}}\nonumber \\
 & \leq h\left(\boldsymbol{g}_{21}\brac{\x_{1,1}\x_{2,i}-\x_{2,1}\x_{1,i}}+\x_{1,1}\z_{1,i}-\x_{1,i}\z_{1,1}\middle|\right.\nonumber \\
 & \hphantom{\leq h\Big(}\left.\boldsymbol{g}_{21}\brac{\x_{1,1}\x_{2,2}-\x_{2,1}\x_{1,2}}+\x_{1,1}\z_{1,2}-\x_{1,2}\z_{1,1},\U_{2},\X_{1}\right)-\expect{\lgbrac{\abs{\x_{1,1}}^{2}}}.
\end{align}
Now we have
\begin{align*}
 & \boldsymbol{g}_{21}\brac{\x_{1,1}\x_{2,i}-\x_{2,1}\x_{1,i}}+\x_{1,1}\z_{1,i}-\x_{1,i}\z_{1,1}\\
 & =\boldsymbol{g}_{21}\brac{\x_{1,1}\u_{2,i}-\u_{2,1}\x_{1,i}}+\brac{\boldsymbol{g}_{21}\brac{\x_{1,1}\x_{\text{p}2,i}-\x_{\text{p}2,1}\x_{1,i}}+\x_{1,1}\z_{1,i}-\x_{1,i}\z_{1,1}}
\end{align*}
in the entropy expression. And in the conditioning, the term
\[
\boldsymbol{g}_{21}\brac{\x_{1,1}\u_{2,2}-\u_{2,1}\x_{1,2}}+\brac{\boldsymbol{g}_{21}\brac{\x_{1,1}\x_{\text{p}2,2}-\x_{\text{p}2,1}\x_{1,2}}+\x_{1,1}\z_{1,2}-\x_{1,2}\z_{1,1}}
\]
and $\U_{2},$ $\boldsymbol{X}_{1}$ are available. Hence, by elimination
we can get
\begin{eqnarray}
\xi & = & \brac{\x_{1,1}\u_{2,2}-\u_{2,1}\x_{1,2}}\brac{\boldsymbol{g}_{21}\brac{\x_{1,1}\x_{\text{p}2,i}-\x_{\text{p}2,1}\x_{1,i}}+\x_{1,1}\z_{1,i}-\x_{1,i}\z_{1,1}}\nonumber \\
 &  & {-}\:\brac{\x_{1,1}\u_{2,i}-\u_{2,1}\x_{1,i}}\brac{\boldsymbol{g}_{21}\brac{\x_{1,1}\x_{\text{p}2,2}-\x_{\text{p}2,1}\x_{1,2}}+\x_{1,1}\z_{1,2}-\x_{1,2}\z_{1,1}}
\end{eqnarray}
in the entropy expression. Using elimination and using the fact that
conditioning reduces entropy, we get
\begin{align}
 & h\brac{\boldsymbol{g}_{11}\x_{1,i}+\boldsymbol{g}_{21}\x_{2,i}+\z_{1,i}\middle|\boldsymbol{g}_{11}\x_{1,1}+\boldsymbol{g}_{21}\x_{2,1}+\z_{1,1},\boldsymbol{g}_{11}\x_{1,2}+\boldsymbol{g}_{21}\x_{2,2}+\z_{1,2},\U_{2},\X_{1}}\nonumber \\
 & \leq h\brac{\xi}-\expect{\lgbrac{\abs{\x_{1,1}\u_{2,2}-\u_{2,1}\x_{1,2}}^{2}}}-\expect{\lgbrac{\abs{\x_{1,1}}^{2}}}\nonumber \\
 & \overset{\brac i}{\eqdof}h\brac{\xi}-\log\brac{P^{3}}\label{eq:h(xi)}
\end{align}
where $\brac i$ is using properties of i.i.d. Gaussian random variables
to evaluate the second moments and Fact \ref{fact:Jensens_gap}. Let
$\xi$ be expanded into a sum of product form
\begin{eqnarray*}
\xi & = & \sum_{i=1}^{L}\xi_{i}\\
 & = & \x_{1,1}\u_{2,2}\boldsymbol{g}_{21}\x_{1,1}\x_{\text{p}2,i}+\brac{-\x_{1,1}\u_{2,2}\boldsymbol{g}_{21}\x_{\text{p}2,1}\x_{1,i}}+\cdots
\end{eqnarray*}
where $\xi_{i}$ is in a simple product form. Now due to triangle
inequality and generalized mean inequality \cite[Ch. 3]{means_inequalities_bullen},
we have
\begin{align}
\abs{\sum_{i=1}^{L}\xi_{i}}^{2} & \leq L\brac{\sum_{i=1}^{L}\abs{\xi_{i}}^{2}}.
\end{align}
Hence we have
\begin{align}
\expect{\abs{\xi}^{2}} & \leq L\brac{\sum_{i=1}^{L}\expect{\abs{\xi_{i}}^{2}}}.
\end{align}
 Now, for example, consider the term $\expect{\abs{\x_{1,1}\u_{2,2}\boldsymbol{g}_{21}\x_{1,1}\x_{\text{p}2,i}}^{2}}$
in the last equation
\begin{eqnarray}
\expect{\abs{\x_{1,1}\u_{2,2}\boldsymbol{g}_{21}\x_{1,1}\x_{\text{p}2,i}}^{2}} & = & \expect{\abs{\x_{1,1}}^{4}}\expect{\abs{\u_{2,2}}^{2}}\expect{\abs{\boldsymbol{g}_{21}}^{2}}\expect{\abs{\x_{\text{p}2,i}}^{2}}\nonumber \\
 & = & 2P^{2}\times\brac{P-1/\inr}\times\inr\times\brac{1/\inr}\leq2P^{3}.
\end{eqnarray}
Each of $\expect{\abs{\xi_{i}}^{2}}$ will be bounded by a constant
since $\boldsymbol{g}_{21}$ always appears coupled with $\x_{\text{p}2,i}$.
Hence, the power scaling $\expect{\abs{\boldsymbol{g}_{21}}^{2}}=\inr$
gets canceled with the scaling $\expect{\abs{\x_{\text{p}2,i}}^{2}}=1/\inr$.
Hence, by analyzing each of $\expect{\abs{\xi_{i}}^{2}}$ together
with maximum entropy results, it can be shown that, $h\brac{\xi}\leqdof\lgbrac{P^{3}}$.
By substituting $h\brac{\xi}\leqdof\lgbrac{P^{3}}$ in (\ref{eq:h(xi)}),
(\ref{eq:h(y1_given_u2_x1)_zeroterm}) is proved and it completes
our proof for the main result.

\section{Training-Based Rate-Splitting Scheme for the Noncoherent IC without
Feedback\label{app: Training Scheme NoFB}}

\noindent \textbf{Encoding:} We consider a fixed distribution $p\brac{\U_{1}}p\brac{\U_{2}}p\brac{\X_{1}\middle|\U_{1}}p\brac{\X_{2}\middle|\U_{2}}$
where $\U_{1},\U_{2},\X_{1},\X_{2}$ are vectors of length $T-2$.
For transmitter 1, generate $2^{NTR_{\text{c}1}}$ codewords $\U_{1}^{N}\brac i$
with $i\in\cbrac{1,\ldots,2^{NTR_{\text{c}1}}}$ according to $\prod_{l=1}^{N}p\brac{\U_{1(l)}}$.
For each $\U_{1}^{N}\brac i$, generate $2^{NTR_{\text{p}1}}$ codewords
$\X_{1}^{N}\brac{i,j}$, with $j\in\cbrac{1,\ldots,2^{NTR_{\text{p}1}}}$,
according to $\prod_{l=1}^{N}p\brac{\X_{1(l)}\middle|\U_{1(l)}}$.
Similarly for transmitter 2, generate $2^{NTR_{\text{c}2}}$ codewords
$\U_{2}^{N}\brac i$, with $i\in\cbrac{1,\ldots,2^{NTR_{\text{c}2}}}$,
according to $\prod_{l=1}^{N}p\brac{\U_{2(l)}}$. For each $\U_{2}^{N}\brac j$,
generate $2^{NTR_{\text{p}2}}$ codewords $\X_{2}^{N}\brac{i,j}$,
with $j\in\cbrac{1,\ldots,2^{NTR_{\text{p}2}}}$, according to $\prod_{l=1}^{N}p\brac{\X_{2(l)}\middle|\U_{2(l)}}$.

Transmitter 1 has uniformly random messages $w_{\text{c}1}\in\cbrac{1,\ldots,2^{NTR_{\text{c}1}}},w_{\text{p}1}\in\cbrac{1,\ldots,2^{NTR_{\text{p}1}}}$
to transmit and transmitter 2 has uniformly random messages $w_{\text{c}2}\in\cbrac{1,\ldots,2^{NTR_{\text{c}2}}},w_{\text{p}2}\in\cbrac{1,\ldots,2^{NTR_{\text{p}2}}}$
to transmit. Transmitter 1 selects $\boldsymbol{X}_{1}^{N}\brac{w_{\text{c}1},w_{\text{p}1}}$
and transmits $\tilde{\X}_{1}^{N}\brac{w_{\text{c}1},w_{\text{p}1}}$
created from it as
\[
\tilde{\X}_{1}^{N}=\sbrac{\sqrt{P},0,\X_{\text{1}}\brac 1},\ldots\sbrac{\sqrt{P},0,\X_{\text{1}}\brac k},\ldots\sbrac{\sqrt{P},0,\X_{\text{1}}\brac N}
\]
where each of $\X_{\text{1}}\brac k$ is a vector of length $T-2$.
Effectively Transmitter 1 is sending a pilot symbol with $\sqrt{P}$
value at the beginning of every $T$ symbols. Similarly Transmitter
2 selects the symbols $\X_{2}^{N}\brac{w_{\text{c}1},w_{\text{p}1}}$
and transmits the symbols $\tilde{\X}_{2}^{N}\brac{w_{\text{c}2},w_{\text{p}2}}$
created from it with pilot symbols added at the beginning as
\[
\tilde{\X}_{2}^{N}=\sbrac{0,\sqrt{P},\X_{\text{2}}\brac 1},\ldots\sbrac{0,\sqrt{P},\X_{\text{2}}\brac k},\ldots\sbrac{0,\sqrt{P},\X_{\text{2}}\brac N}.
\]
At receiver 1, using pilot symbols, in $k^{\text{th}}$ set of $T$
symbols, we get $\boldsymbol{y}_{11,\text{train}}(k)=\sqrt{P}\boldsymbol{g}_{11}(k)+\boldsymbol{z}_{11}(k)$,
$\boldsymbol{y}_{12,\text{train}}(k)=\sqrt{P}\boldsymbol{g}_{21}(k)+\boldsymbol{z}_{21}(k),$
and the minimum mean squared error (MMSE) estimates can be obtained
as
\begin{eqnarray}
\hat{\boldsymbol{g}}_{11} & = & \frac{\sqrt{P}\expect{\abs{\boldsymbol{g}_{11}}^{2}}}{1+P\expect{\abs{\boldsymbol{g}_{11}}^{2}}}\boldsymbol{y}_{11,\text{train}}\nonumber \\
 & = & \sqrt{P}\expect{\abs{\boldsymbol{g}_{11}}^{2}}\frac{\sqrt{P}\boldsymbol{g}_{11}+\boldsymbol{z}_{11}}{1+P\expect{\abs{\boldsymbol{g}_{11}}^{2}}}.
\end{eqnarray}
\begin{eqnarray}
\hat{\boldsymbol{g}}_{21} & = & \frac{\sqrt{P}\expect{\abs{\boldsymbol{g}_{21}}^{2}}}{1+P\expect{\abs{\boldsymbol{g}_{21}}^{2}}}\boldsymbol{y}_{12,\text{train}}\nonumber \\
 & = & \sqrt{P}\expect{\abs{\boldsymbol{g}_{21}}^{2}}\frac{\sqrt{P}\boldsymbol{g}_{21}+\boldsymbol{z}_{21}}{1+P\expect{\abs{\boldsymbol{g}_{21}}^{2}}}.
\end{eqnarray}
and similar estimates $\hat{\boldsymbol{g}}_{22}$, $\hat{\boldsymbol{g}}_{12}$
are obtained at receiver 2. We call $\underline{\hat{\boldsymbol{g}}_{1}}=[\hat{\boldsymbol{g}}_{11},\hat{\boldsymbol{g}}_{21}]$
and $\underline{\hat{\boldsymbol{g}}_{2}}=[\hat{\boldsymbol{g}}_{22},\hat{\boldsymbol{g}}_{12}]$.
We use the notation $\Y_{1}^{N},\Y_{2}^{N}$ to indicate received
symbols containing data and not training symbols:
\begin{equation}
\Y_{1}^{N}=\boldsymbol{g}_{11}^{N}\X_{1}^{N}+\boldsymbol{g}_{21}^{N}\X_{2}^{N}+\boldsymbol{Z}_{1}^{N},
\end{equation}
\begin{equation}
\Y_{2}^{N}=\boldsymbol{g}_{12}^{N}\X_{1}^{N}+\boldsymbol{g}_{22}^{N}\X_{2}^{N}+\boldsymbol{Z}_{2}^{N}.
\end{equation}

\noindent \textbf{Decoding: }For decoding, receiver 1 finds a triplet
$\brac{\hat{i},\hat{j},\hat{k}}$ requiring $\hat{i},\hat{j}$ to
be unique with
\[
\brac{\X_{1}^{N}\brac{\hat{i},\hat{j}},\U_{1}^{N}\brac{\hat{i}},\U_{2}^{N}\brac{\hat{k}},\Y_{1}^{N},\underline{\hat{\boldsymbol{g}}_{1}^{N}}}\in\mathcal{A}_{\epsilon}^{\brac N}.
\]
 Similarly receiver 2 finds a triplet $\brac{\hat{i},\hat{j},\hat{k}}$
requiring $\hat{i},\hat{j}$ to be unique with
\[
\brac{\X_{2}^{N}\brac{\hat{i},\hat{j}},\U_{2}^{N}\brac{\hat{i}},\U_{1}^{N}\brac{\hat{k}},\Y_{2}^{N},\underline{\hat{\boldsymbol{g}}_{2}^{N}}}\in\mathcal{A}_{\epsilon}^{\brac N},
\]
where $A_{\epsilon}^{\brac N}$ indicates the set of jointly typical
sequences. Similar to the analysis in Section \ref{sec:NC_IC_noFB},
we can obtain the following rate region:\begin{subequations}\label{eq:ach_nofb_training}
\begin{eqnarray}
TR_{1} & \leq & I\brac{\X_{1};\Y_{1},\underline{\hat{\boldsymbol{g}}_{1}}\middle|\U_{2}},\\
TR_{2} & \leq & I\brac{\X_{2};\Y_{2},\underline{\hat{\boldsymbol{g}}_{2}}\middle|\U_{1}},\\
T\brac{R_{1}+R_{2}} & \leq & I\brac{\X_{2},\U_{1};\Y_{2},\underline{\hat{\boldsymbol{g}}_{2}}}+I\brac{\X_{1};\Y_{1},\underline{\hat{\boldsymbol{g}}_{1}}\middle|\U_{1},\U_{2}},\\
T\brac{R_{1}+R_{2}} & \leq & I\brac{\X_{1},\U_{2};\Y_{1},\underline{\hat{\boldsymbol{g}}_{1}}}+I\brac{\X_{2};\Y_{2},\underline{\hat{\boldsymbol{g}}_{2}}\middle|\U_{1},\U_{2}},\\
T\brac{R_{1}+R_{2}} & \leq & I\brac{\X_{1},\U_{2};\Y_{1},\underline{\hat{\boldsymbol{g}}_{1}}\middle|\U_{1}}+I\brac{\X_{2},\U_{1};\Y_{2},\underline{\hat{\boldsymbol{g}}_{2}}\middle|\U_{2}},\\
T\brac{2R_{1}+R_{2}} & \leq & I\brac{\X_{1},\U_{2};\Y_{1},\underline{\hat{\boldsymbol{g}}_{1}}}+I\brac{\X_{1};\Y_{1},\underline{\hat{\boldsymbol{g}}_{1}}\middle|\U_{1},\U_{2}}\nonumber \\
 &  & {+}\:I\brac{\X_{2},\U_{1};\Y_{2},\underline{\hat{\boldsymbol{g}}_{2}}\middle|\U_{2}},\\
T\brac{R_{1}+2R_{2}} & \leq & I\brac{\X_{2},\U_{1};\Y_{2},\underline{\hat{\boldsymbol{g}}_{2}}}+I\brac{\X_{2};\Y_{2},\underline{\hat{\boldsymbol{g}}_{2}}\middle|\U_{1},\U_{2}}\nonumber \\
 &  & {+}\:I\brac{\X_{1},\U_{2};\Y_{1},\underline{\hat{\boldsymbol{g}}_{1}}\middle|\U_{1}}.
\end{eqnarray}
\end{subequations}Now similar to that in \cite{etkin_tse_no_fb_IC},
we choose $\U_{k}$ as a vector of length $T-2$ with i.i.d. $\mathcal{CN}\brac{0,\lambda_{\text{c}}}$
elements and $\X_{\text{p}k}$ as a vector of length $T-2$ with i.i.d.
$\mathcal{CN}\brac{0,\lambda_{\text{p}}}$ elements for $k\in\cbrac{1,2}$.
The random variables are chosen independent of each other so that
the set $\cbrac{\U_{1},\U_{2},\X_{\text{p}1},\X_{\text{p}2}}$ is
mutually independent. We use $\X_{1}=\U_{1}+\X_{\text{p}1},\quad\X_{2}=\U_{2}+\X_{\text{p}2}$
where $\lambda_{\text{c}}+\lambda_{\text{p}}=P$ and $\lambda_{\text{p}}=\min\brac{1/\inr,P}$.

We analyze the terms in the rate region in the following subsections.

\subsection{First Term $I\protect\brac{\protect\X_{1};\protect\Y_{1},\underline{\hat{\boldsymbol{g}}_{1}}\middle|\protect\U_{2}}$\label{subsec:training_term1}}

We have
\begin{eqnarray}
I\brac{\X_{1};\Y_{1},\underline{\hat{\boldsymbol{g}}_{1}}\middle|\U_{2}} & = & I\brac{\X_{1};\Y_{1}\middle|\U_{2},\underline{\hat{\boldsymbol{g}}_{1}}}\nonumber \\
 & = & I\brac{\X_{1};\boldsymbol{g}_{11}\X_{1}+\boldsymbol{g}_{21}\X_{2}+\Z_{1}\middle|\U_{2},\underline{\hat{\boldsymbol{g}}_{1}}}\nonumber \\
 & \overset{\brac i}{=} & I\brac{\X_{1};\hat{\boldsymbol{g}}_{11}\X_{1}+\hat{\boldsymbol{g}}_{21}\X_{2}+\Z_{1}+\boldsymbol{\tilde{g}}_{11}\X_{1}+\boldsymbol{\tilde{g}}_{21}\X_{2}\middle|\U_{2},\underline{\hat{\boldsymbol{g}}_{1}}}\nonumber \\
 & \overset{\brac{ii}}{=} & I\brac{\X_{1};\hat{\boldsymbol{g}}_{11}\X_{1}+\hat{\boldsymbol{g}}_{21}\X_{2}+\hat{\Z}_{1}\middle|\U_{2},\underline{\hat{\boldsymbol{g}}_{1}}}\nonumber \\
 & = & h\brac{\X_{1}\middle|\U_{2},\underline{\hat{\boldsymbol{g}}_{1}}}-h\brac{\X_{1}\middle|\hat{\boldsymbol{g}}_{11}\X_{1}+\hat{\boldsymbol{g}}_{21}\X_{p2}+\hat{\Z}_{1},\U_{2},\underline{\hat{\boldsymbol{g}}_{1}}}\nonumber \\
 & = & h\brac{\X_{1}\middle|\underline{\hat{\boldsymbol{g}}_{11}}}-h\brac{\X_{1}\middle|\hat{\boldsymbol{g}}_{11}\X_{1}+\hat{\boldsymbol{g}}_{21}\X_{p2}+\hat{\Z}_{1},\U_{2},\underline{\hat{\boldsymbol{g}}_{1}}}\nonumber \\
 & \geq & h\brac{\X_{1}\middle|\underline{\hat{\boldsymbol{g}}_{11}}}-h\brac{\X_{1}\middle|\hat{\boldsymbol{g}}_{11}\X_{1}+\hat{\boldsymbol{g}}_{21}\X_{p2}+\hat{\Z}_{1},\underline{\hat{\boldsymbol{g}}_{11}}}\nonumber \\
 & = & I\brac{\X_{1};\underline{\hat{\boldsymbol{g}}_{11}}\X_{1}+\hat{\boldsymbol{g}}_{21}\X_{p2}+\hat{\Z}_{1}\middle|\underline{\hat{\boldsymbol{g}}_{11}}}\nonumber \\
 & \overset{\brac{iii}}{\geq} & I\brac{\X_{1};\underline{\hat{\boldsymbol{g}}_{11}}\X_{1}+\hat{\Z}_{worst}\middle|\underline{\hat{\boldsymbol{g}}_{11}}}\label{eq:training_rate_term}
\end{eqnarray}
where in step $\brac i$
\[
\boldsymbol{\tilde{g}}_{11}=\boldsymbol{g}_{11}-\hat{\boldsymbol{g}}_{11}=\boldsymbol{g}_{11}-\sqrt{P}\expect{\abs{\boldsymbol{g}_{11}}^{2}}\frac{\sqrt{P}\boldsymbol{g}_{11}+\boldsymbol{z}_{11}}{1+P\expect{\abs{\boldsymbol{g}_{11}}^{2}}},
\]
\[
\boldsymbol{\tilde{g}}_{21}=\boldsymbol{g}_{21}-\hat{\boldsymbol{g}}_{21}=\boldsymbol{g}_{21}-\sqrt{P}\expect{\abs{\boldsymbol{g}_{21}}^{2}}\frac{\sqrt{P}\boldsymbol{g}_{21}+\boldsymbol{z}_{21}}{1+P\expect{\abs{\boldsymbol{g}_{21}}^{2}}}.
\]
In step $\brac{ii}$, $\hat{\Z}_{1}=\Z_{1}+\boldsymbol{\tilde{g}}_{11}\X_{1}+\boldsymbol{\tilde{g}}_{21}\X_{2}$.
In step $(iii)$, we used the worst case noise result from \cite[(A7)]{Hassibi_worst_noise}
with $\hat{\Z}_{worst}$ being a Gaussian random variable with same
covariance as $\hat{\boldsymbol{g}}_{21}\X_{\text{p}2}+\hat{\Z}_{1}$.
Note that $\X_{1}$ and $\hat{\boldsymbol{g}}_{21}\X_{\text{p}2}+\hat{\Z}_{1}$
are uncorrelated even though they are dependent. Hence, the result
from \cite{Hassibi_worst_noise} can be applied. We have
\begin{align*}
\hat{\boldsymbol{g}}_{21}\X_{p2}+\hat{\Z}_{1} & =\hat{\boldsymbol{g}}_{21}\X_{\text{p}2}+\Z_{1}+\brac{\boldsymbol{g}_{11}-\hat{\boldsymbol{g}}_{11}}\X_{1}+\brac{\boldsymbol{g}_{21}-\hat{\boldsymbol{g}}_{21}}\X_{2}.
\end{align*}
The components of $\hat{\boldsymbol{g}}_{21}\X_{\text{p}2}+\hat{\Z}_{1}$
are identically distributed and are uncorrelated (although they are
dependent through the common random variables $\boldsymbol{g}_{11},\boldsymbol{z}_{11},\boldsymbol{g}_{21},\boldsymbol{z}_{21}$).
Each of the components have variance given by
\begin{align*}
 & \expect{\abs{\hat{\boldsymbol{g}}_{21}\boldsymbol{x}_{\text{p}2,1}+\boldsymbol{z}_{1,1}+\brac{\boldsymbol{g}_{11}-\hat{\boldsymbol{g}}_{11}}\boldsymbol{x}_{1,1}+\brac{\boldsymbol{g}_{21}-\hat{\boldsymbol{g}}_{21}}\boldsymbol{x}_{2,1}}^{2}}\\
 & \overset{\brac i}{=}\lambda\expect{\abs{\hat{\boldsymbol{g}}_{21}}^{2}}+1+P\expect{\abs{\boldsymbol{g}_{11}-\hat{\boldsymbol{g}}_{11}}^{2}}+P\expect{\abs{\boldsymbol{g}_{21}-\hat{\boldsymbol{g}}_{21}}^{2}}\\
 & =\lambda\expect{\abs{\hat{\boldsymbol{g}}_{21}}^{2}}+1\\
 & \hphantom{=}{+}\:P\expect{\brac{\boldsymbol{g}_{11}-\sqrt{P}\expect{\abs{\boldsymbol{g}_{11}}^{2}}\frac{\sqrt{P}\boldsymbol{g}_{11}+\boldsymbol{z}_{11}}{1+P\expect{\abs{\boldsymbol{g}_{11}}^{2}}}}^{2}}\\
 & \hphantom{=}{+}\:P\expect{\brac{\boldsymbol{g}_{21}-\sqrt{P}\expect{\abs{\boldsymbol{g}_{21}}^{2}}\frac{\sqrt{P}\boldsymbol{g}_{21}+\boldsymbol{z}_{21}}{1+P\expect{\abs{\boldsymbol{g}_{21}}^{2}}}}^{2}}\\
 & =\lambda\expect{\abs{\hat{\boldsymbol{g}}_{21}}^{2}}+1\\
 & \hphantom{=}{+}\:P\expect{\abs{\frac{\boldsymbol{g}_{11}}{1+P\expect{\abs{\boldsymbol{g}_{11}}^{2}}}}^{2}}+\frac{P^{2}\expect{\abs{\boldsymbol{g}_{11}}^{2}}^{2}}{\abs{1+P\expect{\abs{\boldsymbol{g}_{11}}^{2}}}^{2}}\\
 & \hphantom{=}{+}\:P\expect{\abs{\frac{\boldsymbol{g}_{21}}{1+P\expect{\abs{\boldsymbol{g}_{21}}^{2}}}}^{2}}+\frac{P^{2}\expect{\abs{\boldsymbol{g}_{21}}^{2}}^{2}}{\abs{1+P\expect{\abs{\boldsymbol{g}_{21}}^{2}}}^{2}}\\
 & =\lambda\expect{\abs{\hat{\boldsymbol{g}}_{21}}^{2}}+\frac{P\expect{\abs{\boldsymbol{g}_{11}}^{2}}}{1+P\expect{\abs{\boldsymbol{g}_{11}}^{2}}}+\frac{P\expect{\abs{\boldsymbol{g}_{21}}^{2}}}{1+P\expect{\abs{\boldsymbol{g}_{21}}^{2}}}+1\\
 & =\lambda\expect{\abs{\hat{\boldsymbol{g}}_{21}}^{2}}+N
\end{align*}
where
\[
N=\frac{P\expect{\abs{\boldsymbol{g}_{11}}^{2}}}{1+P\expect{\abs{\boldsymbol{g}_{11}}^{2}}}+\frac{P\expect{\abs{\boldsymbol{g}_{21}}^{2}}}{1+P\expect{\abs{\boldsymbol{g}_{21}}^{2}}}+1.
\]
 In step $\brac i$, we used the facts that $\boldsymbol{x}_{\text{p}2,1}$
has power $\lambda=1/\inr$; the power settings and independency of
$\boldsymbol{x}_{1,1},\boldsymbol{x}_{2,1},\boldsymbol{z}_{1,1}$;
$\hat{\boldsymbol{g}}_{21}$ and $\brac{\boldsymbol{g}_{21}-\hat{\boldsymbol{g}}_{21}}$
are uncorrelated due to the orthogonality property of MMSE estimate.
Using the above simplifications in (\ref{eq:training_rate_term}),
we get
\begin{align}
I\brac{\X_{1};\Y_{1},\underline{\hat{\boldsymbol{g}}_{1}}\middle|\U_{2}}\geq & \brac{T-2}\Big(\expect{\lgbrac{P\abs{\hat{\g}_{11}}^{2}+\lambda\expect{\abs{\hat{\boldsymbol{g}}_{21}}^{2}}+N}}\nonumber \\
 & \hphantom{\brac{T-2}\Big(}-\lgbrac{\lambda\expect{\abs{\hat{\boldsymbol{g}}_{21}}^{2}}+N}\Big).
\end{align}

\subsection{Second Term $I\protect\brac{\protect\X_{1},\protect\U_{2};\protect\Y_{1},\underline{\hat{\boldsymbol{g}}_{1}}}$\label{subsec:training_term2}}

Following similar analysis as for the previous term, by replacing
$\hat{\boldsymbol{g}}_{21}\X_{\text{p}2}+\hat{\Z}_{1}$ with the worst
case noise, we get:
\begin{eqnarray}
I\brac{\X_{1},\U_{2};\Y_{1},\underline{\hat{\boldsymbol{g}}_{1}}} & = & I\brac{\X_{1},\U_{2};\hat{\boldsymbol{g}}_{11}\X_{1}+\hat{\boldsymbol{g}}_{21}\boldsymbol{U}_{2}+\hat{\boldsymbol{g}}_{21}\X_{\text{p}2}+\hat{\Z}_{1}\middle|\underline{\hat{\boldsymbol{g}}_{1}}}\nonumber \\
 & \geq & \brac{T-2}\Big(\expect{\lgbrac{P\abs{\hat{\g}_{11}}^{2}+\brac{P-\lambda}\abs{\hat{\g}_{21}}^{2}+\lambda\expect{\abs{\hat{\boldsymbol{g}}_{21}}^{2}}+N}}\nonumber \\
 &  & \hphantom{\brac{T-2}\Big(}-\lgbrac{\lambda\expect{\abs{\hat{\boldsymbol{g}}_{21}}^{2}}+N}\Big)
\end{eqnarray}

\subsection{Third Term $I\protect\brac{\protect\X_{1};\protect\Y_{1},\underline{\hat{\boldsymbol{g}}_{1}}\middle|\protect\U_{1},\protect\U_{2}}$\label{subsec:training_term3}}

We have
\begin{eqnarray}
I\brac{\X_{1};\Y_{1},\underline{\hat{\boldsymbol{g}}_{1}}\middle|\U_{1},\U_{2}} & = & I\brac{\boldsymbol{U}_{1}+\X_{\text{p}1};\hat{\boldsymbol{g}}_{11}\boldsymbol{U}_{1}+\hat{\boldsymbol{g}}_{21}\boldsymbol{U}_{2}+\hat{\boldsymbol{g}}_{11}\X_{\text{p}1}+\hat{\boldsymbol{g}}_{21}\X_{\text{p}2}+\hat{\Z}_{1}\middle|\underline{\hat{\boldsymbol{g}}_{1}},\U_{1},\U_{2}}\nonumber \\
 & \geq & I\brac{\X_{\text{p}1};\hat{\boldsymbol{g}}_{11}\X_{\text{p}1}+\hat{\boldsymbol{g}}_{21}\X_{\text{p}2}+\hat{\Z}_{1}\middle|\underline{\hat{\boldsymbol{g}}_{1}}}\nonumber \\
 & \geq & \brac{T-2}\Big(\expect{\lgbrac{\lambda\abs{\hat{\g}_{11}}^{2}+\lambda\expect{\abs{\hat{\boldsymbol{g}}_{21}}^{2}}+N}}\nonumber \\
 &  & \hphantom{\brac{T-2}\Big(}-\lgbrac{\lambda\expect{\abs{\hat{\boldsymbol{g}}_{21}}^{2}}+N}\Big)
\end{eqnarray}
where in the last step, we used the worst case noise result.

\subsection{Fourth Term $I\protect\brac{\protect\X_{2},\protect\U_{1};\protect\Y_{2},\underline{\hat{\boldsymbol{g}}_{2}}\middle|\protect\U_{2}}$\label{subsec:training_term4}}

We have
\begin{eqnarray}
I\brac{\X_{1},\U_{2};\Y_{1},\underline{\hat{\boldsymbol{g}}_{1}}\middle|\U_{1}} & = & I\brac{\boldsymbol{U}_{1}+\X_{\text{p}1},\U_{2};\hat{\boldsymbol{g}}_{11}\boldsymbol{U}_{1}+\hat{\boldsymbol{g}}_{21}\boldsymbol{U}_{2}+\hat{\boldsymbol{g}}_{11}\X_{\text{p}1}+\hat{\boldsymbol{g}}_{21}\X_{\text{p}2}+\hat{\Z}_{1}\middle|\underline{\hat{\boldsymbol{g}}_{1}},\U_{1}}\nonumber \\
 & = & I\brac{\X_{\text{p}1},\U_{2};\hat{\boldsymbol{g}}_{21}\boldsymbol{U}_{2}+\hat{\boldsymbol{g}}_{11}\X_{\text{p}1}+\hat{\boldsymbol{g}}_{21}\X_{\text{p}2}+\hat{\Z}_{1}\middle|\underline{\hat{\boldsymbol{g}}_{1}}}\nonumber \\
 & \geq & \brac{T-2}\Big(\expect{\lgbrac{\lambda\abs{\hat{\g}_{11}}^{2}+\brac{P-\lambda}\abs{\hat{\g}_{21}}^{2}+\lambda\expect{\abs{\hat{\boldsymbol{g}}_{21}}^{2}}+N}}\nonumber \\
 &  & \hphantom{\brac{T-2}\Big(}-\lgbrac{\lambda\expect{\abs{\hat{\boldsymbol{g}}_{21}}^{2}}+N}\Big)
\end{eqnarray}
where in the last step, we again used the worst case noise result.

\subsection{Simplified Rate Region}

We collect the results from the previous four subsections in the following
table with $r_{1}'=\log\big(\lambda\mathbb{E}\big[\big|\hat{\boldsymbol{g}}_{21}\big|^{2}\big]+N\big)$
and subsequently obtain an achievable rate region.
\begin{table}[H]
\centering{}\caption{Lower bounds for the terms in the achievability region\label{tab:lower-bounds-training-nofb}}
\begin{tabular}{|c|c|}
\hline
Term & Lower bound\tabularnewline
\hline
\hline
$I\brac{\X_{1};\Y_{1},\underline{\hat{\boldsymbol{g}}_{1}}\middle|\U_{2}}$ & $\brac{T-2}\brac{\mathbb{E}\Big[\log\big(P\abs{\hat{\g}_{11}}^{2}+\lambda\mathbb{E}\big[\big|\hat{\boldsymbol{g}}_{21}\big|^{2}\big]+N\big)\Big]-r_{1}'}$\tabularnewline
\hline
$I\brac{\X_{1},\U_{2};\Y_{1},\underline{\hat{\boldsymbol{g}}_{1}}}$ & $\brac{T-2}\brac{\mathbb{E}\Big[\log\big(P\abs{\hat{\g}_{11}}^{2}+\brac{P-\lambda}\abs{\hat{\g}_{21}}^{2}+\lambda\mathbb{E}\big[\big|\hat{\boldsymbol{g}}_{21}\big|^{2}\big]+N\big)\Big]-r_{1}'}$\tabularnewline
\hline
$I\brac{\X_{1};\Y_{1},\underline{\hat{\boldsymbol{g}}_{1}}\middle|\U_{1},\U_{2}}$ & $\brac{T-2}\brac{\mathbb{E}\Big[\log\big(N+\lambda\abs{\hat{\g}_{11}}^{2}+\lambda\mathbb{E}\big[\big|\hat{\boldsymbol{g}}_{21}\big|^{2}\big]\big)\Big]-r_{1}'}$\tabularnewline
\hline
$I\brac{\X_{1},\U_{2};\Y_{1},\underline{\hat{\boldsymbol{g}}_{1}}\middle|\U_{1}}$ & $\brac{T-2}\brac{\mathbb{E}\Big[\log\big(\lambda\abs{\hat{\g}_{11}}^{2}+\brac{P-\lambda}\abs{\hat{\g}_{21}}^{2}+\lambda\mathbb{E}\big[\big|\hat{\boldsymbol{g}}_{21}\big|^{2}\big]+N\big)\Big]-r_{1}'}$\tabularnewline
\hline
\end{tabular}
\end{table}

Using the lower bounds from Table \ref{tab:lower-bounds-training-nofb}
in (\ref{eq:ach_nofb_training}) and using symmetry of the terms,
we obtain that the following rate region is achievable with $r_{1}'=\log\big(\lambda\mathbb{E}\big[\big|\hat{\boldsymbol{g}}_{21}\big|^{2}\big]+N\big)$,
$r_{2}'=\log\big(\lambda\mathbb{E}\big[\big|\hat{\boldsymbol{g}}_{12}\big|^{2}\big]+N\big)$:\begin{subequations}\label{eq:ach_nofb_training_simplified}

\begin{eqnarray}
R_{1} & \leq & \brac{1-\frac{2}{T}}\brac{\mathbb{E}\Big[\log\big(P\abs{\hat{\g}_{11}}^{2}+\lambda\mathbb{E}\big[\big|\hat{\boldsymbol{g}}_{21}\big|^{2}\big]+N\big)\Big]-r_{1}'},\\
R_{2} & \leq & \brac{1-\frac{2}{T}}\brac{\mathbb{E}\Big[\log\big(P\abs{\hat{\g}_{22}}^{2}+\lambda\mathbb{E}\big[\big|\hat{\boldsymbol{g}}_{12}\big|^{2}\big]+N\big)\Big]-r_{2}'},\\
R_{1}+R_{2} & \leq & \brac{1-\frac{2}{T}}\Big(\mathbb{E}\Big[\log\big(P\abs{\hat{\g}_{22}}^{2}+\brac{P-\lambda}\abs{\hat{\g}_{12}}^{2}+\lambda\mathbb{E}\big[\big|\hat{\boldsymbol{g}}_{12}\big|^{2}\big]+N\big)\Big]\nonumber \\
 &  & \hphantom{\brac{1-\frac{2}{T}}\Big(}+\mathbb{E}\Big[\log\big(N+\lambda\abs{\hat{\g}_{11}}^{2}+\lambda\mathbb{E}\big[\big|\hat{\boldsymbol{g}}_{21}\big|^{2}\big]\big)\Big]-r_{1}'-r'_{2}\Big),\\
R_{1}+R_{2} & \leq & \brac{1-\frac{2}{T}}\Big(\mathbb{E}\Big[\log\big(P\abs{\hat{\g}_{11}}^{2}+\brac{P-\lambda}\abs{\hat{\g}_{21}}^{2}+\lambda\mathbb{E}\big[\big|\hat{\boldsymbol{g}}_{21}\big|^{2}\big]+N\big)\Big]\nonumber \\
 &  & \hphantom{\brac{1-\frac{2}{T}}\Big(}+\mathbb{E}\Big[\log\big(N+\lambda\abs{\hat{\g}_{22}}^{2}+\lambda\mathbb{E}\big[\big|\hat{\boldsymbol{g}}_{12}\big|^{2}\big]\big)\Big]-r_{1}'-r'_{2}\Big),\\
R_{1}+R_{2} & \leq & \brac{1-\frac{2}{T}}\Big(\mathbb{E}\Big[\log\big(\lambda\abs{\hat{\g}_{11}}^{2}+\brac{P-\lambda}\abs{\hat{\g}_{21}}^{2}+\lambda\mathbb{E}\big[\big|\hat{\boldsymbol{g}}_{21}\big|^{2}\big]+N\big)\Big]-r_{1}'\nonumber \\
 &  & \hphantom{\brac{1-\frac{2}{T}}\Big(}+\mathbb{E}\Big[\log\big(\lambda\abs{\hat{\g}_{22}}^{2}+\brac{P-\lambda}\abs{\hat{\g}_{12}}^{2}+\lambda\mathbb{E}\big[\big|\hat{\boldsymbol{g}}_{12}\big|^{2}\big]+N\big)\Big]\nonumber \\
 &  & \hphantom{\brac{1-\frac{2}{T}}\Big(}-r_{2}'\Big)\\
2R_{1}+R_{2} & \leq & \brac{1-\frac{2}{T}}\Big(\mathbb{E}\Big[\log\big(P\abs{\hat{\g}_{11}}^{2}+\brac{P-\lambda}\abs{\hat{\g}_{21}}^{2}+\lambda\mathbb{E}\big[\big|\hat{\boldsymbol{g}}_{21}\big|^{2}\big]+N\big)\Big]\nonumber \\
 &  & \hphantom{\brac{1-\frac{2}{T}}\Big(}+\mathbb{E}\Big[\log\big(N+\lambda\abs{\hat{\g}_{11}}^{2}+\lambda\mathbb{E}\big[\big|\hat{\boldsymbol{g}}_{21}\big|^{2}\big]\big)\Big]\nonumber \\
 &  & \hphantom{\brac{1-\frac{2}{T}}\Big(}+\mathbb{E}\Big[\log\big(\lambda\abs{\hat{\g}_{22}}^{2}+\brac{P-\lambda}\abs{\hat{\g}_{12}}^{2}+\lambda\mathbb{E}\big[\big|\hat{\boldsymbol{g}}_{12}\big|^{2}\big]+N\big)\Big]\nonumber \\
 &  & \hphantom{\brac{1-\frac{2}{T}}\Big(}-2r_{1}'-r'_{2}\Big),\\
2R_{2}+R_{1} & \leq & \brac{1-\frac{2}{T}}\Big(\mathbb{E}\Big[\log\big(P\abs{\hat{\g}_{22}}^{2}+\brac{P-\lambda}\abs{\hat{\g}_{12}}^{2}+\lambda\mathbb{E}\big[\big|\hat{\boldsymbol{g}}_{12}\big|^{2}\big]+N\big)\Big]\nonumber \\
 &  & \hphantom{\brac{1-\frac{2}{T}}\Big(}+\mathbb{E}\Big[\log\big(N+\lambda\abs{\hat{\g}_{22}}^{2}+\lambda\mathbb{E}\big[\big|\hat{\boldsymbol{g}}_{12}\big|^{2}\big]\big)\Big]\nonumber \\
 &  & \hphantom{\brac{1-\frac{2}{T}}\Big(}+\mathbb{E}\Big[\log\big(\lambda\abs{\hat{\g}_{11}}^{2}+\brac{P-\lambda}\abs{\hat{\g}_{21}}^{2}+\lambda\mathbb{E}\big[\big|\hat{\boldsymbol{g}}_{21}\big|^{2}\big]+N\big)\Big]\nonumber \\
 &  & \hphantom{\brac{1-\frac{2}{T}}\Big(}-2r_{2}'-r'_{1}\Big).
\end{eqnarray}
\end{subequations} Using Fact \ref{fact:Jensens_gap}, the above
rate region yields the following prelog region \begin{subequations}\label{eq:gdof_no_fb_with_training}
\begin{eqnarray}
d_{1} & \leq & \brac{1-2/T},\\
d_{2} & \leq & \brac{1-2/T},\\
d_{1}+d_{2} & \leq & \brac{1-2/T}\brac{\max\brac{1,\alpha}+\max\brac{1-\alpha,0}},\\
d_{1}+d_{2} & \leq & 2\brac{1-2/T}\max\brac{1-\alpha,\alpha},\\
2d_{1}+d_{2} & \leq & \brac{1-2/T}\brac{\max\brac{1,\alpha}+\max\brac{1-\alpha,\alpha}+\max\brac{1-\alpha,0}}\\
d_{1}+2d_{2} & \leq & \brac{1-2/T}\brac{\max\brac{1,\alpha}+\max\brac{1-\alpha,\alpha}+\max\brac{1-\alpha,0}}.
\end{eqnarray}
\end{subequations} Note that $\lgbrac{\mathbb{E}\big[\big|\hat{\boldsymbol{g}}_{ij}\big|^{2}\big]}\eqdof\lgbrac{\mathbb{E}\big[\big|\boldsymbol{g}_{ij}\big|^{2}\big]}$
for $i,j\in\cbrac{1,2}$ and $\lgbrac N\eqdof0$. The region (\ref{eq:gdof_no_fb_with_training})
can be simplified to obtain the region described in Table \ref{tab:gdof_no_FB_with_training}
of Theorem \ref{thm:training_noFB}.

\section{Training-Based Rate-Splitting Scheme for the Noncoherent IC with
Feedback\label{app:Trainin_Scheme_FB}}

We use a block Markov scheme similar to that in Theorem \ref{thm:noncoh_IC_FB}
and \cite[Lemma 1]{suh_tse_fb_gaussian}, but we include training
symbols in the scheme described here.

\noindent \textbf{Encoding: }Fix a joint distribution $p\brac{\U_{1}}p\brac{\U_{2}}p\brac{\X_{1}\middle|\U_{1}}p\brac{\X_{2}\middle|\U_{2}}$
where $\U_{1},\U_{2},\X_{1},\X_{2}$ are vectors of length $T-2$.
Generate $2^{NT\brac{2R_{\text{c}1}+R_{\text{c}2}}}$ codewords $\U_{1}^{N}\brac{i,j,k}$
with $i,k\in\cbrac{1,\ldots,2^{NTR_{\text{c}1}}}$, $j\in\cbrac{1,\ldots,2^{NTR_{\text{c}2}}}$
according to $\prod_{l=1}^{N}p\brac{\U_{1(l)}}$. For each codeword
$\U_{1}^{N}\brac{i,j,k}$, generate $2^{NTR_{\text{p}1}}$ codewords
$\X_{1}^{N}\brac{i,j,k,l}$ with $l\in\cbrac{1,\ldots,2^{NTR_{\text{p}1}}}$
according to $\prod_{l=1}^{N}p\brac{\X_{1(l)}\middle|\U_{1(l)}}$.

Similarly generate $2^{NT\brac{2R_{\text{c}2}+R_{\text{c}1}}}$ codewords
$\U_{2}^{N}\brac{i,j,r}$ with $i,r\in\cbrac{1,\ldots,2^{NTR_{\text{c}2}}}$,
$j\in\cbrac{1,\ldots,2^{NTR_{\text{c}1}}}$. For each codeword $\U_{2}^{N}\brac{i,j,r}$,
generate $2^{NTR_{\text{p}2}}$ codewords $\X_{2}^{N}\brac{i,j,r,s}$
with $s\in\cbrac{1,\ldots,2^{NTR_{\text{p}2}}}$ according to $\prod_{s=1}^{N}p\brac{\X_{2(s)}\middle|\U_{2(s)}}$.

At block 1, transmitter 1 has uniformly random messages $w_{\text{c}1}^{\brac 1}\in\cbrac{1,\ldots,2^{NTR_{\text{c}1}}},w_{\text{p}1}^{\brac 1}\in\cbrac{1,\ldots,2^{NTR_{\text{p}1}}}$
to transmit and transmitter 2 has uniformly random messages $w_{\text{c}2}^{\brac 1}\in\cbrac{1,\ldots,2^{NTR_{\text{c}2}}},w_{\text{p}2}^{\brac 1}\in\cbrac{1,\ldots,2^{NTR_{\text{p}2}}}$
to transmit. Transmitter 1 obtains $\X_{1}^{N}\big(1,1,w_{\text{c}1}^{(1)},w_{\text{p}1}^{(1)}\big)$
and transmits $\tilde{\X}_{1}^{N}$ created from it with
\[
\tilde{\X}_{1}^{N}=\sbrac{\sqrt{P},0,\X_{\text{1}}\brac 1},\ldots\sbrac{\sqrt{P},0,\X_{\text{1}}\brac k},\ldots\sbrac{\sqrt{P},0,\X_{\text{1}}\brac N}
\]
where each of $\X_{\text{1}}\brac k$ is a vector of length $T-2$.
Effectively transmitter 1 is sending a pilot symbol with $\sqrt{P}$
value at the beginning of every $T$ symbols. Similarly transmitter
2 obtains $\X_{2}^{N}\big(1,1,w_{\text{c}2}^{b},w_{\text{p}2}^{b}\big)$
and transmits the symbols $\tilde{\X}_{2}^{N}$ created from it with
pilot symbols added at the beginning with
\[
\tilde{\X}_{2}^{N}=\sbrac{0,\sqrt{P},\X_{\text{2}}\brac 1},\ldots\sbrac{0,\sqrt{P},\X_{\text{2}}\brac k},\ldots\sbrac{0,\sqrt{P},\X_{\text{2}}\brac N}.
\]

At receiver 1, using pilot symbols, in $k^{\text{th}}$ set of $T$
symbols we get $\boldsymbol{y}_{11,\text{train}}(k)=\sqrt{P}\boldsymbol{g}_{11}(k)+\boldsymbol{z}_{11}(k)$,
$\boldsymbol{y}_{12,\text{train}}(k)=\sqrt{P}\boldsymbol{g}_{21}(k)+\boldsymbol{z}_{21}(k),$
and the MMSE estimates $\hat{\boldsymbol{g}}_{11}$, $\hat{\boldsymbol{g}}_{21}$
can be obtained at receiver 1. Similarly $\hat{\boldsymbol{g}}_{22}$,
$\hat{\boldsymbol{g}}_{12}$ can be obtained at receiver 2. The details
of MMSE estimation is same as that in Appendix \ref{app: Training Scheme NoFB}.
We call $\underline{\hat{\boldsymbol{g}}_{1}}=[\hat{\boldsymbol{g}}_{11},\hat{\boldsymbol{g}}_{21}]$
and $\underline{\hat{\boldsymbol{g}}_{2}}=[\hat{\boldsymbol{g}}_{22},\hat{\boldsymbol{g}}_{12}]$.
We use the notation $\Y_{1}^{N},\Y_{2}^{N}$ to indicate received
symbols containing data and not training symbol:
\begin{equation}
\Y_{1}^{N}=\boldsymbol{g}_{11}^{N}\X_{1}^{N}+\boldsymbol{g}_{21}^{N}\X_{2}^{N}+\boldsymbol{Z}_{1}^{N},
\end{equation}
\begin{equation}
\Y_{2}^{N}=\boldsymbol{g}_{12}^{N}\X_{1}^{N}+\boldsymbol{g}_{22}^{N}\X_{2}^{N}+\boldsymbol{Z}_{2}^{N}.
\end{equation}

At block $b>1$, transmitter 1 has uniformly random messages $w_{\text{c}1}^{\brac b}\in\sbrac{1,2^{NTR_{\text{c}1}}},w_{\text{p}1}^{\brac b}\in\sbrac{1,2^{NTR_{\text{p}1}}}$
to transmit and transmitter 2 has uniformly random messages $w_{\text{c}2}^{\brac b}\in\sbrac{1,2^{NTR_{\text{c}2}}},w_{\text{p}2}^{\brac b}\in\sbrac{1,2^{NTR_{\text{p}2}}}$
to transmit. Transmitter 1 obtains the feedback $\Y_{1}^{N,\brac{b-1}},\underline{\hat{\boldsymbol{g}}_{1}}^{N,(b-1)}$
from receiver 1. Transmitter 1 tries to decode $\hat{w}_{2c}^{\brac{b-1}}=\hat{k}$
from transmitter 2 by finding unique $\hat{k}$ such that
\begin{align*}
 & \Big(\U_{1}^{N}\big(w_{\text{c}1}^{\brac{b-2}},w_{\text{c}2}^{\brac{b-2}},w_{\text{c}1}^{\brac{b-1}}\big),\X_{1}^{N}\big(w_{\text{c}1}^{\brac{b-2}},w_{\text{c}2}^{\brac{b-2}},w_{\text{c}1}^{\brac{b-1}},w_{\text{p}1}^{\brac{b-1}}\big),\\
 & \quad\U_{2}^{N}\big(w_{\text{c}2}^{\brac{b-2}},w_{\text{c}1}^{\brac{b-2}},\hat{k}\big),\Y_{1}^{N,\brac{b-1}},\underline{\hat{\boldsymbol{g}}_{1}}^{N,(b-1)}\Big)\in\mathcal{A}_{\epsilon}^{\brac N}.
\end{align*}
 where $A_{\epsilon}^{\brac N}$ indicates the set of jointly typical
sequences. Transmitter 1 already knows $w_{\text{c}1}^{\brac{b-2}},w_{\text{c}1}^{\brac{b-1}},w_{\text{p}1}^{\brac{b-1}}$.
Also $w_{\text{c}2}^{\brac{b-2}}$ is assumed to be correctly decoded
in the previous block at transmitter 1 and $w_{\text{c}1}^{\brac{b-2}}$
is assumed to be correctly decoded in the previous block at transmitter
2. Based on $\hat{w}_{2c}^{\brac{b-1}}$, transmitter 1 obtains $\X_{1}^{N}\big(w_{\text{c}1}^{\brac{b-1}},\hat{w}_{2c}^{\brac{b-1}},w_{\text{c}1}^{(b)},w_{\text{p}1}^{(b)}\big)$
and transmits $\tilde{\X}_{1}^{N}$ created from it as
\[
\tilde{\X}_{1}^{N}=\sbrac{\sqrt{P},0,\X_{\text{1}}\brac 1},\ldots\sbrac{\sqrt{P},0,\X_{\text{1}}\brac k},\ldots\sbrac{\sqrt{P},0,\X_{\text{1}}\brac N}
\]
where each of $\X_{\text{1}}\brac k$ is a vector of length $T-2$.
Similarly transmitter 2 decodes $\hat{w}_{1c}^{\brac{b-1}}$, obtains
$\X_{2}^{N}\big(w_{\text{c}2}^{\brac{b-1}},\hat{w}_{1c}^{\brac{b-1}},w_{\text{c}2}^{(b)},w_{\text{p}2}^{(b)}\big)$
and transmits the symbols $\tilde{\X}_{2}^{N}$ created from it with
pilot symbols added at the beginning as
\[
\tilde{\X}_{1}^{N}=\sbrac{0,\sqrt{P},\X_{\text{1}}\brac 1},\ldots\sbrac{0,\sqrt{P},\X_{\text{1}}\brac k},\ldots\sbrac{0,\sqrt{P},\X_{\text{1}}\brac N}.
\]
 The messages transmitted at $b=B$ can be set to be fixed and known
to the receivers to facilitate decoding.

\noindent \textbf{Decoding:} After receiving $B$ blocks, each receiver
performs backward decoding. At receiver 1, block $b$ is decoded assuming
block $b+1$ is correctly decoded. It finds unique triplet $\brac{\hat{i},\hat{j},\hat{l}}$
such that
\[
\brac{\U_{1}^{N}\big(\hat{i},\hat{j},w_{\text{c}1}^{\brac b}\big),\X_{1}^{N}\big(\hat{i},\hat{j},w_{\text{c}1}^{\brac b},\hat{l}\big),\U_{2}^{N}\big(\hat{j,}\hat{i},w_{\text{c}2}^{\brac b}\big),\Y_{1}^{N,\brac b},\underline{\hat{\boldsymbol{g}}_{1}}^{N,(b)}}\in\mathcal{A}_{\epsilon}^{\brac N}.
\]
Similarly receiver 2 finds unique triplet $\brac{\hat{j},\hat{i},\hat{s}}$
such that
\[
\brac{\U_{2}^{N}\big(\hat{j},\hat{i},w_{\text{c}1}^{\brac b}\big),\X_{2}^{N}\big(\hat{j},\hat{i},w_{\text{c}2}^{\brac b},\hat{s}\big),\U_{1}^{N}\big(\hat{i},\hat{j,}w_{\text{c}1}^{\brac b}\big),\Y_{2}^{N,\brac b},\underline{\hat{\boldsymbol{g}}_{2}}^{N,(b)}}\in\mathcal{A}_{\epsilon}^{\brac N}.
\]

We obtain the following rate region similar to that in Theorem \ref{thm:noncoh_IC_FB}:\begin{subequations}\label{eq:ach_fb_with_training}
\begin{eqnarray}
TR_{1} & \leq & I\brac{\X_{1},\U_{2};\Y_{1},\underline{\hat{\boldsymbol{g}}_{1}}},\\
TR_{1} & \leq & I\brac{\U_{1};\Y_{2},\underline{\hat{\boldsymbol{g}}_{2}}\middle|\X_{2}}+I\brac{\X_{1};\Y_{1},\underline{\hat{\boldsymbol{g}}_{1}}\middle|\U_{1},\U_{2}},\\
TR_{2} & \leq & I\brac{\X_{2},\U_{1};\Y_{2},\underline{\hat{\boldsymbol{g}}_{2}}},\\
TR_{2} & \leq & I\brac{\U_{2};\Y_{1},\underline{\hat{\boldsymbol{g}}_{1}}\middle|\X_{1}}+I\brac{\X_{2};\Y_{2},\underline{\hat{\boldsymbol{g}}_{2}}\middle|\U_{1},\U_{2}},\\
T\brac{R_{1}+R_{2}} & \leq & I\brac{\X_{1};\Y_{1},\underline{\hat{\boldsymbol{g}}_{1}}\middle|\U_{1},\U_{2}}+I\brac{\X_{2},\U_{1};\Y_{2},\underline{\hat{\boldsymbol{g}}_{2}}},\\
T\brac{R_{1}+R_{2}} & \leq & I\brac{\X_{2};\Y_{2},\underline{\hat{\boldsymbol{g}}_{2}}\middle|\U_{1},\U_{2}}+I\brac{\X_{1},\U_{2};\Y_{1},\underline{\hat{\boldsymbol{g}}_{1}}}.
\end{eqnarray}
\end{subequations}

We choose $\U_{k}$ as a vector of length $T-2$ with i.i.d. $\mathcal{CN}\brac{0,\lambda_{\text{c}}}$
elements and $\X_{\text{p}k}$ as a vector of length $T-2$ with i.i.d.
$\mathcal{CN}\brac{0,\lambda_{\text{p}}}$ elements for $k\in\cbrac{1,2}$.
The random variables are chosen independent of each other so that
the set $\cbrac{\U_{1},\U_{2},\X_{\text{p}1},\X_{\text{p}2}}$ is
mutually independent. We use $\X_{1}=\U_{1}+\X_{\text{p}1},\quad\X_{2}=\U_{2}+\X_{\text{p}2}$
where $\lambda_{\text{c}}+\lambda_{\text{p}}=P$ and $\lambda_{\text{p}}=\min\brac{1/\inr,P}$.
The random variables in the single-letter form in (\ref{eq:ach_fb_with_training})
has the same distribution as that in (\ref{eq:ach_nofb_training}).
Hence, we can use the lower bounds for the terms in (\ref{eq:ach_nofb_training})
from Table \ref{tab:lower-bounds-training-nofb} directly to the terms
in (\ref{eq:ach_fb_with_training}). The term $I\brac{\U_{2};\Y_{1},\underline{\hat{\boldsymbol{g}}_{1}}\middle|\X_{1}}$
is not available in Table \ref{tab:lower-bounds-training-nofb}, and
hence we analyze it further in the following subsection.

\subsection{Analysis of the Term $I\protect\brac{\protect\U_{2};\protect\Y_{1},\underline{\hat{\boldsymbol{g}}_{1}}\middle|\protect\X_{1}}$}

We have
\begin{eqnarray*}
I\brac{\U_{2};\Y_{1},\underline{\hat{\boldsymbol{g}}_{1}}\middle|\X_{1}} & = & I\brac{\U_{2};\Y_{1}\middle|\X_{1},\underline{\hat{\boldsymbol{g}}_{1}}}\\
 & = & I\brac{\boldsymbol{U}_{2};\hat{\boldsymbol{g}}_{11}\boldsymbol{X}_{1}+\hat{\boldsymbol{g}}_{21}\boldsymbol{U}_{2}+\hat{\boldsymbol{g}}_{21}\X_{\text{p}2}+\hat{\Z}_{1}\middle|\underline{\hat{\boldsymbol{g}}_{1}},\X_{1}}\\
 & = & I\brac{\boldsymbol{U}_{2};\hat{\boldsymbol{g}}_{21}\boldsymbol{U}_{2}+\hat{\boldsymbol{g}}_{21}\X_{\text{p}2}+\hat{\Z}_{1}\middle|\underline{\hat{\boldsymbol{g}}_{1}},\X_{1}}\\
 & \geq & I\brac{\boldsymbol{U}_{2};\hat{\boldsymbol{g}}_{21}\boldsymbol{U}_{2}+\hat{\boldsymbol{g}}_{21}\X_{\text{p}2}+\hat{\Z}_{1}\middle|\underline{\hat{\boldsymbol{g}}_{1}}}\\
 & \geq & \brac{T-2}\Big(\expect{\lgbrac{\brac{P-\lambda}\abs{\hat{\g}_{21}}^{2}+\lambda\expect{\abs{\hat{\boldsymbol{g}}_{21}}^{2}}+N}}\\
 &  & \hphantom{\brac{T-2}\Big(}-\lgbrac{\lambda\expect{\abs{\hat{\boldsymbol{g}}_{21}}^{2}}+N}\Big)
\end{eqnarray*}
where the last step is by replacing $\hat{\boldsymbol{g}}_{21}\X_{\text{p}2}+\hat{\Z}_{1}$
with the worst noise and
\[
N=\frac{P\expect{\abs{\boldsymbol{g}_{11}}^{2}}}{1+P\expect{\abs{\boldsymbol{g}_{11}}^{2}}}+\frac{P\expect{\abs{\boldsymbol{g}_{21}}^{2}}}{1+P\expect{\abs{\boldsymbol{g}_{21}}^{2}}}+1,
\]
similar to that in Appendix \ref{subsec:training_term1}.

\subsection{Simplified Rate Region}

Using the result from the previous subsection and the results from
Table \ref{tab:lower-bounds-training-nofb} in \ref{eq:ach_fb_with_training},
we obtain that the following rate region is achievable:\begin{subequations}
\begin{eqnarray}
TR_{1} & \leq & \brac{1-\frac{2}{T}}\brac{\mathbb{E}\Big[\log\big(P\abs{\hat{\g}_{11}}^{2}+\brac{P-\lambda}\abs{\hat{\g}_{21}}^{2}+\lambda\mathbb{E}\big[\big|\hat{\boldsymbol{g}}_{21}\big|^{2}\big]+N\big)\Big]-r_{1}'},\\
TR_{1} & \leq & \brac{1-\frac{2}{T}}\Big(\mathbb{E}\Big[\log\big(\brac{P-\lambda}\abs{\hat{\g}_{12}}^{2}+\lambda\mathbb{E}\big[\big|\hat{\boldsymbol{g}}_{12}\big|^{2}\big]+N\big)\Big]\nonumber \\
 &  & \hphantom{\brac{1-\frac{2}{T}}\Big(}+\mathbb{E}\Big[\log\big(N+\lambda\abs{\hat{\g}_{11}}^{2}+\lambda\mathbb{E}\big[\big|\hat{\boldsymbol{g}}_{21}\big|^{2}\big]\big)\Big]-r_{1}'-r'_{2}\Big),\\
TR_{2} & \leq & \brac{1-\frac{2}{T}}\brac{\mathbb{E}\Big[\log\big(P\abs{\hat{\g}_{22}}^{2}+\brac{P-\lambda}\abs{\hat{\g}_{12}}^{2}+\lambda\mathbb{E}\big[\big|\hat{\boldsymbol{g}}_{12}\big|^{2}\big]+N\big)\Big]-r_{2}'},\\
TR_{2} & \leq & \brac{1-\frac{2}{T}}\Big(\mathbb{E}\Big[\log\big(\brac{P-\lambda}\abs{\hat{\g}_{21}}^{2}+\lambda\mathbb{E}\big[\big|\hat{\boldsymbol{g}}_{21}\big|^{2}\big]+N\big)\Big]\nonumber \\
 &  & \hphantom{\brac{1-\frac{2}{T}}\Big(}+\mathbb{E}\Big[\log\big(N+\lambda\abs{\hat{\g}_{22}}^{2}+\lambda\mathbb{E}\big[\big|\hat{\boldsymbol{g}}_{12}\big|^{2}\big]\big)\Big]-r_{1}'-r'_{2}\Big),\\
T\brac{R_{1}+R_{2}} & \leq & \brac{1-\frac{2}{T}}\Big(\mathbb{E}\Big[\log\big(N+\lambda\abs{\hat{\g}_{11}}^{2}+\lambda\mathbb{E}\big[\big|\hat{\boldsymbol{g}}_{21}\big|^{2}\big]\big)\Big]\nonumber \\
 &  & \hphantom{\brac{1-\frac{2}{T}}\Big(}+\mathbb{E}\Big[\log\big(P\abs{\hat{\g}_{22}}^{2}+\brac{P-\lambda}\abs{\hat{\g}_{12}}^{2}+\lambda\mathbb{E}\big[\big|\hat{\boldsymbol{g}}_{12}\big|^{2}\big]+N\big)\Big]\nonumber \\
 &  & \hphantom{\brac{1-\frac{2}{T}}\Big(}-r_{1}'-r'_{2}\Big),\\
T\brac{R_{1}+R_{2}} & \leq & \brac{1-\frac{2}{T}}\Big(\mathbb{E}\Big[\log\big(N+\lambda\abs{\hat{\g}_{22}}^{2}+\lambda\mathbb{E}\big[\big|\hat{\boldsymbol{g}}_{12}\big|^{2}\big]\big)\Big]\nonumber \\
 &  & \hphantom{\brac{1-\frac{2}{T}}\Big(}+\mathbb{E}\Big[\log\big(P\abs{\hat{\g}_{11}}^{2}+\brac{P-\lambda}\abs{\hat{\g}_{21}}^{2}+\lambda\mathbb{E}\big[\big|\hat{\boldsymbol{g}}_{21}\big|^{2}\big]+N\big)\Big]\nonumber \\
 &  & \hphantom{\brac{1-\frac{2}{T}}\Big(}-r_{1}'-r'_{2}\Big).
\end{eqnarray}
\end{subequations}where $r_{1}'=\log\big(\lambda\mathbb{E}\big[\big|\hat{\boldsymbol{g}}_{21}\big|^{2}\big]+N\big)\Big]$,
$r_{2}'=\log\big(\lambda\mathbb{E}\big[\big|\hat{\boldsymbol{g}}_{12}\big|^{2}\big]+N\big)\Big]$.
Using Fact \ref{fact:Jensens_gap}, the above rate region yields the
following prelog region \begin{subequations}\label{eq:training_dof_fb}
\begin{eqnarray}
d_{1} & \leq & \brac{1-2/T}\max\brac{1,\alpha}\\
d_{2} & \leq & \brac{1-2/T}\max\brac{1,\alpha}\\
d_{1}+d_{2} & \leq & \brac{1-2/T}\brac{\max\brac{1,\alpha}+\max\brac{1-\alpha,0}}.
\end{eqnarray}
\end{subequations} Note that $\lgbrac{\mathbb{E}\big[\big|\hat{\boldsymbol{g}}_{ij}\big|^{2}\big]}\eqdof\lgbrac{\mathbb{E}\big[\big|\boldsymbol{g}_{ij}\big|^{2}\big]}$
for $i,j\in\cbrac{1,2}$ and $\lgbrac N\eqdof0$.

The region (\ref{eq:training_dof_fb}) can be simplified to obtain
the region described in Table \ref{tab:gdof_FB_with_training} of
Theorem \ref{thm:training_FB}.

\section{Numerical Calculations for the TDM scheme\label{app:Numerical-Calculation}}

Here we provide the calculations required for numerically evaluating
the achievable rates for the TDM scheme given in Table \ref{tab:Comparison-of-rates}.
We operate the first transmitter-receiver pair during half of the
time, while the second pair one remains OFF. During the other half
of the time, the second transmitter-receiver pair operates and the
first pair remains OFF. Here, we just have point-to-point channels
and we use one symbol for training each point-to-point channel. Note
that in the calculations below, the channels contain the power scaling;\emph{
}the transmit symbols and noise are of unit power. For receiver 1,
we receive $\boldsymbol{y}_{11,\text{train}}=\sqrt{P}\boldsymbol{g}_{11}+\boldsymbol{z}_{11}$
during training and we have the MMSE estimate for the channel as
\begin{eqnarray*}
\hat{\boldsymbol{g}}_{11} & = & \frac{\sqrt{P}\expect{\abs{\boldsymbol{g}_{11}}^{2}}}{1+P\expect{\abs{\boldsymbol{g}_{11}}^{2}}}\boldsymbol{y}_{11,\text{train}}\\
 & = & \sqrt{P}\expect{\abs{\boldsymbol{g}_{11}}^{2}}\frac{\sqrt{P}\boldsymbol{g}_{11}+\boldsymbol{z}_{11}}{1+P\expect{\abs{\boldsymbol{g}_{11}}^{2}}}.
\end{eqnarray*}
The total noise at receiver 1 including MMSE is
\begin{eqnarray*}
N_{\text{1,TDM}}=N_{\text{TDM}} & = & \text{\ensuremath{\expect{\brac{\boldsymbol{g}_{11}-\sqrt{P}\expect{\abs{\boldsymbol{g}_{11}}^{2}}\frac{\sqrt{P}\boldsymbol{g}_{11}+\boldsymbol{z}_{11}}{1+P\expect{\abs{\boldsymbol{g}_{11}}^{2}}}}^{2}P}}+1}\\
 & = & \expect{\abs{\frac{\boldsymbol{g}_{11}}{1+P\expect{\abs{\boldsymbol{g}_{11}}^{2}}}}^{2}}P+\frac{P\expect{\abs{\boldsymbol{g}_{11}}^{2}}^{2}}{\abs{1+P\expect{\abs{\boldsymbol{g}_{11}}^{2}}}^{2}}P+1\\
 & = & \frac{P\expect{\abs{\boldsymbol{g}_{11}}^{2}}}{1+P\expect{\abs{\boldsymbol{g}_{11}}^{2}}}+1.
\end{eqnarray*}
The terms for receiver 2 are similar. Using symmetry of the statistics,
the achievable rates are calculated as
\begin{align*}
R_{\text{1}}=R_{\text{2}}=\frac{1}{2}\brac{1-\frac{1}{T}} & \expect{\lgbrac{1+\frac{P\abs{\hat{\boldsymbol{g}}_{11}}^{2}}{N_{\text{TDM}}}}}.
\end{align*}

\end{document}